\documentclass[preprintnumbers,nofootinbib,
superscriptaddress,amsmath,amssymb,floatfix,prd,10pt]{revtex4}

\usepackage[english]{babel}
\usepackage{graphicx}
\usepackage{bm}
\usepackage{epsf}
\usepackage{amssymb,amsmath}

\usepackage{color}

\usepackage[normalem]{ulem}  

\newcommand{\be}{\begin{equation}}
\newcommand{\ee}{\end{equation}}
\newcommand{\bea}{\begin{eqnarray}}
\newcommand{\eea}{\end{eqnarray}}
\newcommand{\non}{\nonumber\\}
\newcommand{\vg}{\bm{\gamma}}

\newcommand{\Tr}{{\rm Tr}}

\begin{document}

\title{Reaction rates and transport in neutron stars}
\author{Andreas Schmitt}
\affiliation{Mathematical Sciences and STAG Research Centre, University of Southampton, Southampton SO17 1BJ, United Kingdom}
\author{Peter Shternin}
\affiliation{Ioffe Institute, Politekhnicheskaya 26, St. Petersburg, 194021, Russia}

\date{29 October 2018}

\begin{abstract}
Understanding signals from neutron stars requires knowledge about the transport inside the star. We review the transport properties and the underlying reaction rates of dense hadronic and quark matter in the 
crust and the core of neutron stars and point out open problems and future directions.  
\end{abstract}

\maketitle

\tableofcontents

\section{Introduction}\label{S:intro}

\subsection{Context}

Transport describes how conserved quantities such as energy, momentum, particle number, or electric charge are transferred from one region to another. 
Such a transfer occurs if the system is out of equilibrium, for instance through a temperature gradient or a non-uniform chemical composition. Different theoretical methods 
are used to understand transport, depending on how far the system is away from its equilibrium state.  If the system is close to equilibrium locally and perturbations are on large scales in space and time, hydrodynamics  is a  powerful technique. Further away from equilibrium other techniques are required, for example kinetic theory, which can also be used to provide the 
transport coefficients needed in the hydrodynamic equations. In any case, transport is determined by interactions on a microscopic level, and it is the resulting transport 
properties that we are concerned with in this review. 

Signals from neutron stars are sensitive to equilibrium properties such as the equation of state but also, to a large extent, to transport properties
-- here, by neutron stars we mean all objects with a radius of about $10\,{\rm km}$ and a mass of about 1-2 solar masses, including the possibilities
of hybrid stars, which have a quark matter core, and pure quark stars. Therefore, understanding transport is crucial to interpret astrophysical observations, and, turning the argument around,
we can use astrophysical observations to improve our understanding of transport in dense matter and thus ultimately our understanding of the microscopic interactions.

Transport properties are most commonly computed from particle collisions. (Although, in strongly coupled systems, the picture of well-defined particles scattering off each other has to be taken with care.)
These can be scattering processes in which energy and momentum is exchanged without changing the chemical 
composition of the system, or these can be flavor-changing processes from the electroweak interaction. Understanding transport thus amounts to understanding the rates of these processes, as 
a function of temperature and density. Electroweak processes are well understood, but large uncertainties arise if the strong interaction is involved in a reaction that contributes 
to transport. Therefore, approximations such as weak-coupling techniques or one-pion exchange for nucleon-nucleon collisions are being used, and  
efforts in current research aim at improving these approximations. 

In a neutron star, most of the particles involved in these processes are fermions: electrons, muons, neutrinos, neutrons, protons, hyperons, and quarks. Since the Fermi momenta of these fermions are typically much larger than the temperature (neutrinos are an exception), transport probes the excitations in small vicinities of the corresponding 
Fermi surfaces. (It can also probe the values of the Fermi momenta themselves since momentum conservation of a given reaction imposes a constraint on them.) Some of the processes we discuss involve bosonic excitations, for instance the lattice phonons in the crust, the superfluid mode, or mesons such as pions and kaons. Typically, their contribution
is smaller because, well, they do not have a Fermi surface and thus the rates and transport coefficients contain higher powers of temperature. Therefore, purely bosonic 
contributions are usually only relevant if the fermionic ones are suppressed, for example through an energy gap from Cooper pairing.

\subsection{Phenomenological and theoretical motivations}

Computing transport properties of matter inside neutron stars is motivated by phenomenological and theoretical considerations. 
The phenomenological motivation is of course to understand astrophysical data that are sensitive to transport. Our focus in the main part of the 
review is on the transport properties themselves, and we discuss observations only in passing. Therefore, let us now list 
some of the relevant phenomena which are intimately connected with transport. 
(Here we only include very few selected references, which we think are useful for further reading; many more references will be given in the main part.)  

\begin{itemize}

\item Oscillatory modes of the star, most importantly $r$-modes, become unstable with respect to the emission of gravitational waves \cite{Andersson:1997xt,Friedman:1997uh}. We know that these instabilities must be damped because otherwise we would not observe fast rotating stars. Viscous damping plays a major role, and knowledge of both bulk and shear viscosity (which 
are important in different temperature regimes) is required \cite{Haskell2015IJMPE}.

\item Pulsar glitches, sudden jumps in the rotation frequency of the star, are commonly explained through pinning and un-pinning of superfluid vortices in the 
inner crust of the star \cite{Haskell:2015jra}. A quantitative treatment requires the understanding of superfluid transport, including entrainment effects of the superfluid in the crust, and possibly hydrodynamical instabilities.

\item The interpretation of thermal radiation of neutron stars depends on knowledge about heat transport in the outermost layers of the star, the atmosphere and the ocean \cite{Potekhin2014Ufn,Potekhin2007Ap&SS, Potekhin:2015qsa}.

\item Cooling of neutron stars, for instance isolated neutron stars and quiescent X-ray
transients, requires understanding of the microscopic neutrino emission processes. Together with thermodynamic properties such as the specific heat and other transport properties such as heat conductivity, the cooling process can be modeled \cite{YakovlevPethick2004ARA,Page:2005fq}.

\item Understanding the time evolution of magnetic fields in neutron stars and its coupling to the thermal evolution requires magnetohydrodynamical simulations.
As an input from microscopic physics electrical and thermal conductivities are needed \cite{Vigano2013MNRAS}. Additional complications may arise from superconductivity and magnetic flux tubes in the 
core.

\item In accreting neutron stars, the crust is forced out of equilibrium by the accreted matter, and in some cases, for instance `quasi-persistent' sources, 
the subsequent relaxation process can be observed in real time. Nuclear reactions, including pycno-nuclear fusion, contribute to the so-called `deep crustal heating' \cite{HaenselZdunik2008A&A},
and transport properties of the crust such as thermal conductivity are needed to understand the relaxation process \cite{Degenaar:2014nja}. An important role is possibly played by 
transport properties of inhomogeneous phases in the crust/core transition region (`nuclear pasta') \cite{Horowitz2015PhRvL}. Deep crustal heating also plays a pivotal role in maintaining high observed temperatures of X-ray transients \cite{Brown1998ApJ}.

\item Crust relaxation is also important for magnetar flares. Similar to accretion, the crust is disrupted, now by a catastrophic rearrangement of the magnetic field. Crustal transport
properties in the presence of a magnetic field become important \cite{2015RPPh...78k6901T}.  

\item The neutron star in the Cassiopeia A supernova remnant has undergone unusually rapid cooling in the past decade \cite{HeinkeHo2010ApJ,Shternin:2010qi,Elshamouty2013ApJ, Ho2015PhRvC}.
If true (the reliability of this data is under discussion \cite{Posselt2013ApJ,Posselt2018ApJ}) this indicates an unusual neutrino emission process, for instance 
Cooper pair breaking and formation at the critical temperature for neutron superfluidity \cite{Shternin:2010qi,Page:2010aw}.

\item Core-collapse supernovae and the evolution of the resulting proto-neutron star are sensitive to neutrino transport and neutrino-nucleus reactions. The 
phenomenological implications include direct neutrino signals \cite{Janka2017}, nucleosynthesis, the mechanism of the supernova explosion itself, cooling of proto-neutron stars, and pulsar kicks \cite{2012PTEP.2012aA309J}.

\item Neutron star mergers have proved to be multi-messenger events, 
emitting detectable gravitational waves and electromagnetic signals \cite{PhysRevLett.119.161101,2041-8205-848-2-L12}.  Simulations of the merger process within general relativity are being 
performed, using (magneto)hydrodynamics, where viscous effects may be important \cite{Alford:2017rxf}. Merger
events explore 
transport at larger temperatures than neutron stars in (near-)equilibrium. Similar to proto-neutron stars from supernovae, the evolution of merger remnants 
requires understanding of neutrino reactions and transport.

\end{itemize}

The theoretical motivation for understanding transport in neutron star matter can -- 
at least for the ultra-dense regions in the interior of the star -- be put in the wider context of understanding 
transport in matter underlying the theory of Quantum Chromodynamics (QCD) or, even more generally speaking, of understanding transport in relativistic, strongly interacting theories. This perspective connects some of the 
results in this review with questions about the correct formulation of relativistic, dissipative (superfluid) hydrodynamics, about the validity of the quasiparticle picture and thus of 
kinetic theory, about non-perturbative effects in QCD scattering processes, about universal results and bounds for shear viscosity and other transport coefficients and so forth. 
These questions are being discussed extensively in the recent and current literature, be it from an abstract theoretical perspective, e.g., within the gauge/gravity correspondence, 
or in a more applied context such as relativistic heavy-ion collisions or cold atomic gases. Neutron stars may appear to be too specific and too complicated to be viewed as a clean 
laboratory for these questions, but we think it is worth pointing out these connections, and they will be touched in some sections of this review. 

\subsection{Purpose and structure of this review}

\begin{figure}[t]
\begin{center}
\includegraphics[width=0.8\textwidth]{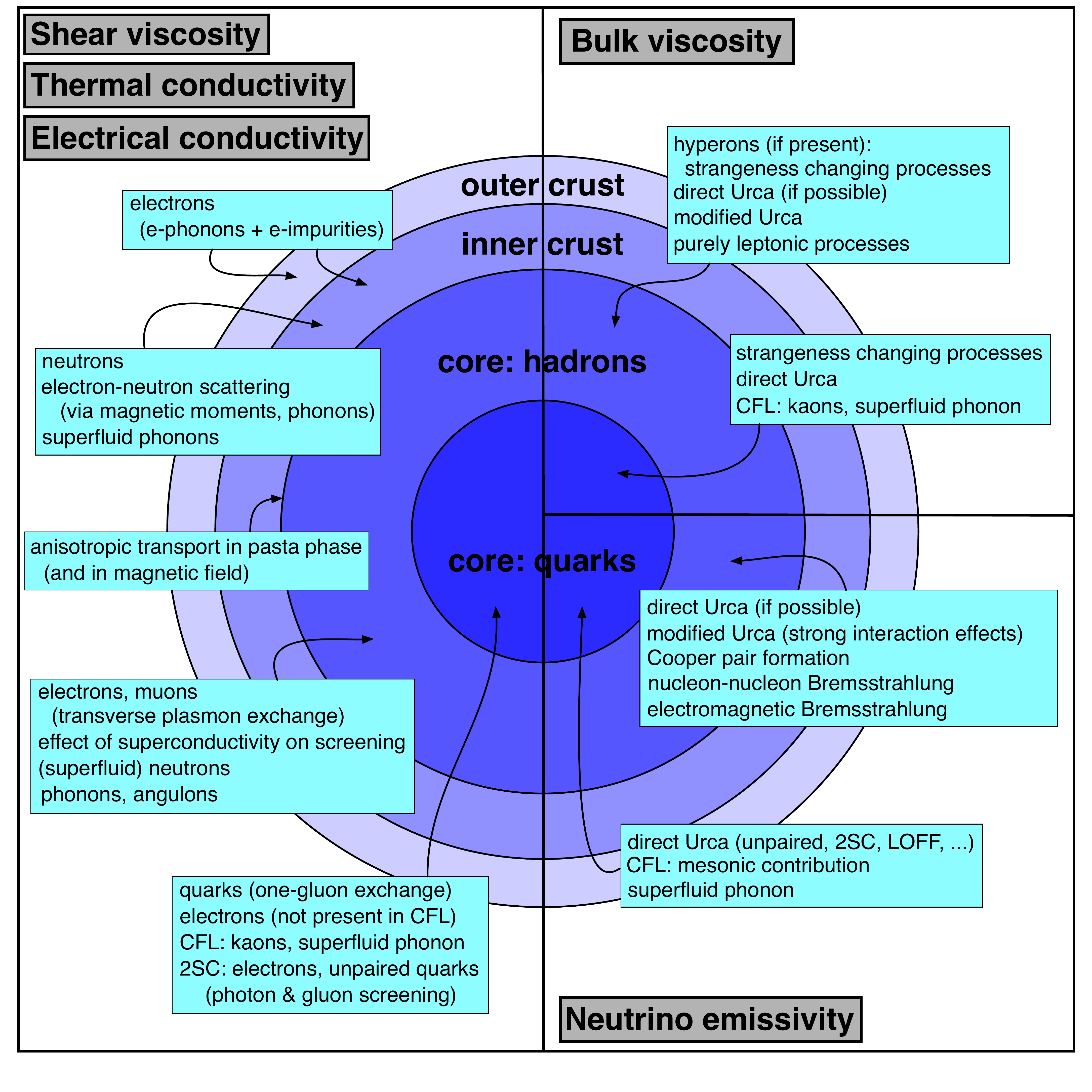}
\caption{Contributions to the main transport coefficients and to  neutrino emission discussed in this review. 
 }
\label{fig:overview}
\end{center}
\end{figure}

We intend to collect and comment on recent results in the literature, pointing out open problems and future directions, with an emphasis on the theoretical, rather than the 
observational, questions. We include pedagogical derivations and explanations in most parts, making this review accessible for non-experts in transport theory and 
neutron star physics. In particular, we start in Sec.\ \ref{S:Gen_kin} by introducing some basic concepts of transport theory and explain how the basic approach must be extended and adjusted to the extreme conditions inside a neutron star. After this introductory section, we have structured the review by moving from the outer layers of the star into the central regions. Since we thereby 
move from low densities to ultra-high densities, we encounter various distinct phases with very distinct transport properties. We start from the crust in Sec.\ \ref{S:crust}, where the matter composition is rather well known: a lattice or a strongly coupled liquid of ions coexists with an electron gas, and, in the inner crust, with a neutron (super)fluid. As we move through the crust/core 
interface, we encounter the so-called nuclear pasta phases, and eventually end up in a region of nuclear matter, composed of neutrons and protons, 
with electrons and muons accounting for charge neutrality.  Additionally, hyperons may be present, and possibly meson condensates. We discuss transport of hadronic matter
in the core in Sec.\ \ref{S:core_nuclear}. At sufficiently large densities, matter becomes deconfined and we enter the quark matter phase. Since the  
density at which this transition happens is unknown, we do not know whether quark matter exists in the 
core of neutron stars (or whether there are pure quark stars). Transport properties of quark matter, which we discuss in Sec.\ \ref{sec:quarks}, are one important ingredient 
to answer that question. For readers unfamiliar with quark matter 
and its possible phases, we have included an introductory section and overview in Secs.\ \ref{sec:quarkremarks} and \ref{sec:quarkover}. 
At the end of Sec.\ \ref{sec:quarks} --  although being a somewhat decoupled topic -- we briefly discuss possible effects of 
quantum anomalies on transport in neutron stars. 
In all sections, our main goal was, besides 
some introductory and pedagogical discussions, to focus on the most recent results and their impact for future research. In some parts, for instance in Sec.\ \ref{sec:quarks}
about quark matter, we have tried to give a more complete overview, including older results, which is possible because of the smaller amount of existing literature compared to 
nuclear matter.
Reaction rates in the core from the weak interaction are discussed in  Secs.~\ref{sec:urca} and \ref{sec:quarkneutrino}.
The rates for these processes are interesting by themselves since they directly feed into the cooling behavior of the star. They are also interesting for the bulk viscosity because bulk viscosity in a neutron star is dominated by chemical re-equilibration and thus by flavor-changing processes. We discuss bulk viscosity, including the rates 
for other leptonic and non-leptonic flavor-changing processes, for hadronic matter in Sec.\ \ref{S:bulk} and for quark matter in Sec.\ \ref{sec:bulkquark}. Shear viscosity, thermal and electrical conductivity, 
are discussed together since they are determined by similar processes, some of which rely on the strong interaction, and we discuss them in 
Secs.\  \ref{S:crust_kincoeff}, \ref{S:core_nuclear_kincoeff}, and \ref{sec:shearquark}. A
schematic overview of the main transport properties and the corresponding microscopic contributions discussed in this review is shown in Fig.\ \ref{fig:overview}. 

\subsection{Related reviews}

There are a number of reviews that (partially) deal with transport properties in neutron stars,
having some overlap with our work, and which we 
recommend for further reading. \citet{PageReddy2012} review transport in the inner crust of the star. A more exhaustive overview of the crust is given by \citet{ChamelHaensel2008LRR},
discussing transport as well as details of the structure and connections to observations. \citet{Potekhin:2015qsa} review cooling of isolated neutron stars and discuss  transport and thermodynamic properties that are needed to understand the cooling process, including the effect of strong magnetic fields. Cooling in proto-neutron stars just after a core collapse supernova explosion has recently been reviewed by \citet{Roberts:2016rsf}. Many of the currently used results for neutrino emissivity in hadronic matter, including superfluid phases, can already be found in the review by \citet{Yakovlev2001physrep}. Superfluidity in neutron stars and some of its effects on transport and reaction rates are reviewed by \citet{Page:2013hxa}. For 
a detailed discussion of many-body techniques for hadronic matter inside neutron stars, including neutrino emission processes, see the review by \citet{Sedrakian2007PrPNP}.
Transport properties of quark matter are discussed in chapter VII of the review about color superconductivity by \citet{Alford:2007xm}, for 
a pedagogical discussion of neutrino emissivity in quark matter see Ref.\ \cite{Schmitt:2010pn}. Our review serves as an update to some of these earlier reviews and has a somewhat different focus than most of them, bringing together theoretical results for transport properties from the crust through nuclear matter in the core up to ultra-dense deconfined quark matter. 

There are several aspects of transport and reaction rates in neutron stars which we do not discuss or only touch very briefly: we will not elaborate on reactions relevant for neutrino transport in supernovae \cite{Burrows2006NuPhA} and  neutrino-nucleus reactions relevant for supernovae nucleosynthesis \cite{Balasi2015PrPNP}.
Nuclear astrophysics in a broader context is discussed by \citet{Wiescher2012ARA&A} and \citet{Schatz2016JPhG},
and we refer the reader to the review \cite{Meisel2018JPG} and more specific literature regarding nuclear reactions in accreting crusts \cite{Yakovlev:2006fi,Gupta:2006fd,Gupta2008PhRvL,HaenselZdunik2008A&A, Steiner:2012bq, Afanasjev2012PhRvC,Schatz2014Natur}. 
Neutrino emission reactions in the crust are summarized by \citet{Yakovlev2001physrep} with more recent updates by \citet{ChamelHaensel2008LRR} and \citet{Potekhin:2015qsa}, and we have nothing to add to these reviews.  
Finally, we will not discuss  transport in the outer layers of the star, including the atmosphere and the heat blanketing envelopes, where radiative transfer, transport of non-degenerate electrons \cite{PotekhinDeLuca2015SSRv,Potekhin:2015qsa}, and diffusion processes \cite{Beznogov2013PhRvL,Beznogov2016MNRAS}, among others, are important.

\section{Basic concepts of transport theory}\label{S:Gen_kin}

\subsection{Basic equations and transport coefficients}
\label{sec:basic}

We start with a brief introduction to the basic concepts that will be used throughout this review. The goal of this section is to provide the definition of the most important 
transport coefficients, to show how they appear in the hydrodynamical framework and how they are computed from kinetic theory. In the present section, we shall present a  general setup for a dilute gas of one non-relativistic fermionic species. Further assumptions and specifications will be made in the subsequent sections. Our starting point is the Boltzmann equation for the non-equilibrium fermionic distribution function $f({\bm x},{\bm p},t)$,
\be \label{boltz}
\frac{\partial f}{\partial t}  + 
\bm{u}\cdot\frac{\partial f}{\partial {\bm x}} + {\bm R}\cdot\frac{\partial f}{\partial {\bm p}} = I[f] \, , 
\ee
where the particle velocity ${\bm u}$ is related to momentum via ${\bm p} = m{\bm u}$ with the particle mass $m$, and ${\bm R}$ is the external force which we do not specify for now, except for 
assuming that $\nabla_{\bm p} \cdot {\bm R} = 0$.  For instance, it can include the gravitational force or, if the particles carry electric charge, the 
Lorentz force.
The collision term is
\be
I[f] = -\int_{{\bm p}_1}\int_{{\bm p}'}\int_{{\bm p}_1'} W({\bm p},{\bm p}_{1};{\bm p}',{\bm p}'_1)[ff_1(1-f')(1-f'_1)-(1-f)(1-f_1)f'f'_1] \, , \label{boltz_collint}
\ee
with the the abbreviations $f_1=f({\bm x}_1,{\bm p}_1,t)$, $f'=f({\bm x}',{\bm p}',t)$, $f'_1=f({\bm x}'_1,{\bm p}'_1,t)$, and
\be
\int_{\bm p}\equiv\int \frac{d^3{\bm p}}{(2\pi\hbar)^3} \, .
\ee 
The collision integral gives the number of collisions per unit time in which a particle with a given momentum ${\bm p}$ is lost in a scattering process with another ingoing particle with momentum ${\bm p}_1$ to produce two outgoing particles with momenta ${\bm p}'$ and ${\bm p}'_1$, plus the number of collisions of the inverse process, in which a particle with momentum ${\bm p}$ is created. The transition rates $W({\bm p},{\bm p}_{1};{\bm p}',{\bm p}'_1)$ depend on the details of the collision process and contain energy and momentum 
conservation of the process. Their specific form is not needed for now; we shall see later how the Boltzmann equation is solved approximately in specific cases. 
For notational convenience, we have omitted the spin variable. One may think of the momentum to actually be a pair of momentum and spin 
and the momentum integral to include the sum over spin. 
We have written the collision term in the simplified form that only contains 
scattering of a given, single particle species with itself. Later, we shall discuss approximate solutions to the Boltzmann equation for more than one particle species, 
for instance electrons and ions in the neutron star crust. 

The Boltzmann equation allows us to derive an equation for the transport of any dynamical variable $\psi(\bm{x},\bm{p},t)$. To this end, we introduce the average value of $\psi$ per particle as
\be\label{distav}
\langle \psi \rangle  = \frac{1}{n}\int_{\bm p} \psi f\,, \qquad n = \int_{\bm p} f \, , 
\ee
where $n$ is the number density. Multiplying the Boltzmann equation with $\psi$ and integrating over momentum then yields 
\be\label{transf}
\frac{\partial n \langle \psi \rangle}{\partial t} + \nabla\cdot( n\langle \psi \bm{u}\rangle) =
n\left(\left\langle\frac{\partial \psi }{\partial t}\right\rangle+ \langle\bm{u}\cdot \nabla \psi\rangle 
+ \left\langle \bm{R}\cdot\frac{\partial \psi}{\partial \bm{p}}\right\rangle\right) + \int_{\bm{p}} \psi I[f] \, ,
\ee
where $\nabla$ is the spatial gradient. 
The first two terms in the parentheses on the right-hand side account for the production of $\psi$ due to its space and time variations, the third term gives the supply from forces, and 
the last term gives the production rate from collisions. 

From the transport equation (\ref{transf}) we derive 
the hydrodynamic equations by choosing $\psi$ to be a quantity 
that is conserved in a collision, such that the momentum integral over 
$\psi$ times the collision term vanishes. These invariants are $\psi=1$, which corresponds to particle number conservation, energy $\psi=p^2/(2m)$, and momentum components $\psi = p_i$. 
Thus we  obtain three  equations (two scalar equations, one vector equation) that do not depend on the collision term explicitly (but contain the non-equilibrium distribution function, which in principle has to be
determined from the full Boltzmann equation). These equations can be written as 
\begin{subequations}\label{cons}
\bea
\frac{\partial \rho}{\partial t} + \nabla\cdot {\bm g}&=& 0 \, , \label{contin}\\[2ex]
\frac{\partial{\cal E}}{\partial t} +\nabla\cdot{\bm j}_{\cal E}&=&
 n \bm{R}\cdot \bm{v} , \label{encons} \\[2ex]
\frac{\partial g_i}{\partial t} +\partial_j(\Pi_{ji}+\pi_{ji})&=&
n R_i
\label{momcons} \, .
\eea
\end{subequations}
Here we have introduced the center-of-mass velocity
${\bm v}$. In the present case of a single fluid, this velocity is identical to the drift velocity of the (single) fluid $\langle {\bm u}\rangle$. For multi-fluid mixtures, there is a drift velocity for each fluid, which of course does not have to be identical to the total velocity ${\bm v}$
of the mixture. This case will become important in the next section, where we discuss electrons in an ion background with a nonzero ${\bm \vartheta} \equiv \langle {\bm u}\rangle - {\bm v}$. In Eq.\ (\ref{cons}) we have also introduced  mass density $\rho = mn$, momentum density ${\bm g} = \rho{\bm v}$,  
energy density 
${\cal E} = {\cal E}_0 + \rho v^2/2$,  and stress tensor $\Pi_{ij} = \rho v_i v_j + \delta_{ij}P$, where  
  the energy density in the co-moving frame of the fluid ${\cal E}_0$ and the pressure $P$ are given by 
\be
{\cal E}_0= n \langle \varepsilon \rangle = \frac{\rho}{2}\langle w^2\rangle \, , \qquad P = \frac{\rho}{3}\langle w^2 \rangle = \frac{2}{3}{\cal E}_0 \, , \label{simplefields}
\ee
where ${\bm w}\equiv {\bm u}-{\bm v}$ is the difference between the single-particle velocity and the macroscopic 
center-of-mass velocity, and 
\be \label{singleeps}
\varepsilon = \frac{mw^2}{2}   
\ee
is the single-particle energy in the 
co-moving frame of the  
fluid. 
The flux terms in the energy conservation (\ref{encons}) and momentum conservation (\ref{momcons}) 
equations are 
\begin{subequations}
\bea
j_{{\cal E},i} = ({\cal E}+P)v_i+\pi_{ij} v_j + j_{T,i} &=&\frac{m}{2}\int_{\bm p} u^2u_if \, , \label{qtot} \\[2ex]
\Pi_{ij} + \pi_{ij} &=&  m\int_{\bm p} u_iu_jf \, , \label{Pitot}
\eea
\end{subequations} 
which include the dissipative contributions, which vanish in equilibrium, 
\be \label{Qpi}
 {\bm j}_T \equiv n\langle \varepsilon \bm{w} \rangle \, , \qquad \pi_{ij}\equiv \rho \langle w_i w_j\rangle  -\delta_{ij}P \, .
\ee
We assume that close to equilibrium we can apply the thermodynamic relations ${\cal E}_0+P=\mu n+Ts$ and $d{\cal E}_0=\mu dn + Tds$ locally, with the $t$ and $\bm{x}$ 
dependent chemical potential $\mu$, entropy density $s$, and temperature $T$. 
Using these relations, together with Eqs.\ (\ref{contin}) and (\ref{momcons}),
the energy conservation (\ref{encons}) can be written as an equation for entropy production. And, using Eq.\ (\ref{contin}), 
the momentum conservation (\ref{momcons}) can be written in the form of the Navier-Stokes equation. Hence, Eqs.\ (\ref{encons}) and (\ref{momcons}) become
\begin{subequations}
\bea
\frac{\partial s}{\partial t}+\nabla\cdot\left(s{\bm v}+\frac{\bm{j}_{T}}{T}\right) &=&-\frac{\pi_{ji}\partial_j v_i+{\bm j}_T \cdot\nabla T /T}{T}\equiv \varsigma \, ,  \label{dsdt}\\[2ex]
\frac{\partial v_i}{\partial t}+({\bm v}\cdot\nabla)v_i&=&-\frac{\partial_i P}{\rho}+\frac{R_i}{m}
-\frac{\partial_j\pi_{ji}}{\rho} \, , \label{NavSto}
\eea
\end{subequations}
where we have defined the entropy production rate $\varsigma$.
Instead of deriving the hydrodynamical equations from the Boltzmann equation, we can also view them as an effective theory where dissipative terms can be added systematically 
with certain transport coefficients. 
These transport coefficients are then an input to hydrodynamics, for instance computed from a kinetic approach. From Eq.~(\ref{dsdt}) we see that the dissipative part is composed of products of the thermodynamic forces  $\nabla T/T$ and $\partial_i v_j$ and the corresponding thermodynamic fluxes $\bm{j}_T$ and $\pi_{ij}$. The usual transport coefficients are then introduced by assuming linear relations between them  with the coefficients being thermal 
conductivity $\kappa$, shear viscosity $\eta$, and bulk viscosity $\zeta$, 
\begin{subequations} \label{Ejpi}
\bea
{\bm j}_{T}&=&-\kappa \nabla T, \label{Qkappa} \\[2ex]
\pi_{ij} &=& -2\eta \left(v_{ij}-\frac{\delta_{ij}}{3}\nabla\cdot{\bm v}\right) - \zeta \delta_{ij}\nabla\cdot{\bm v} \, ,\label{piij}
\eea
\end{subequations}
where we have abbreviated
\be
v_{ij}\equiv\frac{\partial_i v_j + \partial_j v_i}{2} \, .
\ee
In principle, one can systematically
expand the fluxes in powers of derivatives and thus create terms beyond linear order \cite{PLMS:PLMS0385,PLMS:PLMS0382,GARCIACOLIN2008149}. Higher-order hydrodynamical coefficients
 are rarely used in the non-relativistic context (see however 
 Refs.\ \cite{Chao:2011cy,Schaefer:2014xma} for 
a discussion of second-order hydrodynamics, motivated by applications to unitary Fermi gases). In contrast, second-order {\it relativistic} hydrodynamics has been studied much more extensively, motivated by the acausality of the 
first-order equations and by applications to relativistic heavy-ion collisions \cite{Israel:1979wp,Romatschke:2009im,Denicol:2012cn}. Here we will not go beyond first order. 

The simple one-component 
monatomic gas discussed above does not have a bulk viscosity
$\zeta$ because $\zeta$ is proportional to the trace of 
$\pi_{ij}$, as we see from Eq.\ (\ref{piij}), and the trace of 
$\pi_{ij}$ vanishes in our simple example, as Eq.\ (\ref{Qpi}) shows, due to the relation between energy density and pressure in 
Eq.\ (\ref{simplefields}). In more general cases, the hydrostatic pressure is not given by (\ref{simplefields}), and the bulk viscosity is nonzero.  Notice that the three terms in Eqs.~(\ref{Ejpi}) have different spatial symmetry and do not couple. 
We can compute the rate of the total entropy change $\dot{S}$ of the system by integrating Eq.~(\ref{dsdt})
over the volume $V$ of the system. Making use of Eqs.\ (\ref{Ejpi}), we obtain
\bea \label{dotS}
\dot{S} &=&\int_V \frac{d^3{\bm x}}{T}\left[2\eta\left(v_{ij}-\frac{\delta_{ij}}{3}\nabla\cdot{\bm v}\right)^2+\zeta(\nabla\cdot{\bm v})^2 +\frac{\kappa(\nabla T)^2}{T} 
\right]  - \int_{\partial V} d\bm{\sigma} \cdot \frac{\bm{j}_T} {T}  \, .
\eea
The first integral gives the total entropy production by the dissipative processes inside the system, while the surface integral corresponds to the heat exchange with the external thermostat. Due to the second law of thermodynamics, all phenomenological coefficients $\kappa$, $\eta$, and $\zeta$ have to be non-negative.

In more general cases, the entropy production  equation (\ref{dsdt}) contains more terms, for instance related to diffusion in multi-component mixtures. Some of these terms will be discussed in the following sections. When additional dissipative processes are considered, the equations become more cumbersome, but the principal scheme is the same.

\subsection{Calculating transport coefficients in the Chapman-Enskog approach} \label{S:crust_boltz}

Kinetic theory allows us to compute the transport coefficients  on microscopic grounds. The basic idea is to expand the distribution function around the local equilibrium distribution function. The kinetic equation then describes the evolution of the system towards local equilibrium. There exist two elaborate methods for this expansion, namely  Grad's moment method \cite{CPA:CPA3160020403} and the Chapman-Enskog method \cite{ChapmanCowling1999}, see also the textbooks 
by \citet{kremer2010introduction} and \citet{zhdanov2002transport} for extensive 
discussions of both methods. Here we give a brief sketch of the Chapman-Enskog method. We write the distribution function as 
\be \label{ff0}
f({\bm x},{\bm u},t) 
\approx f^{(0)}+\delta f \, , \qquad \delta f = -\frac{\partial f^{(0)}}{\partial \varepsilon} \Phi + {\cal O}(\Phi^2)\approx \frac{f^{(0)}(1-f^{(0)})}{k_B T}\Phi  \, , 
\ee
with a small correction $\Phi({\bm x},t)$ to the Fermi-Dirac function in local equilibrium 
\be
f^{(0)}({\bm x},{\bm u},t) = \left\{\exp\left[\frac{\varepsilon({\bm x},{\bm u},t)-\mu({\bm x},t)}{k_B T(\bm{x},t)}\right]+1\right\}^{-1} \, , \label{dist_leq}
\ee
where $k_B$ is the Boltzmann constant and where $\varepsilon$ from Eq.\ (\ref{singleeps})
is a function of ${\bm w}({\bm x},{\bm u},t) = {\bm u} - {\bm v}({\bm x},t)$.
The idea of the following approximation is to only keep the lowest order in $\Phi$ and also drop higher-order terms in the derivatives of $T$, $\mu$, and ${\bm v}$.
Inserting the ansatz (\ref{ff0}) into the Boltzmann equation (\ref{boltz}) yields the following lowest order equation
\be\label{boltzlin} 
\frac{\partial f^{(0)}}{\partial t}  + {\bm u}\cdot\frac{\partial f^{(0)}}{\partial {\bm x}} + {\bm R}\cdot\frac{\partial f^{(0)}}{\partial {\bm p}} \approx I_{\rm lin}[\Phi] \, , 
\ee
where $I_{\rm lin}[\Phi]$ is the linearized collision term. To be more general than in the previous section we do not specify its expression for now. [Linearizing the collision term (\ref{boltz_collint}) yields Eq.\ (\ref{boltz_coll_lin}).] Note that on the left-hand side the terms proportional to $\Phi$ are counted as higher order since they are multiplied by derivatives of $T$, $\mu$, and ${\bm v}$. Certain integral constraints on the deviation functions $\Phi$ can be obtained 
from the condition that number density, momentum, and energy in a gas volume element must be the same if calculated with the local equilibrium distribution (\ref{dist_leq}) and with the full function $f$ \cite{Landau10eng}.

Let us for now assume the system to be incompressible, which is a good approximation for instance for the neutron star crust. On account of the continuity equation (\ref{contin}), this is equivalent to $\nabla\cdot{\bm v}=0$. (In an incompressible fluid, the density of a fluid element is constant in time, $\partial_t\rho +{\bm v}\cdot\nabla\rho=0$.) As a consequence, there is no dissipation through bulk 
viscosity. We shall come back to bulk viscosity later when we address the core of the star. There, bulk viscosity {\it is} an important source of dissipation. We also focus on static systems, i.e., we shall neglect all time derivatives.
Extending the results of the previous section, we will include the electrical conductivity. To this end, we set ${\bm R} = -e{\bm E}$, where $\bm{E}$ is the electric field and $e$ is the elementary charge.
For now, we do not include a magnetic field and keep the assumption of a single particle species. This assumption deserves a comment. The expression (\ref{dsdt}) does not contain the external force $\bm{R}$, indicating that the force does not create dissipation. Of course, the work done by the force $\bm{R}$ affects the energy conservation (\ref{encons}), but this 
only enters the bulk motion, as Eq.~(\ref{NavSto}) shows. Dissipation from the electric field emerges if there exists a friction force which opposes the diffusive motion. This is not described by the collision integral (\ref{boltz_collint}), but is realized in a multi-component system such as the electron-ion plasma in the neutron star crust or nuclear matter in the core made of neutrons, protons, and leptons. In this case, as already mentioned below Eq.\ (\ref{cons}), the average velocity of the constituents $\langle \bm{u} \rangle$ is different from the 
center-of-mass velocity $\bm{v}$ of the mixture. This
gives rise to an electric (and diffusive) current $\bm{j}=-en\bm{\vartheta}= -en (\langle \bm{u} \rangle-\bm{v})$. In the neutron 
star crust (liquid or solid), due to the small mass ratio $m_e/m_{\rm i}$ of electron and ion masses, the contribution of the ion diffusion to the electric current can be 
neglected. Therefore, the rest frame of the ions is, to a good approximation, identical to the center-of mass frame and we can keep working with a single particle species (the electrons).

With these assumptions, we find for the left-hand side of Eq.\ (\ref{boltzlin}),  
\bea
{\bm u}\cdot\frac{\partial f^{(0)}}{\partial {\bm x}} -e{\bm E}\cdot\frac{\partial f^{(0)}}{\partial {\bm p}} = -\frac{\partial f^{(0)}}{\partial \varepsilon}\left[
\frac{\varepsilon - h}{T}{\bm w}\cdot\nabla T + e{\bm w}\cdot {\bm E}^* +p_iw_j\left(v_{ij}-\frac{\delta_{ij}}{3}\nabla\cdot{\bm v}\right)\right] \, . \label{boltz_lin_lhs}
\eea
Here we work in the co-moving frame of the total fluid, i.e., we have set ${\bm v}=0$ after taking the derivatives, such that from now on we have ${\bm w} = {\bm u} = {\bm p}/m$. We have added a term proportional to $\nabla\cdot{\bm v}$ (which is zero in our approximation)
in order to reproduce the structure needed for the shear viscosity, defined the enthalpy per particle $h=\mu+sT/n$, and the effective electric field
\be
{\bm E}^*  = {\bm E}+\frac{\nabla\mu}{e}+\frac{s}{n}\frac{\nabla T}{e} \, .\label{eq:elfield_eff}
\ee
The enthalpy is included in the thermal conduction term (proportional to $\nabla T$) to eliminate the convective heat flux [cf.\ first term in Eq.\ (\ref{qtot})].

In order to express  the dissipative currents in terms of the deviation function $\Phi$, 
we re-derive the entropy production equation (\ref{dsdt}) as follows. We assume the entropy density of the system close to  equilibrium to be given by the usual statistical expression 
\begin{equation}\label{entr}
s=-k_B \int_{\bm{p}} \left[f\ln f+(1-f)\ln(1-f)\right].
\end{equation}
This suggests to set $\psi=\ln f+(f^{-1}-1)\ln(1-f)$ in the general transport equation (\ref{transf}). The right-hand side of that equation, including the collision term as well as the 
terms from the explicit $({\bm x},{\bm p},t)$-dependence of $\psi$, yields the entropy production 
\bea 
T\varsigma&=&k_B\int_{\bm p} [\ln f - \ln(1-f)]\, I[f]
=-\int_{\bm{p}}\, \Phi I[f] = \bm{j}\cdot {\bm E}^* -{\bm j}_T\cdot\frac{\nabla T}{T}-\pi_{ij}\partial_jv_i \, .\label{eq:entr_rate}
\eea
In the second step we have performed the linearization according to Eq.\ (\ref{ff0}), taking into account that $\varsigma$ vanishes for the local equilibrium function $f^{(0)}$. In the third step, we have used that, according to the Boltzmann equation (\ref{boltzlin}), we can replace the collision integral by Eq.~(\ref{boltz_lin_lhs}), and we have expressed the fluxes in terms of $\Phi$, 
\begin{equation} \label{fluxes}
\bm{j}=e\int_{\bm{p}} 
\Phi\frac{\partial f^{(0)}}{\partial \varepsilon}
{\bm w}\, , \qquad 
\bm{j}_T=-\int_{\bm{p}} \Phi\frac{\partial f^{(0)}}{\partial \varepsilon}  
(\varepsilon - h){\bm w}\, , 
\qquad \pi_{ij} = -\int_{\bm{p}} \Phi\frac{\partial f^{(0)}}{\partial \varepsilon}  p_i w_j \, .
\end{equation}
Now, generalizing Eq.\ (\ref{Qkappa}), we introduce the transport coefficients associated with the electric and heat fluxes, 
\be\label{T-E}
\left(\begin{array}{c} {\bm E}^* \\[2ex] {\bm j}_T \end{array}\right)=\left(\begin{array}{cc} \frac{1}{\sigma} & -Q_T \\[2ex] -Q_T T & -\kappa\end{array}\right)
\left(\begin{array}{c} {\bm j} \\[2ex] \nabla T \end{array}\right) \, ,
\ee
where $\sigma$ is the electrical conductivity and $Q_T$ is the thermopower. The form of the non-diagonal terms is a consequence of Onsager's symmetry principle \cite{Landau10eng}. Notice that due to the same spatial rank-one tensor structure  of the thermodynamic forces $\nabla T/T$ and $\bm{E}^*$, their linear response laws are coupled. The perturbation that drives the shear viscosity is the second-rank tensor (\ref{piij}), hence the corresponding response law decouples. 
In terms of the transport coefficients, the local entropy production rate  (\ref{eq:entr_rate}) becomes
\be
T\varsigma = \kappa \frac{(\nabla T)^2}{T}
+ \frac{j^2}{\sigma} +2\eta\left(v_{ij}-\frac{\delta_{ij}}{3}\nabla\cdot{\bm v}\right)^2 \, , 
\ee
implying the non-negativeness of $\kappa,\eta,$ and $\sigma$. 

The transport coefficients $\eta$, $\kappa$, $\sigma$, $Q_T$ can now be computed as follows. 
To compute the shear viscosity, we make the ansatz
\be  \label{Aeta}
\Phi = -A_\eta(\varepsilon) \left(p_iw_j-\frac{\delta_{ij}}{3}\bm{p}\cdot\bm{w}\right)\left(v_{ij}-\frac{\delta_{ij}}{3}\nabla\cdot{\bm v}\right) \, , 
\ee
where, in an isotropic system, the unknown function $A_\eta$ only depends on the particle energy. This function has to be determined 
by inserting the ansatz for $\Phi$ into the linearized 
Boltzmann equation (\ref{boltzlin}). We can express the  shear viscosity through $A_\eta$ 
as
\be \label{etaA}
\eta = -\frac{2}{15}\int_{\bm p} p^2w^2 A_\eta(\varepsilon) \frac{\partial f^{(0)}}{\partial \varepsilon} \, .
\ee
This relation is obtained by inserting the ansatz (\ref{Aeta}) into $\pi_{ij}$ from Eq.\ (\ref{fluxes}), using the form of the viscous stress tensor (\ref{piij}) and the angular integral in velocity (or momentum) space (remember that $\bm{p}=m\bm{w}$ in the frame we are working in)
\be
\int\frac{d\Omega_{\bm p}}{4\pi} w_iw_j\left(w_kw_\ell-\frac{\delta_{k\ell}}{3}w^2\right) = \frac{w^4}{15}\left(\delta_{ik}\delta_{j\ell}+\delta_{i\ell}\delta_{jk}-\frac{2}{3}\delta_{ij}\delta_{k\ell}\right) \, .
\ee
In Eq.\ (\ref{etaA}) we have multiplied the result by a factor 2 from the sum over the 2 spin degrees of freedom of a spin-$\frac{1}{2}$ fermion (such that now the integral does 
not implicitly include the spin sum anymore).  

To compute electrical and thermal conductivities and the thermopower, we use the ansatz
\be \label{Akappa}
\Phi =- A_\kappa(\varepsilon) \frac{\varepsilon-h}{T}{\bm w}\cdot\nabla T-A_\sigma(\varepsilon) e\,{\bm w}\cdot{\bm E}^* \, , 
\ee
with $A_\kappa$ and $A_\sigma$  computed from the linearized Boltzmann equation, and the transport coefficients are found in an analogous way as just demonstrated 
for the shear viscosity: we insert the ansatz (\ref{Akappa}) into ${\bm j}$ and ${\bm j}_T$ from Eq.\ (\ref{fluxes}), 
perform the  angular integral,
\be
\int\frac{d\Omega_{\bm p}}{4\pi} w_iw_j= \frac{w^2\delta_{ij}}{3} \, , 
\ee
and compare the result with Eq.\ (\ref{T-E}) to obtain (again taking into account the 2 spin degrees of freedom)
\begin{subequations}\label{kincoeff_boltz}
\bea
\sigma &=& -\frac{2e^2}{3}\int_{\bm p} w^2 A_\sigma(\varepsilon)\frac{\partial f^{(0)}}{\partial \varepsilon} \, , \\[2ex]
\sigma Q_T &=& -\frac{2e}{3}\int_{\bm p} w^2 A_{\kappa,\sigma}(\varepsilon)\frac{\varepsilon-h}{T}\frac{\partial f^{(0)}}{\partial \varepsilon} 
\, , \\[2ex]
\kappa + \sigma Q_T^2T &=& -\frac{2}{3}\int_{\bm p} w^2 A_\kappa(\varepsilon)\frac{(\varepsilon-h)^2}{T}\frac{\partial f^{(0)}}{\partial \varepsilon} \, , \label{kincoeff3}
\eea
\end{subequations}
from which $\sigma$, $Q_T$, and $\kappa$ can be computed.  As a consequence of Onsager's symmetry principle, 
we have obtained two expressions for $\sigma Q_T$, using either  $A_\sigma$ or $A_\kappa$ in the integral. 

In general, even the solution of the linearized 
Boltzmann equation is not an easy task and various methods and approximations are used. First, one needs to specify the explicit expression for the collision integral. For instance, the linearization of the collision integral (\ref{boltz_collint}) gives
\be \label{boltz_coll_lin}
I_{\rm lin}[\Phi]=-\frac{1}{k_BT}\int_{{\bm p}_1}\int_{{\bm p}'}\int_{{\bm p}_1'} W({\bm p},{\bm p}_{1},{\bm p}',{\bm p}'_1)f^{(0)}f_1^{(0)}(1-f'^{(0)})(1-f'^{(0)}_{1})(\Phi+\Phi_1-\Phi'-\Phi'_1) \, ,
\ee
where we have used $f^{(0)}f_1^{(0)}(1-f'^{(0)})(1-f'^{(0)}_{1}) = (1-f^{(0)})(1-f_1^{(0)})f'^{(0)}f'^{(0)}_{1}$ due to energy conservation.

One of the simplest cases is realized when the collision integral can be written in the form of the (energy-dependent) relaxation-time approximation,
\begin{equation}
I=-\sum_{lm} \frac{\delta f^{lm}}{\tau^l(\varepsilon)}
Y_{lm}(\Omega_{\bm{p}}) \, ,\label{eq:reltime_approx}
\end{equation}
which takes into account the angular dependence of the deviation
to the equilibrium distribution function by expanding it in spherical harmonics $Y_{l m}$.  Here
$\tau^l(\varepsilon)$ is the relaxation time for the
perturbation of multiplicity $l$. The solution of the Boltzmann
equation is then
\begin{equation}\label{eq:A_reltime}
A_{\sigma}(\varepsilon)=A_{\kappa}(\varepsilon)=\tau^1(\varepsilon) \, ,\qquad
A_{\eta}(\varepsilon)=\tau^2(\varepsilon) \, .
\end{equation}
When the relaxation time approximation is not available, one usually represents the functions $A(\varepsilon)$ in the form of a series expansion in some basis functions. This basis has to be chosen carefully for a satisfactory convergence of the expansion. In some cases the infinite chain of equations for the coefficients can be solved analytically and the exact solution for the transport coefficients is obtained from  (\ref{kincoeff_boltz}) (in form of an infinite series). In practice, the chain of equations is truncated at a finite number of coefficients. The truncation procedure is justified on the basis of the variational principle of kinetic theory  \cite{ZimanBook}. The variational principle uses the fact that the entropy production rate calculated from (\ref{eq:entr_rate}) with the linearized collision integral $I[\Phi]$ is a semi-positive definite functional of $\Phi$. This is readily seen for the binary collision integral (\ref{boltz_coll_lin}) since the probability $W$ is positive, but it holds in general. Suppose that the arbitrary function $\tilde{\Phi}$ is subject to the constraint 
\be\label{varboltz}
\int_{\bm{p}} X\tilde{\Phi} = \int_{\bm{p}} I_{\rm lin}[\tilde{\Phi}] \tilde{\Phi} = - T\varsigma[\tilde{\Phi}] \, ,
\ee
where we have abbreviated (\ref{boltz_lin_lhs}) by $X$. The variational principle states that over the class of such functions, the entropy production is maximal for the solution of the Boltzmann equation $X=I_{\rm lin}[\Phi]$, in other words $\varsigma [\Phi] \geq \varsigma[\tilde{\Phi}]$. Increasing the number of terms in the functional expansion and maximizing the functional $\varsigma[\tilde{\Phi}]$ under the constraint  (\ref{varboltz}), one approaches the exact solution. This principle can be reformulated to give a direct limit on the diagonal coefficients in the Onsager relations. For instance, setting the thermodynamic forces to zero, $\nabla T/T =0$ and $v_{ij}=0$, and keeping only $\bm{E}^*$, one obtains the electrical conductivity by minimizing 
\be
\frac{1}{\sigma} \leq \frac{E^{*2}}{T\varsigma [\tilde{\Phi}]}
\ee
over the functions subject to  (\ref{varboltz}). Notice that the off-diagonal coefficient $Q_T$ cannot be constrained in this way. 

The variational principle discussed here applies for the stationary case in the absence of a magnetic field. The extension of the variational principle beyond this approximation 
is non-trivial and is outside the scope of the present section.

\subsection{Towards neutron star conditions}
In this section we briefly comment on some modifications and extensions of the kinetic theory laid out in the previous sections due to the specific conditions inside neutron stars.
We mention plasma effects, transport in Fermi liquids, relativistic effects, and effects from Cooper pairing. 

\subsubsection{Plasma effects}\label{Sec:plasma}
Electrically charged particles, for instance electrons in the crust and in the core, interact via the long-range Coulomb potential. This seems to be at odds with the concept of instant binary collisions, which forms the basis of the Boltzmann approach to compute transport properties of dilute gases. However, the interaction between charged particles in a plasma is screened and thus is effectively damped on length scales $r> r_D$, where $r_D$ is the Debye screening length. Therefore, the Boltzmann equation becomes appropriate to describe the processes occurring on large scales, provided the screened interaction potential is used in the collision integral \cite{Landau10eng}. The screening itself depends on the distribution functions of the plasma components, which severely complicates the solution. However, for weak deviations from equilibrium, when the linearized Boltzmann equation is used, the screening which enters the collision integral in Eq.\ (\ref{boltzlin}) can be calculated from the equilibrium distribution functions (i.e., in the collisionless limit). 
Additional justification comes from the degeneracy conditions, which are appropriate for electrons in most parts of the star (and other charged particles in the core). In this case, only a small fraction of the thermal excitations contribute to transport phenomena. Moreover, the kinetic energy of the particles increases with density stronger than the Coulomb interaction energy. In other words, the denser the gas is, the closer it is to the ideal Fermi gas \cite{Landau5eng}. All these properties allow us to use the formalism of the linearized Boltzmann equation discussed above. Note that the force term ${\bm R}$ should contain the Lorentz force with the self-consistent electromagnetic field. The generalized Ohm law (\ref{T-E}) then is written in the co-moving frame of the plasma and contains the electric field measured in this frame, ${\bm E}'=\bm{E}+\frac{1}{c}{\bm v}\times {\bm B}$. We will return to this aspect in more details in Sec.~\ref{S:core_nuc_mag}.

The ions in the neutron star crust are non-degenerate and non-ideal. The discussion of their transport phenomena is more involved.  Fortunately, the ion contribution is usually negligible, see Sec.~\ref{S:crust}.

\subsubsection{Transport in Fermi liquids}\label{S:fl_transp}
Nuclear matter in the core of a neutron star is a strongly interacting, non-ideal, multi-component fluid. The kinetic theory of rarefied gases described above cannot be applied directly. However, the relevant temperatures are low and the matter is highly degenerate.  In this case, the framework of Landau's Fermi-liquid theory \cite{BaymPethick} can be used to describe the low-energy excitations of the system. 
The excitations are considered as a dilute gas of quasiparticles which obey the
Fermi-Dirac distribution (\ref{dist_leq}) in momentum space, normalized to give the total local number density $n$ of the real particles. The single-quasiparticle energy $\varepsilon(\bm{p})$ is a functional of the distribution function $f$, the quasiparticle Fermi momentum is $p_F=\hbar (3\pi^2 n)^{1/3}$, and in equilibrium the spectrum of quasiparticles in the vicinity of the Fermi surface is described by the effective mass on the Fermi surface $m^*=p_F/v_F$, where 
\begin{equation}\label{eq:qp_spectrum}
  \varepsilon^{(0)}-\mu= v_F (p-p_F) \,, \qquad v_F=\left(\frac{\partial \varepsilon^{(0)}}{\partial p}\right)_{p=p_F},
\end{equation}
with the Fermi velocity $v_F$, and the superscript $(0)$ indicates 
equilibrium.

The evolution of the quasiparticle distribution function is described by the Landau transport equation
\be \label{landau}
\frac{\partial f}{\partial t}  + 
\bm{u}\cdot\frac{\partial f}{\partial {\bm x}}-\nabla\varepsilon \cdot\frac{\partial f}{\partial {\bm p}} = I[f] \, , 
\ee
where now $\bm{u}=\nabla_{\bm{p}}\,\varepsilon$. The equation (\ref{landau}) is different from the Boltzmann equation  (\ref{boltz}) since the term $\nabla \varepsilon$ is present even in the absence of external forces $\bm{R}$. This is because the energy spectrum -- being a functional of $f$ -- changes from one coordinate point to another. Thus, $\nabla \varepsilon$ contains the combined effects of the external forces and the effective field resulting from interactions between quasiparticles. In addition, the quasiparticle velocity is coordinate-dependent for the same reason. 

Transport coefficients of the Fermi-liquid are computed by considering a small deviation from  local equilibrium and performing the linearization of the Landau equation in a way similar to Sec.~\ref{S:crust_boltz} 
\cite{BaymPethick,Landau10eng}. However, there is an important difference. The local equilibrium distribution function is $f^{(0)}(\varepsilon^{(0)})$, but the conservation laws from the collision integral employ the true quasiparticle energies $\varepsilon$. Hence the collision integral vanishes for the functions $f^{(0)}(\varepsilon)$ instead of true local distribution function. As a consequence, the linearized collision integral depends not on $\delta f=f-f^{(0)}(\varepsilon^{(0)})$ but on $\delta \tilde{f}=f-f^{(0)}(\varepsilon)$, and the definition of the function $\Phi$ (\ref{ff0}) is modified to 
\be\label{phi_fl}
\delta\tilde{f} = -\frac{\partial f^{(0)}}{ \partial \varepsilon} \Phi \, .
\ee
Since the definitions of the fluxes also contain the true quasiparticle energies and velocities, they are given by the  expressions (\ref{fluxes}) with $\Phi$ redefined according to  (\ref{phi_fl}). Therefore, in the stationary case, we obtain formally identical equations as in the above derivation.
Fermi-liquid effects do not appear explicitly. The same is true if a magnetic field is taken into account \cite{Landau10eng}. 
In more general cases, terms containing $\delta f$ can appear on the left-hand side of the linearized Boltzmann equation. This situation is realized for instance when the bulk viscosity of the Fermi liquid is considered \cite{BaymPethick,SykesBrooker1970AnPhy}.

 \subsubsection{Relativistic effects} \label{sec:relativistic}

Neutron stars are ultra-dense objects, and thus relativistic effects are important for 
the transport in the star. They manifest themselves in various forms, and we have to distinguish between effects on a microscopic level (e.g., calculations of transport coefficients) and a macroscopic level (e.g., simulations based on hydrodynamic equations), as well as  between effects from special relativity (large velocities) and general relativity
(spacetime curvature on scales of interest). In this review, we are almost exclusively concerned with microscopic calculations, where we can usually ignore effects from general relativity.     
The reason is the large separation of the scale on which the gravitational field changes inside the star from the microscopic scales on which the equilibration processes (collisions or reactions) operate \cite{TauberWeinbergPhysRev, Thorne1966}. If the mean free paths of the particles are microscopic in this sense, one can study transport processes in the local Lorentz frame, and gravity effectively does not appear in the analysis. If the mean free path, however, becomes comparable to the macroscopic scale of gravity, one has to consider the full general relativistic transport equation \cite{VereshchaginAksenov2017}
\be \label{ref_boltz}
p^\mu\frac{\partial  f}{\partial x^\mu}-\Gamma^\mu_{\;\;\nu\rho} \, p^\nu p^\rho \frac{\partial f}{\partial p^\mu}= I[f] \, ,
\ee
where we have omitted external forces, where $\Gamma^\mu_{\;\;\nu\rho}$ are the Christoffel symbols, $x^\mu$ is the spacetime four-vector, $p^\mu$ the four-momentum, and $I[f]$ is the collision integral (specified in the local reference frame). This situation occurs for instance for neutrino transport in supernovae and proto-neutron stars \cite{Pons1999ApJ}.  In neutron stars, this general approach may be important for example
in superfluid phases if the only available excitations are the Goldstone 
modes, whose mean free path can become of the order of the size of the star, see Secs.~\ref{S:core_nuc_pairing} and \ref{S:quark_colsup}.
Effects from general relativity are also important when transport coefficients -- computed from 
a microscopic approach
-- are used as an input for hydrodynamic equations. These equations, when they concern the structure of the whole star or a significant fraction of it, must be formulated within general relativity.  
An example is the equation for the radial component of the heat flux in a cooling star 
\cite{Thorne1966,Thorne1977ApJ},
\be \label{Fr}
F_r=-\kappa \, e^{-\lambda-\phi} \frac{\partial\tilde{T}}{\partial r} \, ,
\ee
where $\kappa$ is the thermal conductivity, $\lambda$ and $\phi$ appear in the
parametrization of the metric,  
\be 
ds^2= e^{2\phi} d(ct)^2 - e^{2\lambda} d r^2 -r^2(d\theta^2 + \sin^2\theta\, d\varphi^2) \, , 
\ee 
and $\tilde{T}\equiv T e^\phi$ is the redshifted temperature. (It is the redshifted temperature, not the temperature $T$, which is constant in equilibrium.) 

To connect the non-relativistic hydrodynamic equations of Sec.\ \ref{sec:basic} to a covariant 
formalism, one introduces the (special) relativistic stress-energy tensor,
\be \label{Tmunu11}
T^{\mu\nu} = T^{\mu\nu}_{\rm ideal}  + T^{\mu\nu}_{\rm diss} \, ,
\ee
where we have separated the ideal part $T^{\mu\nu}_{\rm ideal}$ from the dissipative contribution $T^{\mu\nu}_{\rm diss}$, 
\begin{subequations}\label{Tmunurel}
\bea
T^{\mu\nu}_{\rm ideal} &=&  (\epsilon+P)v^\mu v^\nu - g^{\mu\nu} P \, , \label{Tmunuideal}\\[2ex]
T^{\mu\nu}_{\rm diss}  &=& \kappa(\Delta^{\mu\gamma} v^\nu +\Delta^{\nu\gamma} v^\mu)[\partial_\gamma T + T(v\cdot\partial)v_\gamma]+ \eta\Delta^{\mu\gamma}\Delta^{\nu\delta}\left(\partial_\delta v_\gamma+\partial_\gamma v_\delta - \frac{2}{3}g_{\gamma\delta} \partial\cdot v\right) +\zeta\Delta^{\mu\nu}\partial\cdot v  \, . \label{Tmunudiss}
\eea
\end{subequations}
Here, $\epsilon$ and $P$ are energy density and pressure measured in the rest frame of the fluid, $g^{\mu\nu}=(1,-1,-1,-1)$ is the metric tensor in flat space, $v^\mu = \gamma(1,{\bm v})$ is the four-velocity with the Lorentz factor $\gamma$ and the 
three-velocity ${\bm v}$ used in Secs.\ \ref{sec:basic} and \ref{S:crust_boltz}. We have abbreviated $\Delta^{\mu\nu}=g^{\mu\nu}-v^\mu v^\nu$, and the transport coefficients $\kappa$, $\eta$,
$\zeta$ are heat conductivity, shear and bulk viscosity, as in the non-relativistic formulation (\ref{Ejpi}). In the non-relativistic limit, using the notation from Sec.\ \ref{sec:basic}, $T^{00}_{\rm ideal} \to {\cal E}$ is the energy density, $T^{0i}_{\rm ideal}\to g_i$ is the momentum density, $T^{i0}_{\rm ideal}\to ({\cal E}+P) v_i$ is the non-dissipative part of the energy flux ${\bm j}_{\cal E}$, and $T^{ij}_{\rm ideal}\to \Pi_{ij}$ is the non-relativistic stress
tensor. The dissipative terms are formulated in the so-called Eckart frame \cite{PhysRev.58.919}, where -- in contrast 
to the Landau frame \cite{Landau1987Fluid} -- the conserved four-current $j^\mu = nv^\mu$ does not receive dissipative corrections \cite{Kovtun:2012rj}. 
The hydrodynamic equations are then obtained from the conservation laws for the stress-energy tensor and the current, 
\be
\partial_\mu T^{\mu\nu} = \partial_\mu j^\mu = 0 \, .
\ee
They reduce to Eqs.\ (\ref{cons}) in the non-relativistic limit.
We will briefly return to this relativistic formulation in Sec.\ \ref{S:quark_colsup1}, but otherwise we will 
not discuss any of the effects illustrated by Eqs.\ (\ref{ref_boltz}), (\ref{Fr}), and (\ref{Tmunurel}). In particular, since we do not discuss neutrino 
transport in supernovae, no effects from general relativity will be further discussed.
Therefore, when we use `relativistic' in the rest of the review, we mean effects 
from special relativity in the following simple sense: relativistic effects are important 
if the rest mass  (times the speed of light) of a given particle species is not overwhelmingly larger than its 
Fermi momentum. (In this case, the Fermi velocity introduced in the previous section, i.e., the slope of the dispersion relation at the Fermi surface, becomes a sizable fraction of the speed of light.) With this 
criterion, the ions in the crust and the nucleons in the core are often treated non-relativistically (for ultra-high densities in the core, this treatment becomes questionable), 
while the lighter electrons and quarks are relativistic (except for electrons at very 
low densities in the outer crust).

\subsubsection{Transport with Cooper pairing}\label{S:cooper}

The effect of Cooper pairing on reaction rates and transport will be discussed specifically in various sections throughout the review. As a preparation and a simple overview, we now 
give some general remarks that may be helpful to understand and put into perspective the more detailed discussions and results. For a pedagogical introduction, bringing together elements from non-relativistic and relativistic approaches to Cooper pairing in superfluids and superconductors see Ref.\ \cite{Schmitt:2014eka}.

Cooper pairing in neutron stars is expected to occur in the inner crust for neutrons 
 and in the core for neutrons, protons, and, if present, for hyperons and quarks.
The critical temperatures of these systems vary over several orders of magnitude, depending on the form of matter, on density, and on the particular pairing channel. Moreover, it is prone to large uncertainties because the attractive force needed for Cooper pairing originates from the strong interaction. Nevertheless, a rough benchmark to keep in mind is $T_c\sim 1\, {\rm MeV}$, which 
is the maximal critical temperature reached for nuclear matter\footnote{In units where $k_B=1$, temperature and energy have the same units, 1 MeV corresponds to $1.160 \times 10^{10}\, {\rm K}$.\label{fo:kB}} (with significantly smaller values for neutron triplet pairing) 
and which is exceeded by about an order of magnitude, maybe even two,
by quark matter, where $T_c \sim (10 -100)\, {\rm MeV}$ (also in quark matter, there are pairing patterns with significantly lower critical temperatures). In any case, we conclude that the temperatures inside the star -- except for very young neutron stars -- are sufficiently low to allow for Cooper pairing. The resulting stellar superfluids and superconductors \cite{Alford:2007xm,Page:2013hxa,Haskell:2017lkl} are similar to their relatives in the laboratory, but the situation in the star is typically more complicated. For instance, the 
neutron superfluid in the inner crust coexists with a lattice of ions, the core might be a superconductor and a superfluid at the same time, and quark matter might introduce 
effects of color superconductivity. In addition, the star rotates and has a magnetic field, which suggests the presence of superfluid vortices and possibly magnetic flux tubes, which 
may coexist and interact with each other. Therefore, understanding superfluid transport in the environment of a neutron star is a difficult task, and some care is required in using results from ordinary superfluids. 

One obvious effect of Cooper pairing is the suppression of reaction rates and scattering processes of the fermions that pair. 
This effect is very easy to understand. Cooper pairing induces an energy gap $\Delta$ in the quasiparticle dispersion relation (one needs a finite amount of energy to break up a pair), and thus, for temperatures much smaller than the gap, quasiparticles are not available for a given process. As a consequence, if at least one of the participating fermions is gapped, the rate is 
exponentially suppressed by a factor $\exp(-\Delta/T)$ for $T\ll\Delta$.  The suppression is milder if the pairing is not isotropic and certain directions 
in momentum space are left ungapped. This is conceivable for some forms of neutron pairing and in certain color-superconducting quark matter phases. In this case, 
if for instance only one- or zero-dimensional regions of the Fermi surface contribute (as opposed to the full two-dimensional Fermi surface in the unpaired case), the rate is suppressed by a power of the small parameter $T/\Delta$. Except for these special cases, at low temperatures we can usually neglect the processes 
suppressed by Cooper pairing and can restrict ourselves to contributions from ungapped fermions or other low-energy excitations, if present. 

At larger temperatures, as we move towards the critical temperature $T_c$, the form of the exponential suppression no longer holds and the rate in the Cooper-paired phase 
has to be evaluated numerically. Since particle number conservation is broken spontaneously, particles can be deposited into or created from the Cooper pair condensate. 
This effect induces subprocesses that are called Cooper pair breaking and formation processes. They are particularly 
interesting in nuclear matter, where more efficient processes, such as the direct Urca process, are suppressed. Then, somewhat counterintuitively, an enhancement of the neutrino emission is possible as the system cools through the critical temperature for 
neutron superfluidity. 

While Cooper pairing removes fermionic degrees of freedom from transport at low temperatures, it introduces one or several massless bosonic excitations if a {\it global} 
symmetry is spontaneously broken 
by the formation of a Cooper pair condensate. This is due to the Goldstone theorem, and the corresponding Goldstone mode for superfluidity 
is, following the terminology of superfluid helium, usually called  phonon (or `superfluid mode', or `superfluid phonon' to distinguish it from the lattice phonons in the neutron star crust). In this case, the broken global symmetry is the $U(1)$ associated with particle number conservation. Superfluid neutron matter and the color-flavor locked (CFL) quark 
matter phase both have a phonon. Transport through phonons is mostly computed with the help of an effective theory, and we will quote some of the resulting transport properties in hadronic and quark matter. If Cooper pairing breaks additional global symmetries, such as rotational symmetry, additional Goldstone modes appear. 
This is possible in $^3P_2$ neutron pairing \cite{Bedaque:2012bs,Bedaque:2013fja} and in spin-one color superconductivity \cite{Pang:2010wk}. 

If instead a {\it local} symmetry is spontaneously broken, there is no 
Goldstone mode. This is the case for Cooper pairing of protons and for quark matter phases other than CFL such as the so-called 2SC phase (although, due to the presence of electrons and
the resulting screening effects, the Goldstone mode in a proton 
superconductor can be `resurrected' \cite{Baldo2011PhRvC}). As in ordinary superconductivity, the would-be Goldstone boson is replaced by an additional degree of freedom
of the gauge field, which acquires a magnetic mass. One obvious consequence is the well-known Meissner effect, which is of relevance for the magnetic field evolution in neutron stars. Magnetic screening can also indirectly affect transport properties if a certain transport 
property is dominated by one (unpaired) particle species that is charged under the gauge symmetry which is spontaneously broken by Cooper pairing of a different species (even though the species that pairs does not 
contribute to transport itself because it is gapped). This situation occurs in nuclear matter  when electrons experience a modified electromagnetic interaction due to pairing of protons, and in the 2SC phase of quark matter, where the different particle species are electrons and the different colors and flavors of quarks, which are not all paired in this specific phase, and the relevant gauge bosons are the gluons and the photon. 
  
As we know from some of the earliest experiments with superfluid helium, a superfluid at nonzero temperature (below $T_c$) behaves as a two-fluid system
\cite{tisza38,landau41} (for the connection of the two-fluid picture to an underlying microscopic theory see for instance Ref.\ \cite{Alford:2012vn}). This means that, in a hydrodynamic approach, there are two independent velocity fields: one for the superfluid component, which is the Cooper pair condensate in a fermionic superfluid (or the Bose-Einstein condensate in a bosonic superfluid such as $^4$He), and one for the so-called normal component, which corresponds to the phonons and possibly a fraction of the fermions which have remained unpaired. Since only the 
normal component carries entropy, the two-fluid nature has obvious consequences for heat transport, which now can occur through a counterflow of the 
two fluid components. While this mechanism proves extremely efficient in laboratory experiments with superfluid helium, it may be less effective in the more complicated 
situation in a neutron star. For instance, in the inner crust of the star the counterflow of the normal and superfluid components becomes dissipative due to the presence of electrons which damp the motion of the normal fluid through induced electron-phonon interactions \cite{PageReddy2012}. Another consequence of the 
two-fluid behavior is the existence of second sound. (The phonon, first and second sound are in general three different excitations. At low temperatures, the phonon excitation 
is identical to first sound, while close to the critical temperature it is identical to second sound \cite{Alford:2013koa}.) In superfluid helium, first and second sound are predominantly 
density and temperature oscillations, respectively, for all temperatures $T<T_c$. This is not necessarily true for other superfluids and it has been shown that first and second sound may exchange their roles \cite{Alford:2013koa}. 

Two-fluid systems allow for additional transport coefficients. For instance, in the hydrodynamics of a superfluid, usually three independent bulk viscosity coefficients are taken into 
account \cite{Khalatbook1965}. In a neutron star, the situation might become even more complicated due to the presence of additional fluid components, e.g., a nonzero-temperature neutron superfluid coexisting with electrons and protons, such that we have to deal with an involved multi-fluid system. One interesting feature of multi-fluids with relevance for the physics of neutron stars 
is the possibility of hydrodynamical instabilities due to a counterflow between the fluids. Such an instability may occur for the neutron superfluid in the inner crust, if 
it moves (locally) with a sufficiently large nonzero velocity relative to the ion lattice.
In this review, we shall not further discuss multi-fluid 
transport in detail (except for the transport coefficients of a single superfluid at nonzero $T$) and refer the reader to the recent literature and references therein \cite{Gusakov:2009kc,Gusakov:2009mb,Glampedakis:2011yw,2013PhRvL.110a1101C,2013CQGra..30w5025A,Haber:2015exa,2016arXiv161000445A}.

Finally, let us mention another very important consequence of Cooper pairing, which has been related to various astrophysical observations such as pulsar 
glitches \cite{Haskell:2015jra}, namely 
the formation of rotational vortices in a superfluid and of magnetic flux tubes in a superconductor. (A magnetic field enters a type-II superconductor through quantized 
magnetic flux tubes if its magnitude lies between the upper and lower critical magnetic fields. The presence of a superfluid, to which the superconductor couples, may change the 
textbook-like behavior of type-II superconductors qualitatively \cite{Haber:2016ljn,Haber:2017kth}.) Besides ordinary vortices in hadronic matter, quark matter in the core of neutron stars may contain so-called semi-superfluid 
vortices \cite{Balachandran:2005ev,Alford:2016dco} in the CFL phase and/or color magnetic flux tubes \cite{Alford:2010qf,Glampedakis:2012qp} in 
the CFL or 2SC phases (the latter are not protected by topological arguments and it is unknown if they are energetically stable objects in the neutron star environment). 
As for most of the multi-fluid aspects, we will not review the transport properties of superfluids in the presence of vortices. For various aspects of the hydrodynamics 
of these systems, including the possibility of superfluid turbulence and possible boundaries between phases with and without (or with a different kind of) vortices, see Refs.\ \cite{hall1956rotation,Khalatbook1965,donnelly1999cryogenic,Gusakov:2016eom,Gusakov:2016ftg,Graber:2016imq}.

\section{Transport in the crust and the crust/core transition
region}\label{S:crust}

\subsection{Thermal and electrical conductivity and shear
viscosity}\label{S:crust_kincoeff}
The main carriers which determine the transport processes in the neutron star crust are electrons.  The electrons in the crust form 
an almost ideal, degenerate gas. The degeneracy temperature $T_F$ for electrons is 
\begin{equation}\label{eq:FermiTemp}
T_{Fe}=\frac{\mu_e-m_ec^2}{k_B}=5.9\times 10^{9}{\rm K} \left(\sqrt{1+x_r^2}-1\right) \, ,
\end{equation}
where $x_r=p_{Fe}/(m_e c)$ is the electron relativistic parameter, with  the electron Fermi momentum $p_{Fe}$, the electron rest mass $m_e$, and the electron chemical potential (including the rest mass) $\mu_e=m_ec^2 \sqrt{1+x_r^2}\equiv m_e^* c^2$. In a one-component plasma with ion charge number $Z$ and total nucleon number per ion\footnote{In the inner crust, unbound neutrons exist and the ion mass number $A_{\rm nuc}$ is less than $A$. The ion mass is then $m_{\rm i}=A_{\rm nuc} m_{\rm u}$, with $m_{\rm u}$ being the atomic unit mass \cite{ChamelHaensel2008LRR}.} $A$, $x_r\approx (\rho_{6}Z/A)^{1/3}$, where $\rho_6$ is the mass density $\rho$ in units of  $10^6\, {\rm g}\, \rm{cm}^{-3}$. In most of the crust, $\rho_6\gg 1$ and the electrons are ultra-relativistic.  
We will not discuss  electrons in  non-degenerate or partially degenerate conditions $T\gtrsim T_{Fe}$. The effects of non-degenerate electrons are important when the thermal structure of the stellar heat blanket is calculated. In non-degenerate regions the radiative contribution to heat transport is relevant, which we also do not discuss here, for details see Refs.\ \cite{PotekhinDeLuca2015SSRv,Potekhin:2015qsa}.

For degenerate electrons ($T\ll T_{Fe}$) the analysis of the
Boltzmann equation is simplified since the transport is mainly
provided by those electrons whose energies lie in a narrow thermal
band near the Fermi surface $|\varepsilon-\mu_e|\lesssim k_B T$. When using  Eqs.\ (\ref{Aeta}) and (\ref{kincoeff_boltz}), it is safe to set $h=\mu_e$ and neglect the thermopower correction in Eq.\ (\ref{kincoeff3}). As a result, it is convenient to present the transport coefficients of interest in the form
\begin{subequations}\label{eq:kins_deg} \allowdisplaybreaks
\begin{eqnarray}
\sigma&=& \frac{e^2 n_e \tau_\sigma}{m_e^*}\, ,\label{eq:sigma_deg}\\[2ex]
\kappa&=&\frac{\pi^2k_B^2 T n_e \tau_\kappa}{3m_e^*} \, ,\label{eq:kappa_deg}\\[2ex]
Q_T&=&\frac{\pi^2k_B^2T m_e^*}{3ep^2_{Fe}}(3+\xi) \, ,\label{eq:QT_deg}\\[2ex]
 \eta&=&\frac{n_e p^2_{Fe}\tau_\eta}{5
m_e^*} \,  , \label{eq:eta_deg}
\end{eqnarray}
\end{subequations}
where $\tau_\sigma$, $\tau_\kappa$, and $\tau_\eta$ are the
effective relaxation times, and $\xi\sim 1$ is a dimensionless
factor which can change sign depending on the electron scattering
mechanism. For brevity, we will not consider the thermopower coefficient further. The inverse quantities
$\nu_{\sigma,\kappa,\eta}=\tau^{-1}_{\sigma,\kappa,\eta}$ are
called the effective collision frequencies. 
If the relaxation time approximation (\ref{eq:reltime_approx}) is applicable, the effective relaxation times become the actual relaxation times evaluated at the Fermi surface,
$\tau_\sigma=\tau_\kappa=\tau^1_e(\mu_e)$, $\tau_\eta=\tau^2_e(\mu_e)$, 
cf.~Eq.\ (\ref{eq:A_reltime}), since one approximates $\frac{\partial f^{(0)}}{\partial \varepsilon}\approx - \delta(\varepsilon - \mu_e)$.
In this case, we obtain the standard Wiedemann-Franz rule for conductivities,
\begin{equation}\label{eq:Wiedemann}
  \frac{\kappa}{\sigma} = \frac{\pi^2 k_B^2 T}{3e^2}.
\end{equation}
The relaxation time approximation holds when electron-ion collisions are the
dominant scattering mechanism and the energy $\omega$ transferred
in these collision is small $\omega\ll k_B T$. When this is not the case, the variational calculations outlined in Sec.~\ref{S:crust_boltz} are usually employed. It turns out that already the simplest variational approximation gives a satisfactory estimate for astrophysical conditions.
Moreover, the violation of the Wiedemann-Franz rule is not as dramatic as in ordinary metals at low temperature \cite{YakovlevUrpin1980SvA}.

When there are different relaxation mechanisms for the electron distribution function, for instance collisions with different particle species, the respective collision integrals must be added on the right-hand side of the Boltzmann equation. In practice, one usually considers different mechanisms separately  to obtain the effective collision frequency $\nu_{ej}$ for each scattering
process. Due to the strong degeneracy of electrons, the cumulated collision frequency $\nu_{\rm tot}=\sum_j \nu_{ej}$ obtained in this way is a good approximation to the solution of the Boltzmann equation with all mechanisms included. 
This is known as Matthiessen's rule \cite{ZimanBook}. The variational principle of kinetic theory allows us to estimate the error introduced by this approximation \cite{ZimanBook}, see also Ref.\ \cite{Potekhin:2015qsa}. Below we consider the most important processes that  determine the electron transport.

\subsubsection{Electron-ion collisions}\label{S:crust_ei}
The main process for electron transport is their scattering off ions. The ions in the neutron star crust
form a strongly coupled non-ideal plasma, whose state is defined
by an ion coupling parameter $\Gamma$. 
For a one-component plasma (in the sense that only one sort of ions is present)
\begin{equation}\label{eq:Gamma}
\Gamma=\frac{Z^2 e^2}{a_{\rm WZ} k_B T}\approx 153\, x_r \left(\frac{Z}{50}\right)^{5/3} \left(\frac{T}{10^8~{\rm K}}\right)^{-1}\, ,
\end{equation}
where the ion Wigner-Seitz cell radius $a_{\rm WZ}$ is defined by the relation 
\begin{equation}\label{eq:aWZ}
\frac{4\pi}{3}a_{\rm WZ}^3 n_{\rm i} = 1 \, .
\end{equation}
When $\Gamma\ll 1$, ions are in the gaseous phase, at $\Gamma\gtrsim 1$ in the liquid phase, 
and at $\Gamma=\Gamma_{\rm m}\approx 175$ \cite{Potekhin2000PhRvE}
the ion liquid crystallizes  and is thought to form a body-centered cubic lattice \cite{ChamelHaensel2008LRR}. This condition and Eq.~(\ref{eq:Gamma}) define the (density-dependent) melting temperature $T_{\rm m}$. Notice that the melting point can shift substantially if the electron polarization or magnetic field effects are taken into account \cite{Potekhin2000PhRvE,PotekhinChabrier2013A&A}. 
Another important parameter is the ion plasma temperature
\begin{equation}\label{eq:Tp}
T_{\rm pi}=\frac{\hbar}{k_B} \left(\frac{4\pi Z^2 e^2 n_{\rm i}}{m_{\rm i}}\right)^{1/2},
\end{equation}
above which the thermodynamic properties are classical, and below which quantum effects should be taken into account. 
In the context of electron transport, the important point is that at 
$T<T_{\rm pi}$ the typical energy transferred in the electron-ion collisions is $\omega\sim k_ BT$ and the relaxation time approximation cannot be used \cite{YakovlevUrpin1980SvA}. 
If $T_{\rm pi}< T_{\rm m}$, quantum effects are only important in the crystalline phase. A temperature regime where quantum effects are relevant in the liquid phase can in principle be realized for light elements and high densities.
In this case, the properties of the liquid -- including transport properties -- are modified, but also the crystallization point itself (the value $\Gamma_{\rm m}\approx 175$ is obtained from a classical estimate, not taking into account zero-point vibrations). Calculations show that at some density the crystallization temperature starts to decrease and reaches zero at a certain critical density, above which no crystallization occurs \cite{Chabrier1993ApJ,Jones1996PRL}. 
However, the importance of a quantum liquid regime for neutron star envelopes is questionable since nuclear reactions (electron captures and pycno-nuclear burning) would not allow light elements to exist at sufficiently large densities, see Sec.\ 2.3.5 of Ref.~\cite{HPY2007Book} for more details. Therefore, here we discuss quantum corrections only for the solid phase (see footnote \ref{fn4} for a brief remark about results for the quantum liquid regime).

For any phase state of the ions,  the effective electron-ion collision
frequency, to be used in (\ref{eq:kins_deg}), is usually written in terms of the effective 
Coulomb logarithm $\Lambda_{e{\rm i}}$,
\begin{equation}\label{eq:sigma_Coulomb}
  \nu_{e{\rm i}}=\frac{4\pi Z^2 e^4 n_{\rm i}}{p_{Fe}^2 v_{Fe}}
  \Lambda_{e{\rm i}} \approx 8.8\times 10^{17}\, \frac{Z}{50} \sqrt{1+x_r^2}\,\Lambda_{e{\rm i}}~{\rm s}^{-1}\, ,
\end{equation}
where $v_{Fe}=p_{Fe}/m^*_e$ and we have omitted the transport indices
$\sigma,\kappa,\eta$ for brevity.
The Coulomb logarithm is a central quantity in the transport theory of 
electromagnetic plasmas. In the (classical) liquid regime, $1\lesssim
\Gamma<\Gamma_{\rm m}$, we have $\Lambda_{e{\rm i}}\sim 1$, while in the solid regime
$\Lambda_{e{\rm i}}\propto T/T_{\rm m}$ at $T_{\rm m}> T\gtrsim 0.15\,T_{\rm pi}$ and $\Lambda_{e{\rm i}}\propto T^2/(T_{\rm m} T_{\rm pi})$ at $T\lesssim0.15 \, T_{\rm pi}$  \cite{Potekhinetal1999AA,Chugunov2005ARep,
Potekhin:2015qsa}.
For a one-component plasma it was calculated
by \citet{Potekhinetal1999AA} and \citet{Chugunov2005ARep}, including various effects such as
electron screening, non-Born and relativistic corrections, ion-ion
correlations in the liquid regime, and multi-phonon processes in the solid regime. 
The main complication in the calculation of the Coulomb logarithm is to properly take into account the ion-ion
correlations that are important in a strongly non-ideal Coulomb liquid. In
the conditions of the neutron star  crust, the typical electron
kinetic energy is much larger than the electron-ion interaction
energy (as mentioned in Sec.\ \ref{Sec:plasma}), and electrons can be treated as quasi-free 
particles scattering off the static electric potential created by
charge density fluctuations in the ion system. The resulting
expression in the first-order Born approximation, which is equally
applicable in liquid and solid states can be written as \cite{Baym1964}
\begin{equation}\label{eq:Lambdaei}
\Lambda_{e{\rm i}} = \int\limits_{q_0}^{2k_{Fe}} \frac{dk}{k} |k^2
U(k)|^2 \left[1-\beta_r^2\frac{k^2}{4 k_{Fe}^2}\right] R(k)
\int\limits_{-\infty}^{+\infty} d \omega \frac{z}{e^z-1}
G(k,z) S(\omega,k) \, ,
\end{equation}
where $z=\hbar \omega/(k_B T)$, $k_{Fe}=p_{F e}/\hbar$, $\beta_r=v_{Fe}/c$, $U(k)$ is the Fourier transform of the effective potential describing single electron-ion scattering\footnote{The long-range nature of the Coulomb interaction leads to a logarithmic divergence of the integral in (\ref{eq:Lambdaei}) since $U(k)\propto k^{-2}$ at small $k$, which is regularized by plasma screening, see Sec.~\ref{Sec:plasma}. Therefore, very roughly, $\Lambda_{e{\rm i}}\sim\log [2k_{Fe}/\max(q_0,r_D^{-1})]$, and hence the name `Coulomb logarithm'.}
\begin{equation}\label{eq:pseudopot}
U(k)=\frac{F(k)}{k^2 \epsilon(k)} \, ,
\end{equation}
which includes electron screening via the static dielectric function $\epsilon(k)$ and finite-size corrections for nuclei through the form-factor term $F(k)$, 
and the term in square brackets describes the relativistic
suppression of the backward scattering. In the liquid phase, $q_0=0$,
while in the solid phase, $q_0=q_{\rm BZ}=(6\pi^2 n_{\rm i})^{1/3}$, see below. The functions
$R(k)$ and $G(k,z)$ are kinematic factors depending on the transport property that is calculated, namely $R_{\sigma,\kappa}(k)=1$,
$R_\eta(k)=3[1-k^2/(4k_{F e}^2)]$, $G_{\sigma,\eta}(k,z)=1$, and
\begin{equation}\label{eq:Gkappa}
  G_\kappa(k,z)=1+\frac{z^2}{\pi^2}\left(3\frac{k_{Fe}^2}{k^2}-\frac{1}{2}\right) \, .
\end{equation}
Finally, $S(\omega,k)$ is the dynamical structure
factor which describes the ion density fluctuations,
\begin{equation}\label{eq:struct_fact_gen}
  S(\omega,\bm{k})=\frac{1}{2\pi N_{\rm
  i}}\int_{-\infty}^{+\infty} dt\,\int d^3 \bm{x} \, d^3 \bm{x}' \, e^{i\bm{k}\cdot(\bm{x}-\bm{x}')-i\omega
  t}
  \left\langle \delta\hat{n}^{\dag} (\bm{x},t)
  \delta\hat{n}(\bm{x}',0)\right\rangle_{\rm eq} \, ,
\end{equation}
where $\langle\dots\rangle_{\rm eq}$ stands for average over the Gibbs
ensemble of ions (thermal average),  $N_\mathrm{i}$ is the total number of ions, and
\begin{equation}\label{eq:dens_fluct}
\delta\hat{n}(\bm{x},t)=\hat{n}_{\rm
i}(\bm{x},t)-\langle\hat{n}_{\rm i}(\bm{x},t)\rangle_{\rm eq} \, ,
\end{equation}
with the ion number density
operator $\hat{n}_{\rm i}(\bm{x},t)$. 

Let us first consider a liquid with a temperature reasonably far above the melting temperature, $T>T_{\rm m}$.  Then $\langle\hat{n}_{\rm
i}(\bm{x},t)\rangle_{\rm eq}=n_{\rm i}$ takes into account
the uniform compensating background. 
Ignoring quantum effects in the liquid, as argued above, the $z\to 0$ limit can be used in the integrand of the 
$\omega$-integration in Eq.~(\ref{eq:Lambdaei}),
and one is left with the static structure
factor $S(k)$. This case corresponds to the relaxation time
approximation, and one obtains the Ziman formula known from transport theory of liquid metals \citep{Ziman1961PMag}. The Wiedemann-Franz rule
(\ref{eq:Wiedemann}) is also fulfilled. The static structure
factor can be calculated from numerical simulations of the
Coulomb plasma. In the absence of correlations, $S(k)\to 1$. \citet{Potekhinetal1999AA} used static structure
factors obtained by \citet{Young1991PhRvA}
and
provided a useful analytical fit for the Coulomb logarithm that
can be readily used in simulations.

Now consider the case $T<T_{\rm m}$, when ions are assumed to form a
perfect one-component body-centered cubic (bcc) crystal. The high symmetry of the cubic lattice implies that the transport properties are isotropic \cite{Harrison1980}. In this case, the electrons are
scattered off phonons, i.e., lattice vibrations.
The Coulomb logarithm is still given by
Eq.~(\ref{eq:Lambdaei}), where an expression for the structure
factor can now be obtained using a multi-phonon expansion. For 
temperatures not too close to the melting temperature the
single-phonon contribution to the structure factor is sufficient
\citep{FlowersItoh1976ApJ,YakovlevUrpin1980SvA}. 
In this regime, useful approximate expressions for the collision frequencies (that however do not include various corrections already mentioned above) are \cite{YakovlevUrpin1980SvA,BaikoYakovlev1995AstL,Chugunov2005ARep}
\begin{equation}\label{eq:ei_freq_simple}
\nu_{\rm ei}^{\kappa,\sigma}=\alpha_f u_{-2} \beta_r^{-1} \frac{k_B T}{\hbar}\left(2-\beta_r^2\right)F_{\kappa,\sigma}\left(\frac{T}{T_{\rm pi}}\right),\qquad \nu_{\rm ei}^{\eta}=\alpha_f u_{-2} \beta_r^{-1} \frac{k_B T}{\hbar} \left(3-\beta_r^2\right)F_{\eta}\left(\frac{T}{T_{\rm pi}}\right),
\end{equation}
where $\alpha_f$ is the fine structure constant, $u_{-2}=13.0$ is one of the frequency moments of the bcc lattice, and the functions $F(T/T_{\rm pi})$ describe quantum corrections,
 \begin{subequations}\label{eq:ei_lowtemp}
 \begin{eqnarray}
 F_\sigma(t)=F_\eta(t)&=&\frac{t}{\sqrt{t^2+a_0^2}}, \label{eq:ei_lowtemp_sigma}\\
 F_\kappa(t)&=&F_\sigma(t)+\frac{t}{\pi^2 u_{-2} \left(t^2+a_2^2\right)^{3/2}}\frac{\ln(4Z) - 1 -\beta_r^2}{2-\beta_r^2},\label{eq:ei_lowtemp_kappa}
 \end{eqnarray}
 \end{subequations}
where $a_0=0.13$ and $a_2=0.11$. Accordingly, when $T\gtrsim 0.15 \,T_{\rm pi}$ one can set $F_{\sigma,\kappa,\eta}=1$ in Eq.~(\ref{eq:ei_freq_simple}). In this classical limit, the relaxation time approximation still works fairly well and the Wiedemann-Franz  rule $\nu^\sigma_{\rm ei}=\nu^\kappa_{\rm ei}$ applies. The difference between $\nu^\eta_{\rm ei}$ and $\nu^{\kappa,\sigma}_{\rm ei}$ is due to the difference in the kinematic factor $R$ in Eq.~(\ref{eq:Lambdaei}). At low temperatures, $T\lesssim 0.15 \, T_{\rm pi}$, the relaxation time approximation breaks down and quantum effects are important. Since $F_{\kappa,\sigma,\eta}(t)\propto t$, the quantum corrections suppress the electron-ion collisions in this limit. Because of the second term in Eq.~(\ref{eq:ei_lowtemp_kappa}), which is a consequence of the factor (\ref{eq:Gkappa}), $\nu^\sigma_{\rm ei}\neq\nu^\kappa_{\rm ei}$ and the Wiedemann-Franz rule is violated. This violation is, however, not as dramatic as for terrestrial solids \cite{YakovlevUrpin1980SvA}.

It is important
to stress that the electron-phonon interaction in Coulomb crystals
in the astrophysical environment is very different from that in
terrestrial metals. For the latter, normal processes within
one Brillouin zone are dominant, $k\lesssim q_{\rm BZ}$, while in the astrophysical context, since electrons are quasi-free, with $k_{Fe}\gg q_{\rm BZ}$, the typical 
momentum transfer is large compared to $q_{\rm BZ}$, and
Umklapp processes, which transfer an electron from one Brillouin zone to
another, play the major role. 
At  very low temperatures,
the picture of quasi-free electrons is modified, since the distortion
of the quasi-spherical Fermi surface by band gaps
becomes important. This suppresses the Umklapp processes. However, 
\citet{Chugunov2012AstL} has shown that this `freezing' of the
Umklapp processes is only important at 
$T\lesssim 10^{-2} T_{\rm pi}$
and
is relatively slow, see also Ref.~\cite{PageReddy2012}. In practice, at
these temperatures the transport is dominated by other processes
(see below), and the freezing of 
Umklapp processes can be safely
neglected in practical calculations.

As the temperature of the Coulomb solid approaches the melting
temperature, $T\to T_{\rm m}$, the single-phonon picture is no longer
valid. \citet{Baiko1998PhRvL} calculated the multi-phonon
contribution to the structure factor $S(\omega,k)$ in the harmonic
approximation; these results were later incorporated in analytical
fits by \citet{Potekhinetal1999AA}. Recent quantum Monte Carlo
simulations have shown that the harmonic approximation works well
up to the vicinity of the melting temperature \cite{Abbar2015PhRvC}. Note that in a pure
perfect lattice, only the inelastic part $S'(\omega,k)$ of the
total structure factor
$S(\omega,k)=S'(\omega,k)+S''(k)\delta(\omega)$ contributes to
transport properties. The elastic term $S''(k)$ describes Bragg
diffraction (zero-phonon process). It does not contribute to
scattering, but it leads to a renormalization of the electron
ground state (which are the Bloch waves) and the appearance of the
electron band structure. Notice that the elastic component is
automatically taken out by $\langle\hat{n}_{\rm
i}(\bm{x},t)\rangle_{\rm eq}$ in Eq.~(\ref{eq:dens_fluct})
\cite{Baym1964,RosenfeldStott1990PhRvB}. The Bragg elastic
contribution to an unmodified density (charge) correlator $\langle
\hat{n}^\dag \hat{n}\rangle$ is
\begin{equation}\label{eq:S_el_DW}
S''({\bm k})= e^{-2W(k)} (2\pi)^3 n_{\rm i} \sum_{\bm
G}\delta({\bm k}-{\bm G})\,,
\end{equation}
where the summation is taken over the reciprocal
lattice vectors ${\bm G}$ and the exponent $W(k)$ is the Debye-Waller
factor 
\cite{Harrison1980}, 
which describes thermal damping of the Bragg
peaks. In addition, \citet{Baiko1998PhRvL} have proposed that in
the liquid regime, sufficiently close to the melting point, an
incipient long-range order exists, which is preserved during the
typical electron scattering time. Solid-like features such as a
shear mode are observed in a strongly coupled system in the liquid
regime both in numerical experiments and in laboratory. Thus,
\citet{Baiko1998PhRvL} suggested that the electrons obey the local
band structure which is preserved during the electron relaxation. As a consequence, in order to account for this ion local
ordering in the electron transport, they proposed to subtract an
`elastic' contribution given by Eq.~(\ref{eq:S_el_DW}) averaged
over the orientations of ${\bm k}$ from the total liquid structure
factor. This procedure removes the large jumps of the Coulomb
logarithm and hence of the transport coefficients at the melting
point. This prescription allowed \citet{Potekhinetal1999AA} and
\citet{Chugunov2005ARep} to construct a single fit for $\Lambda_{e{\rm i}}$
valid in both liquid and solid regimes. An interesting feature of
the approach by \citet{Potekhinetal1999AA} is that they do not fit
the numerical results for the Coulomb logarithms. Instead, they
introduce a fitting expression for the effective potential which
encapsulates the contributions from non-Born terms, electron screening, ion correlations, the Debye-Waller factor, and the structure factor.
The
Coulomb logarithms are then found by analytical integration in
Eq.~(\ref{eq:Lambdaei}).\footnote{This fit has also been applied to  transport coefficients in a liquid at $T\lesssim T_\mathrm{pi}$, where quantum effects become important. It is supposed \cite{Potekhinetal1999AA,Chugunov2005ARep} 
to give a more reliable estimate than the use of direct numerical calculations based on the classical structure factors.  This is reasonable since  a unified analytical expression in both liquid and solid phase is used and in the latter phase quantum effects are properly included, see Ref.~\cite{Potekhinetal1999AA} for a detailed argumentation. Robust results for transport coefficients in the quantum liquid domain are not present in the literature up to our knowledge since the structure factors in the quantum liquid regime are unknown.\label{fn4}}

This approach is attractive but it was criticized in Refs.~\cite{Itoh2008ApJ,DaligaultGupta2009ApJ}. The main argument is
that in the simple terrestrial metals the jump in resistivity at
the melting point is a well-established indication of a
solid-liquid transition \cite[e.g.,][]{Schaeffer2012PhRvL}.
It seems that a convincing way to describe electron transport in the
disordered state of the strongly coupled Coulomb melt is missing.
It is, in principle, possible to extract the behavior of the
crustal thermal conductivity from studies of the crustal cooling
in X-ray transients after the outburst stages \cite{BrownCumming2009,PageReddy2013PhRvL,Meisel2018JPG}.
However, in this case, effects related to the multi-component
composition of the accreted crust will probably dominate \cite{Mckinven2016ApJ}.

\subsubsection{Impurities and mixtures}\label{S:kincoeff_imp}
The crustal lattice is not expected to be strictly
perfect. Like terrestrial crystalline solids, it
can possess various defects, which are jointly
called impurities. One usually considers impurities in the form of
charge fluctuations and introduces the impurity parameter
\begin{equation}\label{eq:Qimp}
  {\cal Q}=\sum_j Y_j (Z_j-\langle Z\rangle)^2 \, ,
\end{equation}
where the summation is taken over the different ion species,
$Y_j$ and $Z_j$ are number fraction and charge number 
of each species, respectively, and $\langle Z\rangle$ is the mean
charge. If the impurities are relatively rare and weakly
correlated, electron-phonon interactions and electron-impurity scatterings can be considered as different transport relaxation mechanisms. Employing Matthiessen's rule, the total electron-ion collision frequency is expressed as
$\nu_{e{\rm i}}=\nu_{e-{\rm ph}}+\nu_{e-{\rm imp}}$. The electron-impurity effective
collision frequency $\nu_{e-{\rm imp}}$ is calculated form Eq.~(\ref{eq:sigma_Coulomb})
by substituting $Z^2\to{\cal Q}$ and using the Coulomb logarithm from
Eq.~(\ref{eq:Lambdaei}) with the elastic structure factor $S(k)=1$. 
Since the
elastic scattering is temperature-independent, it limits the
collision frequencies at low temperatures. In the simplest model of Debye screening, $U(k)\propto(k^2+k_D^2)^{-1}$, and the integration in Eq.~(\ref{eq:Lambdaei}) gives
\begin{subequations}\label{eq:L_imp}
\begin{eqnarray}
\Lambda_\mathrm{imp}^{\kappa,\sigma}&=&\frac{1}{2}\left[1+4\beta_r^2\xi_S^2\right]\ln\left(1+\xi_S^{-2}\right)-\frac{\beta^2_r}{2} - \frac{1+\beta_r^2\xi_S^2}{2+2\xi_S^2} \, ,
\label{eq:L_impkappa}\\ 
\Lambda_\mathrm{imp}^{\eta}&=&\frac{3}{2}\left[1+3\beta_r^2\xi_S^4+2\xi_S^2(1+\beta_r^2)\right]\ln\left(1+\xi_S^{-2}\right)-\frac{9}{2}\beta_r^2 \xi_S^2 -\frac{3}{4}\beta_r^2-3 \, ,
\label{eq:L_impeta}
\end{eqnarray}
\end{subequations}
where $\xi_S=k_D/(2 k_{Fe})$. The screening wavenumber $k_D$ in principle  acquires contributions from Thomas-Fermi screening of degenerate electrons and impurity screening, $k_D^2=k_\mathrm{TF}^2+k_\mathrm{imp}^2$, however $k_\mathrm{imp}$ can usually be neglected (e.g., \cite{Chugunov2005ARep}).

In the opposite case, 
when no crystal is formed in a multi-component
plasma (in a liquid, or in a glassy solid), the so-called plasma
additivity rule can be used \citep{Potekhinetal1999AA},
and $Z^2 n_{\rm i} \Lambda_{e{\rm i}}$ is replaced by $\sum_j
Z_j^2 n_j \Lambda^j_{e{\rm i}}$, where $\Lambda_{e{\rm i}}^j$ is the Coulomb
logarithm for scattering off the ion species $j$. A
modification of this rule was proposed by
\citet{DaligaultGupta2009ApJ} based on large scale molecular
dynamical simulations. They suggest that it is more accurate to
use $\langle Z\rangle ^{1/3} Z_j^{5/3}$ instead of $Z_j^2$.

The intermediate case is more complicated. Molecular dynamics
simulations strongly suggests that the crystallization of the
multi-component Coulomb plasma occurs even in the case of large
impurity parameter ${\cal Q}$
\citep{Horowitz2009PhRvE,Horowitz2009PhRvC}. An amorphous crust
structure was also proposed, see for 
instance Ref.~\citep{DaligaultGupta2009ApJ}.
Some studies show that the diffusion in the solid phase is
relatively rapid and quickly relaxes amorphous structures to a
regular lattice \citep{Hughto2011PhRvE}. In addition, an amorphous
crustal structure is in contradiction with observations \cite{Shternin2007MNRAS, BrownCumming2009}. 
Already in the case of a moderate 
impurity parameter, ${\cal Q}\sim 1$,  the
simple prescription of electron scattering as a sum of phonon
contribution and uncorrelated impurity scattering is
questionable. In fact, all information about electron-ion
scattering (from lattice vibrations or impurities) is 
encoded
in the structure factor, which naturally takes into account
correlations in the minority species on the same footing
as the correlations in the majority species. The structure factor of a multi-component solid can
be obtained from numerical simulations. To calculate the transport properties it is necessary
to correctly separate the Bragg contribution, which does not
contribute to scattering, from the total structure factor. This is not as simple as in case of one-component plasma
\citep{Horowitz2009PhRvE}. The remaining part of the structure
factor is then used to calculate the Coulomb logarithms. As a
result, both classical molecular dynamics simulations
\citep{Horowitz2009PhRvE, Horowitz2009PhRvC} and recent quantum
path integral Monte Carlo approach \citep{Abbar2015PhRvC, Roggero2016PhRvC} show
that the simple impurity expression based on the parameter ${\cal Q}$
 underestimates the Coulomb logarithm and hence
overestimates the corresponding values of transport coefficients.
Moreover,  \citet{Roggero2016PhRvC} found that their results for a
broad range of ${\cal Q}$ can be approximated by the
standard lattice + impurity formalism, where the effective impurity
parameter $\tilde{{\cal Q}}=L(\Gamma) {\cal Q}$ is
used\footnote{An appropriate average of individual species $\Gamma$'s calculated from the first equality in Eq.~(\ref{eq:Gamma}) is used as the mixture $\Gamma$ parameter.}. The factor $L(\Gamma)$ is generally larger than one and
increases with $\Gamma$.  \citet{Roggero2016PhRvC} find
$L(\Gamma)\approx 2-4$ for the conditions they consider. Note that
classical simulations can treat only the high-temperature case
$T>T_{\rm pi}$, while the quantum simulations of  \citet{Roggero2016PhRvC}
were the first to investigate the multi-component solid for $T<T_{\rm pi}$,
where the dynamical effects in Eq.~(\ref{eq:Lambdaei}) are
important.

\subsubsection{Other processes}\label{S:crust_ee}
Let us briefly describe other processes which contribute to
transport in neutron star crusts. Electrons in the crust can scatter off
electrons, not only off ions. For degenerate electrons, Matthiessen's rule  is a good approximation, and the 
electron-electron collision frequency $\nu_{ee}$ is simply added to the
electron-ion collision frequency $\nu_{e{\rm i}}$. The impact of the contribution from electron-electron scattering 
on thermal conductivity $\kappa$ and shear viscosity $\eta$ was analyzed in Refs.~\cite{Shternin2006PhRvD,Shternin2008JPhA}.
Note that in this approximation electron-electron scattering 
does not change the charge current and therefore does not contribute to
the electrical conductivity\footnote{This is not the case in the non-degenerate plasma, where Matthiessen's rule does not hold, and both $ee$ and $e$i collisions need to be considered on the right-hand side of the Boltzmann equation. The impact of $ee$ collisions is then especially pronounced at small $Z$ \cite{Braginskii1957}.}. 

In most part of the neutron star crust, electrons
are relativistic and their collisions are mediated by the
current-current (magnetic) interaction, in contrast to the
electron-ion Coulomb interaction. The current-current interaction
occurs through exchange of transverse plasmons, which leads to a
peculiar temperature and density dependence of the transport
coefficients, as we describe in detail in the context of lepton 
and quark transport in the core of the star, see Secs.~\ref{S:core_nuc_lepton} and \ref{S:kincoeff_uqm}.
However, except for a very low-temperature, pure one-component
plasma, the electron-electron collisions are found to be
unimportant. (They can be important in a low-Z plasma, i.e.,
in white dwarfs and degenerate cores of the red giants. In fact,
the correct inclusion of the electron-electron collisions have
important consequences for the position of the red giant branch
tip in the Hertzsprung-Russell diagram \citep{Cassisi2007ApJ}.)

Ions in the liquid phase (or phonons in the crystalline solid phase) can also
contribute to transport properties of neutron star crusts. The ion
contribution to shear viscosity was considered by
\citet{Caballero2008PhRvC} and is found to be negligible. A
similar conclusion for the thermal conductivity was reached by
\citet{Chugunov2007MNRAS}, see also \citep{Perez2006AA}. However,
in a certain parameter region, the ion contribution can be
significant in the magnetized crust for the heat conduction
across the field lines. Still, simulations suggest that its
importance is limited also in this case, see for example \cite{Potekhin:2015qsa}.

\subsubsection{Inner crust: free neutron
transport}\label{S:crust_neutrons}
In the inner crust of a neutron star the density becomes sufficiently 
high for neutrons to detach from nuclei. The structure of the inner crust
then consists of a lattice of nuclear clusters (where charged
protons are localized) alongside with the gas of unbound (or
`free') neutrons, see, e.g., \citet{ChamelHaensel2008LRR}. In
addition, the neutrons are believed to form Cooper pairs in the
$^1S_0$ channel.
The charge distribution in the nuclear clusters in the inner crust
differs from the point-like nuclei in the outer crust. This is taken
into account by introducing nuclear form factors in the
electron-nuclei scattering potential. These corrections have been 
included by \citet{Gnedin2001MNRAS} for Coulomb logarithms
relevant to thermal and electrical conductivities and by
\citet{Chugunov2005ARep} for shear viscosity. The finite size of the
charge distribution generally reduces the collision frequencies and
hence increases the values of electron transport coefficients.

In the presence of a large amount of free neutrons, electron-neutron
scattering can become important.  The relativistic electrons
interact with the neutron spins (magnetic moments). This
contribution was analyzed by \citet{FlowersItoh1976ApJ}. Recently,
an induced interaction between electrons and neutrons was
proposed \citep{Bertoni2015PhRvC}, which can be effectively
understood as occurring via exchange of lattice phonons. However,
\citet{Bertoni2015PhRvC} found that the contribution from this
interaction is never relevant when calculating kinetic
coefficients in the inner crust. In contrast, a similar interaction
can be important in the core (see Sec.~\ref{S:core_nuc_pairing}). If neutrons are
superfluid, both these contributions are further suppressed.

Since the gas of unbound neutrons is present in the inner crust, they can also contribute themselves to the transport properties. For instance, the thermal conductivity becomes a sum of
electron and neutron contributions, $\kappa=\kappa_e+\kappa_n$. The
neutron contribution for normal neutrons was discussed by
\citet{Bisnovaty1982} and more recently by \citet{Deibeletal2017ApJ}, 
and it was
found to be negligible  compared to the electron contribution, except probably the region near the crust-core boundary \cite{Deibeletal2017ApJ}.
 We are not aware of any calculations for the shear viscosity of the 
 neutron fluid in the inner crust. The
potential importance of the free neutron transport is further
reduced if one takes into account that the unbound neutrons move
in the periodic potential of the nuclear lattice, hence their spectrum 
shows a band structure. \citet{Chamel2012PhRvC} has argued that due to 
Bragg scattering of neutrons the actual density of conducting
neutrons that participate in transport is much smaller than the
total density of unbound neutrons, which further reduces the role
of neutrons.

When neutrons are superfluid, a collective superfluid mode (`superfluid phonons') can contribute to transport, as explained in Sec.\ \ref{S:cooper}. Initial estimates suggested
that the superfluid phonon contribution to the thermal
conductivity can be important in magnetized stars
\citep{Aguilera2009PhRvL}. However, more detailed considerations
which include the neutron band structure have shown that this 
contribution is always less than the contribution of lattice phonons
\citep{Chamel2013PhRvC, Chamel2016JPhCS}. We will come back to 
collective modes in the discussion of the core, see Secs.\ \ref{S:core_nuc_pairing} and \ref{sec:nu_Cooper} for superfluid phonons in nuclear matter, and Secs.\ \ref{S:quark_colsup1} and \ref{S:quark_colsup} for superfluid phonons in quark matter.

\subsubsection{Transport in a magnetic
field}\label{S:crust_mag}
The magnetic field $\bm{B}$ in the crust modifies the motion of
charged particles in the directions perpendicular to the direction
of the magnetic field $\bm{b}\equiv\bm{B}/B$. It can be strong
enough to have an influence on the transport properties. Electrons
are light and thus lower fields affect their transport (compared to the 
fields needed to affect ions). We start from the situation where the electron motion across the magnetic field is not quantized. In this 
case, magnetic field effects are characterized by
the Hall magnetization parameter
\begin{equation}\label{eq:Hallmag}
\omega_g
\tau=1760\frac{B_{12}}{\sqrt{1+x_r^2}}\frac{\tau}{10^{-16}~{\rm
s}},
\end{equation}
where $\tau$ is the characteristic relaxation time, $B_{12}\equiv B/(10^{12}\, {\rm G})$, and $\omega_g=|e|B/(m^*_e c)$ is the electron gyrofrequency, which is
related to the electron cyclotron  frequency $\omega_c=|e|B/(m_e
c)$ by $\omega_g=\omega_c/\sqrt{1+x_r^2}$. If
$\omega_g\tau\gtrsim 1$, the electron transport becomes
anisotropic. Let us first consider the electrical and thermal
conductivities (the perturbation of multiplicity $l=1$, see Sec.~\ref{S:crust_boltz}). The
general expressions for the currents (\ref{T-E}) 
are modified such
that the kinetic coefficients become tensors instead of scalars
$\kappa,\, \sigma,\, Q_T \to
\hat{\kappa},\,\hat{\sigma},\,\hat{Q}_T$. Accordingly, one
introduces  the effective relaxation time tensors
$\hat{\tau}^{\sigma,\kappa}$ via Eq.\ (\ref{eq:kins_deg}). 
The symmetry relations for the kinetic
coefficients in isotropic media suggest that these tensors have only three
independent components. If one aligns the $z$-axis along $\bm{b}$,
these are longitudinal $\hat{\tau}_{zz}\equiv \tau_\parallel$,
transverse $\hat{\tau}_{yy}=\hat{\tau}_{xx}\equiv \tau_\perp$, and
Hall terms $\hat{\tau}_{xy}=-\hat{\tau}_{yx}\equiv\tau_\Lambda$.
These tensors are found from the solution of the linearized
Boltzmann equation in an external magnetic field. The procedure is
similar as described in Sec.~\ref{S:crust_boltz}. However, now the  force $\bm{R}$
contains the magnetic field contribution, and the term
$\frac{e}{c}\bm{w}\times\bm{B}\frac{\partial \delta f}{\partial
\bm{p}}$ must be retained in Eq.\ (\ref{boltz_lin_lhs}) \cite{ZimanBook}. One usually adopts the
relaxation time approximation (\ref{eq:reltime_approx}), 
where the  relaxation time $\tau^1(\varepsilon)$ is taken from the non-magnetic
problem. In this approximation, the solution to the linearized
Boltzmann equation gives
\begin{equation}\label{eq:taus_mag}
  \tau_{\parallel} = \tau \, ,\qquad \tau_{\perp}=\frac{\tau}{1+(\omega_g
  \tau)^2} \, ,\qquad \tau_\Lambda = \frac{\omega_g \tau^2}{1+(\omega_g
  \tau)^2} \, .
\end{equation}
In fact, an averaging of these relaxation times should be performed following Eq.\ (\ref{kincoeff_boltz}). However, in degenerate matter it is sufficient to set $\tau=\tau^1(\mu_e)$, 
like in the non-magnetized case. 
In the limit of weak magnetization,  $\omega_g \tau \ll
1$, one has $\tau_\parallel=\tau_\perp=\tau$ and $\tau_\Lambda=0$.
In the opposite case of a large Hall magnetization parameter, the
electron  transport across the magnetic field becomes strongly
suppressed and ion or neutron contributions can become important.

If the magnetic field is sufficiently strong, the quantization of the
transverse electron motion can no longer be neglected. This happens when $\hbar
\omega_g \gtrsim k_B T$.  The electrons then occupy several  Landau levels (weakly quantizing field) or only the lowest Landau level 
(strongly quantizing field). In either case, the magnetic field also modifies the
thermodynamic properties of the system.  Transport along and
across the magnetic field must be considered separately. The
thermal and electrical conductivities in a quantizing magnetic field
in different regimes were investigated by many authors
\cite{Kaminker1981TMP,Yakovlev1984ApSS,Hernquist1984ApJS,Potekhin1996AA,PotekhinYakovlev1996AA}.
The results for both quantizing and non-quantizing fields in
the relaxation time approximation were reconsidered and summarized by
\citet{Potekhin1999AA}. He suggested
that in the case of strongly degenerate electrons the
form of Eq.~(\ref{eq:taus_mag}) holds, but two different
relaxation times $\tau$ must be used in the expressions for the parallel
component $\tau_\parallel$ and the transverse
components $\tau_\perp$ and $\tau_\Lambda$. In the weakly
quantizing limit, these two relaxation times oscillate around
$\tau$, approaching it in the non-quantizing limit. Based on the model
of the effective electron-ion potential \cite{Potekhinetal1999AA}
(see Sec.~\ref{S:crust_ei}), \citet{Potekhin1999AA} constructed
useful fitting expressions to calculate Coulomb logarithms
appropriate for thermal and electrical conductivities of  magnetized
electrons in both quantizing and non-quantizing limits in liquid
or solid neutron star crusts, as long as quantum effects on the ion
motion can be ignored, i.e., at $T\gtrsim T_{\rm pi}$. By
construction, these expressions provide transport coefficients
which behave smoothly across the liquid-solid phase transition
(recall the discussion in Sec.~\ref{S:crust_ei}).

Recently, finite-temperature effects on the electrical
conductivity of warm magnetized matter in the neutron star crust were discussed 
by \citet{HarutyunyanSedrakian2016PhRvC}. These authors used the
relaxation time approximation, but included also the transverse
plasmon exchange channel when calculating the electron-ion
transport cross-section. This channel was found
to be suppressed by a small factor $k_BT/(m_{\rm i} c^2)$ and does not
contribute to the relaxation time. 

All results for transport coefficients in magnetized matter
described above were based on various sorts of the relaxation time
approximation. This approach is justified if the scattering
probability does not depend on $\bm{B}$ and if the scattering is
elastic. Both these approximations fail in general at low
temperatures, when the crust is solid
\citep[e.g.,][]{Baiko2016MNRAS, Chugunov2007MNRAS}. In this case,
a more general expression for the collision integral must be used,
and the solution of the Boltzmann equation becomes more
complicated. Unfortunately, the construction of the variational principle in the magnetized case is challenging \citet{ZimanBook}. In the standard approaches, the solution of the Boltzmann equation corresponds only to a stationary point of the variational functional among the class of the trial functions, not to its maximum\footnote{A promising variant of the variational principle was recently suggested by \citet{ReinholzRoepke2012PhRvE}, where the positive-definite variational functional was proposed.}.  
 However, for degenerate matter, relying on the experience from the non-magnetized case, one expects the lowest-order expansion of the deviation function $\Phi$ to give appropriate results.
Based on this expectation, \citet{Baiko2016MNRAS} studied electron electrical
and thermal conductivities in the magnetized, solid crust employing the \citet{ZimanBook}  approach. 
He used
the one-phonon approximation for the electron-lattice interaction and
took into account the phonon spectra distortion due to the magnetic
field. The magnetic field leads to the appearance of a soft phonon
mode, with quadratic dispersion at small wavenumbers. This mode is easier to excite than the usual
non-magnetized acoustic phonon, therefore the electrical and
thermal resistivities increase. Employing the lowest order of the
variational method, and aligning the magnetic field along
one of the symmetry axes of the crystal, \citet{Baiko2016MNRAS} found
that the thermal and electrical conductivity tensors are expressed
via effective relaxation times as in Eq.~(\ref{eq:taus_mag}),
but like for a quantizing magnetic field, two different relaxation
times enter the longitudinal and transverse
parts. The difference between these
effective relaxation times increases with magnetic field. At low
temperatures, $T\lesssim T_{\rm pi}$, both  relaxation times are
appreciably larger than in the field-free case. The results of
Ref.~\cite{Baiko2016MNRAS} are strictly valid in the
non-quantizing case. In this case, phonons are weakly magnetized.
However, the results are also relevant for weakly quantized fields,
when electrons populate several Landau levels, and yield estimates
of the transport coefficients averaged over the quantum
oscillations. The most relevant case of highly magnetized
phonons, where the influence of the magnetic field on $\kappa$ and
$\sigma$ is largest, corresponds to the strongly quantizing
magnetic field, where electrons populate only the lowest Landau
level and the approach used by \citet{Baiko2016MNRAS} is
inappropriate. An accurate analysis of the transport properties
of quantized electrons in strongly magnetized Coulomb
crystals has yet do be done.

The effects of a magnetic field on the shear viscosity of the crust has
not received as much attention as the thermal and electrical
conductivities. The electron shear viscosity was considered by
\citet{Ofengeim2015EL} for the non-quantizing magnetic field,
taking into account only electron-ion collisions in the
relaxation-time approximation. In an anisotropic medium, where the
anisotropy is for instance caused by an external magnetic field, the 
viscous stress tensor contains the fourth-rank tensor $\eta_{\alpha\beta\gamma\delta}$ instead of the scalar
coefficient $\eta$. Symmetry constraints leave five
independent shear viscosity coefficients $\eta_0\dots \eta_4$ \cite{Landau10eng}. 
For degenerate electrons, the
expressions for the five shear viscosity coefficients in the 
relaxation time approximation are rather simple
\cite{Ofengeim2015EL}. The coefficients $\eta_0$, $\eta_2$, and
$\eta_4$ are given by Eq.~(\ref{eq:eta_deg}) where
$\tau_\parallel$, $\tau_\perp$, and $\tau_\Lambda$ from
Eq.~(\ref{eq:taus_mag}) are used, respectively. The coefficient
$\eta_0$ is independent of $B$ and can be called longitudinal
viscosity in analogy to longitudinal conductivities. The two
remaining coefficients can be found from the relations
$\eta_1(B)=\eta_2(2B)$ and $\eta_3(B)=\eta_4(2B)$. The
`Hall' viscosity coefficients $\eta_3$ and $\eta_4$ do not enter
the expression for the energy dissipation rate. More accurate
calculations of the shear viscosity of the magnetized neutron star crust
should deal with the various effects outlined in the previous
discussion on conductivities. This remains for future studies.

\subsection{Transport in the pasta phase}\label{S:kincoeff_pasta}
 As the density in the inner crust increases, 
 the size of the nuclei -- or better: the size of the nuclear clusters -- 
increases until the clusters start to overlap. The density at the crust-core interface $\rho_{cc}$, where nuclei are
fully dissolved in uniform nuclear matter, is about $\rho_{cc}\approx \rho_0/2=  1.4\times 10^{14}\ {\rm g\,cm}^{-3}$, where $\rho_0$ is the mass density at nuclear saturation. 
It is now generally believed that the transition region hosts several 
phases that are characterized by peculiar shapes of the nuclear clusters,
reminiscent of various shapes of pasta. Hence the term `nuclear pasta' for these phases. 
Loosely speaking, when the spherical nuclear clusters start
to touch, as a result of the competition between the nuclear
attraction and Coulomb repulsion of protons, it may become
energetically favorable for them to rearrange and form elongated
structures like rods, or two-dimensional slabs. This was first
pointed out by \citet{Ravenhall1983PhRvL} and
\citet{Hashimoto1984PThPh}. In a simple picture, five subsequent
phases appear as we increase density, i.e., as we move from the crust into the core of the star: first, usual large
spherically shaped clusters (`gnocchi'), then
cylindrical rods (`spaghetti'), then plane-parallel slabs (`lasagna'), followed by the inverted phases, with rod-like voids, then spherical voids
(`anti-gnocchi' or `swiss cheese') in nuclear matter.
At high temperature, the pasta is in the liquid state, but at low
temperatures it is thought to freeze in ordered or disordered
structures, for example in a regular lattice of slabs. The pasta
region is estimated to exist between densities of about $10^{14}\, {\rm g}\,
{\rm cm}^{-3}$ and $\rho_{cc}$, being about 100~m thick; the total mass of the pasta layer can be as large as the mass of the rest of the crust.
The appearance of the pasta phases in simulations
depends on the details of the interaction and implementation, and
there are models that predict less pasta phases, mixtures of
different phases, or do not predict the pasta phases at all \cite{Douchin2000PhLB,Oyamatsu2007PhRvC}. In
modern models, where large-scale simulations are employed, there
is a rich variety of possibilities for pasta phases, see for instance
Refs.~\cite{Alcain2014PhRvC,Schneider2014PhRvC, Horowitz2015PhRvL,
Berry2016PhRvC}, and Refs.~\cite{Caplan2016arXiv} and
\cite{Yakovlev2015bMNRAS} for more detailed reviews.

The complexity of the nuclear pasta naturally suggests that its transport
properties can be very different from the rest of the crust. The
main contribution to the conductivities and shear viscosity comes
from electrons which now scatter off the non-trivial charge
density fluctuations of the pasta phase. In applications, it is not
unreasonable to treat the transport coefficients for the pasta
phase as phenomenological quantities. In analogy to the
treatment of disorder in the crust, one can introduce an
effective impurity parameter $\tilde{\cal Q}$ to
parametrize the transport coefficients. For instance, assuming that the
pasta layer has much lower electrical conductivity (with
$\tilde{\cal Q}\approx 100$) than the rest of the crust,
\citet{Pons2013NatPh} were able to explain the existence of the
maximal spin period of X-ray pulsars (see also
Ref.~\cite{Vigano2013MNRAS} for the effect of the resistive pasta
layer on the magnetic field evolution of isolated neutron
stars). In a similar way, assuming that the pasta is a thermal
insulator with $\tilde{\cal Q}\approx 40$,
\citet{Horowitz2015PhRvL} were able to explain the late-time
crustal cooling in the quasi-persistent X-ray transient
MXB~1659--29 \cite{Cackett2013ApJ}. Notice that the effective
impurity parameter, of course, does not have to be the same when 
different transport coefficients
$\kappa$, $\sigma$, or $\eta$ are considered.

\citet{Horowitz2008PhRvC} computed shear viscosity and
thermal conductivity of the pasta phase based on 
classical molecular dynamics simulations. The electrons scatter off
the charged protons, whose correlated dynamics is described by the
proton structure factor $S_p$. Thus, one can use the expressions
given in the previous section for electron-ion scattering with
$S_p$ replacing $S$, and using the proton charge $Z=1$. As pointed
out by \citet{Horowitz2008PhRvC}, this approach applies also if
nuclei form spherical clusters (ions) of a charge $Z$, which will
be reflected in the proton structure factor. The results show
that the transport coefficients obtained in this way do not
change dramatically when non-spherical pasta phases are
considered. In fact, \citet{Horowitz2008PhRvC} obtained the same
order-of-the magnitude values as can be inferred from
Ref.~\cite{Chugunov2005ARep}, where spherical nuclei were considered in the same density range. Since classical molecular dynamics
simulations are used, these results are applicable only for high
temperatures. \citet{Horowitz2008PhRvC} set $T=1$~MeV and the
proton fraction $Y_p=0.2$. These values do not apply directly to
 neutron star crusts, and the authors discuss how smaller
proton fractions and smaller temperatures might modify their conclusions. Similar conclusions were reached recently by \citet{NandiSchramm2018ApJ} based on quantum molecular dynamics simulations for a wider range of parameters than in Ref.~\cite{Horowitz2008PhRvC}.

\citet{Horowitz2008PhRvC} and \citet{NandiSchramm2018ApJ} used the expression (\ref{eq:Lambdaei}) for
calculating the Coulomb logarithm. This expression is based on the
angular-averaged structure factor $S_p(q)$, which assumes 
isotropic, or nearly isotropic, scattering. It is clear that this
is not the case in nuclear pasta. One can imagine that 
electron scattering should be much stronger in directions across
the pasta clusters than in directions along them. This is indeed
reflected in the strong dependence of the structure factor
$S_p(\bm{q})$ on the direction of the vector $\bm{q}$
\cite{Schneider2014PhRvC,Schneider2016PhRvC}. Transport in nuclear pasta
is essentially anisotropic,  and the transport theory
of anisotropic solids must be applied. The solution of the
Boltzmann equation becomes more complicated since the collision
integral involves anisotropic scatterings.
The transport coefficients in anisotropic materials become tensor
quantities, just like for the magnetized case considered above, where
the anisotropy (gyrotropy) was induced by the magnetic field.

If the scattering is still elastic, but anisotropic, the relaxation-time
approximation (\ref{eq:reltime_approx}) generalizes to
\begin{equation}
I_e=-\sum_{lml'm'} \delta
f^{lm}(\varepsilon)\left[\hat{\nu}_e(\varepsilon)\right]^{l'm'}_{lm}
Y_{l'm'}(\Omega_{\bm{p}})\, ,\label{eq:reltime_anisotropic}
\end{equation}
where $\hat{\nu}_e(\varepsilon)$ is the inverse relaxation time
(collision frequency) matrix. In the isotropic case, one has
$\left[\hat{\nu}_e(\varepsilon)\right]^{l'm'}_{lm}=\left[\tau^l_e(\varepsilon)\right]^{-1}
\delta_{ll'}\delta_{mm'}$, and we recover Eq.~(\ref{eq:reltime_approx}). In
principle, the expression for the matrix elements
$\left[\hat{\nu}_e(\varepsilon)\right]^{l'm'}_{lm}$ can be
expressed in integral form employing the proton structure factor
$S_p(\bm{q})$ in a similar way to
Eqs.~(\ref{eq:sigma_Coulomb})--(\ref{eq:Lambdaei}).

An essential property of the general anisotropic case is that the
perturbations of different multiplicities $l$ can mix. It is
customary to assume, however, that this mixing is small and can be
neglected, so that $l=l'$ in Eq.~(\ref{eq:reltime_anisotropic}),
see for instance Ref.~\cite{AskerovBook}. \citet{Yakovlev2015bMNRAS} employed this
approximation and considered electrical and thermal conductivities (i.e., 
$l=1$) of the anisotropic pasta, including also 
a magnetic field, and assuming that the pasta phase has 
a symmetry axis (not necessarily aligned with the magnetic field). 
The
$l=1$ perturbation of the distribution function can be written as
$\Phi = -\bm{w}\cdot \bm{\vartheta}$, where the vector $\bm{\vartheta}$ has to be determined. If we orient the $z$-axis of the laboratory system along the
pasta symmetry axis, the generalized relaxation time approximation
can be written as \cite{Yakovlev2015bMNRAS}
\begin{equation}\label{eq:collint_anis_nus}
  I_e=-\frac{\partial f^{(0)}}{\partial \varepsilon}\left[\nu_a(\varepsilon) w_z \vartheta_z + \nu_p(\varepsilon) \bm{w}_p \cdot \bm{\vartheta}_p\right] \, ,
\end{equation}
where the collision
frequencies $\nu_a$ and $\nu_p$ describe relaxation along and across the
symmetry axis, respectively,
where $\bm{w}_p$ is the electron velocity component transverse to
the symmetry axis, and $\bm{\vartheta}_p$ is the corresponding component of
$\bm{\vartheta}$. Using this expression for the collision integral,
one solves the Boltzmann equation containing electric field,
temperature gradient, and external magnetic field 
to find the vector $\bm{\vartheta}$. Then, the thermal and electrical
conductivity tensors (and the thermopower) can be found from the
expressions for the currents. They remain tensor quantities even in
the absence of an external magnetic field. Since the relaxation time
approximation is used, the thermal and electrical conductivity tensors
are still related via the Wiedemann-Franz law
(\ref{eq:Wiedemann}). \citet{Yakovlev2015bMNRAS} discussed the
general structure of the solutions and their qualitative properties
for the case where one of the principal collision frequencies is much larger than the other, say $\nu_p\gg\nu_a$. That
means that the heat or charge transport is much more efficient along the symmetry direction of the pasta phase than across it.
The net effect of this anisotropy on the transport in the inner crust
of the neutron star will depend on the predominant orientation of
the nuclear clusters (they can be aligned with the radius, be
predominantly perpendicular, or form a disordered domain-like
structure). 
We
refer the reader to the original work \cite{Yakovlev2015bMNRAS} for 
a discussion of the rich variety of possibilities.

Microscopic calculations of the relaxation time tensor for 
nuclear pasta remain a task for future studies. \citet{Schneider2016PhRvC}
made a step towards this goal by running a large classical
molecular dynamics simulation. They find a pasta slab phase with a
number of topological defects and calculate the static proton
structure factor $S_p(\bm{q})$, including  the full angular
dependence. While they do not present a full calculation of 
the transport properties, they perform simple estimates for the angular
dependence of the kinetic coefficients. They find that the
relaxation along two symmetry axes can differ by an order of
magnitude, thus supporting the assumptions of
\citep{Yakovlev2015bMNRAS}. As before, the molecular dynamics
simulations were performed at high temperatures and proton
fractions and thus cannot be directly applied to the neutron star crust. However,
they found that topological defects present in the pasta
decrease the values of transport coefficients, and this decrease
can be described by the effective impurity parameter $\tilde{{\cal
Q}} \sim 30$, a value having the same order of magnitude
as inferred from astrophysical observations
\cite{Pons2013NatPh, Horowitz2015PhRvL}.  This suggests that 
detailed investigations of the transport properties in the pasta phase 
along these lines are promising directions for the future.

\section{Transport in the core: hadronic
matter}\label{S:core_nuclear}

At densities above 
$\rho_{cc}$, the nuclear clusters dissolve completely and the
matter in neutron stars is uniform and neutron-rich. The simplest
composition is $npe\mu$ matter, where muons ($\mu$) appear when 
the difference between neutron and proton chemical potentials becomes larger than the muon mass, which occurs at
densities around $\rho_0$. The matter is usually thought to be in (or close to) equilibrium with respect to weak processes. The condition of beta-decay and the inverse process of lepton capture to proceed at the same rate then imposes the following relation between the chemical potentials,
\begin{equation}\label{eq:beta_equil}
  \mu_n=\mu_p+\mu_\ell,
\end{equation}
where $\ell$ stands for electrons or muons, such that
$\mu_e=\mu_\mu$. We have omitted the neutrino chemical potential $\mu_\nu$ because at typical neutron star temperatures neutrinos 
leave the system once they are created (exceptions are the hot cores of proto-neutron stars, binary neutron star mergers, and supernovae interiors, where neutrinos can be trapped and thus $\mu_\nu \neq 0$). This beta-equilibrated matter, together with the condition 
of electric charge neutrality, is highly
asymmetric, or neutron-rich: the typical proton fraction in neutron star  cores is $x_p\lesssim 15\%$ (e.g., \cite{HPY2007Book}).\ 
Electrons and muons form almost ideal degenerate gases (electrons are ultra-relativistic, while muons
become relativistic soon after their threshold). The nucleons, however, 
form a highly non-ideal, strongly-interacting liquid, where
nuclear many-body effects are of utmost importance. In this
section we discuss the transport properties of this high-density nuclear  matter, starting from the non-superfluid case, and including effects
of superfluidity later. We will also briefly discuss some of the effects of hyperons, in particular in Sec.\ \ref{S:bulk}, where we address the bulk viscosity of hadronic matter. At even larger densities, 
it is conceivable that a 
transition to deconfined quark matter occurs. The transport properties of
various possible phases of quark matter are discussed separately in  Sec.\ \ref{sec:quarks}. 

\subsection{Shear viscosity, thermal and electrical
conductivity}\label{S:core_nuclear_kincoeff}

\subsubsection{General formalism}\label{S:core_boltz}
Transport coefficients of nuclear matter in neutron star cores are calculated within the transport theory for Landau Fermi liquids, outlined in Sec.~\ref{S:fl_transp}, adapted to multi-component systems. 
The response to external perturbations is described by 
a system of Landau transport equations for quasiparticles (\ref{landau}), whose solution, as discussed in Sec.~\ref{S:fl_transp}, is equivalent to the solution of the system of linearized Boltzmann equations for the transport coefficients we are interested in. For a given quasiparticle species `$c$', the collision term of the linearized Boltzmann equation $I_c=\sum_i I_{ci}$ contains a sum of collision integrals for collisions with other species `$i$', each of 
the form  (\ref{boltz_coll_lin}), with binary transition probabilities $W_{ci}$. 
Depending on the antisymmetrization of the particle states in the calculation of $W_{ci}$, the symmetry factor $(1+\delta_{ci})^{-1}$ must be included in  (\ref{boltz_coll_lin}) to
avoid double counting of collisions within the same particle species.

Quasiparticle scattering occurs within the
thermal width of the Fermi surface, and thus the typical energy
transfer in the collision event is of the order of temperature.
Therefore, the collisions cannot be considered elastic, and the
relaxation time approximation  -- frequently used in the previous
section, where electron-ion collisions were considered -- is generally
not applicable. The system of transport equations must be solved
retaining the full form of the collision integrals on the right-hand side
of the Boltzmann equation, for instance with variational
methods. 

Some general properties of the transport coefficients can be deduced immediately from
Eq.~(\ref{boltz_coll_lin}). Due to the strong degeneracy, the Pauli blocking factors $f^{(0)}f_1^{(0)}(1-f'^{(0)})(1-f_1'^{(0)})$ effectively place all quasiparticles on their respective Fermi surfaces, and for 
each momentum integration in Eq.~(\ref{boltz_coll_lin}) we can write 
\begin{equation}\label{eq:phase_space_decomp}
d^3 \bm{p}\approx p_F m^* d \varepsilon \, d \Omega_{\bm{ p}}\, ,
\end{equation}
 where the change from  momentum to energy integration has produced the 
density of states on the Fermi surface $\propto p_F^2/v_F=m^*p_F$. 
 It is customary to describe the deviation function
$\Phi(\varepsilon)$ in terms of a series expansion over the dimensionless
excitation energy $x=(\varepsilon-\mu)/(k_BT)$. Moreover, in traditional
Fermi liquids, the rate $W_{ci}$ is considered to be independent of the
energy transfer in the collisions. This leads to a $T^2$
behavior of the collision integral
(\ref{boltz_coll_lin}) irrespective  of the details of
collisions, which only reorient the quasiparticle momenta, leaving their absolute values intact.

In the simplest variational solution,
the deviation functions are assumed to have the form
(\ref{Aeta}) or (\ref{Akappa}), appropriate for the perturbation in
question, with a constant effective relaxation time
(\ref{eq:A_reltime}) for each quasiparticle species. Then the Boltzmann
equation reduces to a system of algebraic equation for the effective relaxation
times 
\begin{equation}\label{eq:collfreq_system}
  1=\sum_i \nu_{ci} \tau_{c}+\sum_{i\neq c}\nu_{ci}'\tau_i,
\end{equation}
where, according to the discussion above, the effective collision
frequencies are $\nu_{ci}\propto T^2 m_c^* m_i^{*2} \langle
W_{ci}\rangle_{\rm tr}$, with a slightly different effective mass dependence for the mixing terms, $\nu_{ci}'\propto T^2 m_c^{*2}m_i^{*} \langle
W_{ci}\rangle_{\rm tr}$. 
Here, $\langle W_{ci}\rangle_{\rm tr}$ is an effective transport scattering cross-section, which is the
angular average of $W_{ci}$ at the Fermi surface with appropriate
kinematic factors [cf.\
Eqs.~(\ref{eq:sigma_Coulomb})--({\ref{eq:Gkappa}})]. Hence, the
effective relaxation times $\tau_c\propto T^{-2}$, and this
temperature dependence is reflected in the transport coefficients.
Thus in a normal Fermi liquid one obtains
\begin{equation}\label{eq:kin_temp_FL}
\eta\propto T^{-2},\quad \sigma \propto T^{-2},\quad \kappa \propto T^{-1}.
\end{equation}
Notice that the
effective collision frequencies are not the same for $\kappa$,
$\eta$, and $\sigma$.

The exact result
obeys the same general properties as the variational
solution. The correction to the variational solution for any
transport coefficient, say $\kappa$, can be written as
$\kappa=C_\kappa \kappa_{\rm var}$  where $C_\kappa$ is
a temperature-independent correction factor. This factor is found from
the solution of a system of dimensionless integral equations for
$\Phi_c(x)$. For a one-component Fermi liquid, the exact
solution was constructed in
Refs.~\cite{BrookerSykes1968PhRvL,SykesBrooker1970AnPhy,JensenSmith1968PhLA}
in the form of a rapidly converging series, see
Ref.~\cite{BaymPethick} for details. The integral equation for
$\Phi_c(x)$ can be also solved numerically by iterative methods.
In any case, the correction constants $C_{\kappa,\sigma,\eta}$
were found to be in the range $1-1.4$, which is unimportant for 
practical purposes in astrophysical applications. An exact analytic expression for the spin response of the Fermi liquid in an external (oscillating) magnetic field was recently constructed from the transport equation by \citet{PethickSchwenk2009PhRvC}. The exact solution of the
transport equation was generalized to multi-component
Fermi liquids in the neutron star context by
\citet{FlowersItoh1979ApJ} and then analyzed in a general form by
\citet{Anderson1987}. Owing to large uncertainties present in
the various parameters describing the neutron star matter, the
simplest variational result seems to be a sufficient approximation in all
cases.

For $npe\mu$ matter in neutron star cores, the collision frequencies in
Eq.~(\ref{eq:collfreq_system}) are determined by electromagnetic
interactions between charged particles, and by strong interactions
between baryons. It turns out that the lepton and nucleon subsystems in
Eq.~(\ref{eq:collfreq_system}) decouple and can be considered
separately \cite{FlowersItoh1979ApJ}.
Then, the thermal
conductivity (or shear viscosity) can be written as
$\kappa=\kappa_{e\mu}+\kappa_{np}$. The situation is different for the
electrical conductivity, which is relevant in the presence of a magnetic field, see
Sec.~\ref{S:core_nuc_mag}.

\subsubsection{Lepton sector}\label{S:core_nuc_lepton}

The lepton (electron and muon) transport coefficients are mediated
by the collisions within themselves and with charged protons,
which now can be considered as passive scatterers. Since the electromagnetic collisions are long-range (and hence small-angle), the corresponding collision frequencies are determined by the character of plasma screening (see Sec.~\ref{Sec:plasma}). Explicitly, the differential transition rate $W_{ci}$ 
is proportional to the squared matrix element for electromagnetic
interaction, which,  in an isotropic plasma, can be written as
\begin{equation}\label{eq:CoulombMatel}
M_{ci}\propto \frac{J^{(0)}_1 J^{(0)}_2 }{q^2+\Pi_l(\omega,q)}-\frac{\bm{J}_{1t}\cdot\bm{J}_{2t}}{q^2-\omega^2+\Pi_t(\omega,q)}\, ,
\end{equation}
where $\omega$ and $q$ are the energy and momentum transferred in the collision, respectively, $J^{(0)}$ and $\bm{J}_t$ are time-like and transverse (with respect to $\bm{q}$) space-like components of the transition current, respectively, and $\Pi_l$ and $\Pi_t$ are the longitudinal and transverse polarization functions. In conditions present in neutron star cores, the long-wavelength $q\ll p_F$ and static $\omega v_F \ll q$ limits are appropriate since the transferred energy is of the order of the temperature, $\omega\sim T$. The first term in Eq.~(\ref{eq:CoulombMatel}) corresponds to the electric (Coulomb) interaction, while the second term corresponds to the magnetic (Amp\`{e}re) part of the interaction. The second term is essentially relativistic and is suppressed for non-relativistic particles by the ratio $J_t/J^{(0)}\propto u/c$. 
Therefore,  the magnetic term is not that important in the crust (see Sec.~\ref{S:crust_kincoeff}), where the dominant contribution to transport coefficients comes from electron collisions with heavy non-relativistic ions,
but it becomes significant in the core. In the context of plasma physics, the relativistic collision integral (\ref{boltz_coll_lin}), taking into account longitudinal and transverse screening as in Eq.~(\ref{eq:CoulombMatel}), was first derived by \citet{Silin1961JETP}. Alternatively, the two terms in Eq.~(\ref{eq:CoulombMatel}) can be viewed as resulting from interaction via longitudinal  and transverse virtual plasmon exchange. 

The dominance of the transverse plasmon exchange in the transport properties of relativistic plasmas was realized by \citet{Heiselberg1992NuPhA} and worked out by \citet{Heiselberg:1993cr} in the context of unpaired quark matter, see Sec.~\ref{S:kincoeff_uqm}. The reason is as follows. To lowest order, the  longitudinal screening is static, $\Pi_l=q_l^2$, where $q_l^2$ is the Thomas-Fermi screening wavenumber. Therefore, the dominant contribution to the part of the collision frequency that is mediated by the longitudinal interaction comes from $q\lesssim q_l$. In contrast, the transverse plasmon (photon) screening is essentially dynamical in the from of Landau damping, so that 
\begin{equation}\label{eq:Pi_T_nonsf}
\Pi_t=i \frac{\pi}{4} \frac{\omega}{qc} q_t^2 \, ,
\end{equation} 
where $q_t\sim q_l$ is a characteristic transverse wavenumber ($q_l=q_t$ if all charged particles are ultra-relativistic). Hence, the dominant contribution to the `transverse' part of the collision frequency comes from $q\lesssim [\pi\omega/(4c q_t)]^{1/3} q_t\ll q_l$. The latter inequality is due to the low temperature ($\omega \sim T$) and has two important consequences. First, the transverse plasmon exchange dominates the collisions between the relativistic particles in a degenerate plasma, and second, the scattering probability depends on the energy transfer of the  collision. As a consequence, the temperature behavior of lepton transport coefficients in neutron star cores is essentially non-Fermi liquid (in contrast to the general theory outlined in the previous section). The modification of the temperature behavior depends on the kinematics of the problem in question and on the relation between the `longitudinal' and `transverse' contributions. For the thermal conductivity, the effective collision frequency becomes $\nu_{ci}\propto T$ if transverse plasmon exchange fully dominates the interaction, while for shear viscosity and electrical conductivity in the same limit, $\nu_{ci}\propto T^{5/3}$ \cite{Heiselberg:1993cr}. Note that the energy dependence of the scattering rate does not change the conclusion that the  simple variational solution described by Eq.~(\ref{eq:collfreq_system}) remains a sufficient approximation.
It can be shown that the correction to the variational solution for the transverse-dominated collisions does not exceed 10\% \cite{ShterninYakovlev2007,ShterninYakovlev2008}. 

The lepton transport coefficients for $npe\mu$ matter with correct account for the transverse plasmon exchange were analyzed in 
\cite{ShterninYakovlev2007, ShterninYakovlev2008,Shternin2008JETP}.
The low-temperature result (when the transverse plasmon exchange dominates) for the thermal conductivity is \cite{ShterninYakovlev2007} 
\begin{eqnarray}\label{eq:kappa_emu_trans_num}
\kappa_{e\mu}&=&\kappa_{e}+\kappa_\mu=\frac{\pi^2}{54\zeta(3)}\frac{k_B c (p_{Fe}^2+p_{\rm F\mu}^2)}{\hbar^2\alpha_f} \nonumber\\[2ex]
&=& 2.43\times 10^{22}\ \left(\frac{n_{\rm B}}{n_0}\right)^{2/3} \left(x_e^{2/3}+x_\mu^{2/3}\right)\ \frac{\rm erg}{\mbox{cm s K}}\, ,
\end{eqnarray}
where $n_0=0.16$~fm$^{-3}$ is the number density at nuclear saturation, $n_{\rm B}$ is the total baryon number density, $x_e$ and $x_\mu$ are the electron and muon number density fractions, respectively.
The expression for the electron and muon contributions to the shear viscosity \cite{ShterninYakovlev2008} $\eta_{e\mu}=\eta_{e}+\eta_\mu$ is more cumbersome in analytical form, and we  only give the numerical result
\begin{equation}\label{eq:eta_emu_trans}
\eta_{e\mu}=8.43\times10^{20}\ \left(\frac{n_{\rm B}}{n_0}\right)^{14/9} \left(\frac{T}{10^8~{\rm K}}\right)^{-5/3} 
\frac{x_{e}^2+x_\mu^2}{(x_{e}^{2/3}+x_\mu^{2/3}+x_{p}^{2/3})^{2/3}}
\ \frac{\rm g}{\mbox{cm s}} \, .
\end{equation}
We also give the expression for the electrical conductivity of non-magnetized $npe\mu$ matter \cite{Shternin2008JETP},
\begin{equation}\label{eq:sigma_parallel_trans}
\sigma = 1.86\times 10^{30}\ \left(\frac{n_{\rm B}}{n_0}\right)^{8/9} \left(\frac{T}{10^8~{\rm K}}\right)^{-5/3} 
\frac{x_{e}^{1/3}+x_\mu^{1/3}+x_{p}^{1/3}}{(x_{e}^{2/3}+x_\mu^{2/3}+x_{p}^{2/3})^{2/3}}
\ { {\rm s}^{-1}}.
\end{equation}
This result for the electrical conductivity is already the full result 
for $npe\mu$ matter (and thus we have not added the subscript `$e\mu$'), because the baryon sector (neutrons) does not contribute to the electrical conductivity in the non-magnetized case. The result (\ref{eq:sigma_parallel_trans}) is of the same (very large) order of magnitude as the classical estimate of \citet{BaymPethikPines1969Natur224}, rendering Ohmic dissipation in neutron star cores insignificant. In magnetized matter, the  situation changes dramatically, as we will discuss in  Sec.~\ref{S:core_nuc_mag}.

The thermal conductivity is temperature-independent and depends only on the carrier Fermi momentum. The result (\ref{eq:kappa_emu_trans_num}) is valid for all practically relevant temperatures and densities in (non-superfluid) neutron star cores \cite{ShterninYakovlev2007}. In contrast, the result (\ref{eq:eta_emu_trans}) can significantly overestimate the shear viscosity  since the dominance of transverse collisions is not always strict (especially at lower densities), for details see Refs.~\cite{ShterninYakovlev2008, Kolomeitsev2015PhRvC}. The same is true for the electrical conductivity in Eq.~(\ref{eq:sigma_parallel_trans}). A relatively compact expression obtained from a fit for $\eta_{e\mu}$ that is valid in a broad temperature and density range can be found in Ref.~\cite{Kolomeitsev2015PhRvC}.

We conclude this subsection by noting the advantage of the lepton kinetic coefficients. Since they are mediated by electromagnetic collisions, the final analytical expressions can be  used for any equation of state (since they depend only on the effective masses of charged particles and their Fermi momenta). In addition, they can easily be updated to include other charged particles acting as passive scatterers, for instance hyperons.

\subsubsection{Baryon sector}\label{S:core_nuc_baryon}
The nucleon transport coefficients in $npe\mu$ cores are governed by collisions between neutrons and protons mediated by the strong interaction.
Since nuclear matter in the core of neutron stars is highly asymmetric (in other words, the proton fraction $x_{p}$ is small), 
the proton contribution to transport coefficients is small, and it is enough to treat them only as a passive scatterers for neutrons. In this case, the system of equations (\ref{eq:collfreq_system}) reduces to one equation for the effective neutron relaxation time $\tau_{n}$ (e.g., \cite{Baiko2001AA}). It is sometimes assumed that due to $x_{p}$ being small the results for pure neutron matter are appropriate for neutron star cores, at least at low densities. However, pure neutron matter turns out to be a bad approximation 
for assessing the transport coefficients, since the protons cannot be ignored even if $x_p\approx 0.01$ \cite{Baiko2001AA, Shternin2013PhRvC}. The reason is that the effective transport cross-section for neutron-proton collisions is larger than that for neutron-neutron collisions due to inclusion of the $T_z=0$ isospin channel in scattering, and different kinematics of these collisions \cite{Shternin2013PhRvC}. Applying the general theory outlined in Sec.~\ref{S:core_boltz} for neutrons scattering off neutrons and protons, one obtains the following results for thermal conductivity and shear viscosity, 
\begin{subequations}\label{eq:kappaeta_n_num}
\begin{eqnarray}\label{eq:kappa_n_num}
\kappa_{n}&=& 1.03\times 10^{22}\ \frac{n_{n}}{n_0} \left(\frac{m_{n}^*}{m_N}\right)^{-2} \left(\frac{T}{10^8~{\rm K}}\right)^{-1} \left[\left(\frac{m_{n}^*}{m_N}\right)^2 m_\pi^2 S_{\kappa nn}+\left(\frac{m_{p}^*}{m_N}\right)^{2}m_\pi^2 S_{\kappa np}\right]^{-1}\frac{\rm erg}{\mbox{cm s K}} \, , \\[2ex]
\label{eq:eta_n_num}
\eta_{n}&=& 2.15\times 10^{17}\ \left(\frac{n_{n}}{n_0}\right)^{5/3} \left(\frac{m_n^*}{m_N}\right)^{-2} \left(\frac{T}{10^8~{\rm K}}\right)^{-2} \left[\left(\frac{m_{n}^*}{m_N}\right)^2 m_\pi^2 S_{\eta nn}+\left(\frac{m_{p}^*}{m_N}\right)^{2}m_\pi^2 S_{\eta np}\right]^{-1}\ \frac{\rm g}{\mbox{cm s}} \, ,
\end{eqnarray}
\end{subequations}
where $m_N=939$~MeV$/c^2$ is the nucleon mass (neglecting the  mass difference between neutron and proton), and   we have used the same notations as in Refs.~\cite{Baiko2001AA,ShterninYakovlev2008,Shternin2013PhRvC,Kolomeitsev2015PhRvC}. The quantities $S_{\kappa/\eta NN}$ ($N=n,p$) are the quasiparticle scattering rates $W_{ci}$ averaged with certain phase factors, for details see for example Refs.~\cite{Shternin2013PhRvC,Shternin2017JPhCS}. They have the meaning of effective transport cross-sections and are normalized by the relevant nuclear force scale -- the inverse pion mass squared, $m_\pi^{-2}\approx 20$~mb in natural units.
 Note that the numerical prefactors in Eqs.~(\ref{eq:kappaeta_n_num}) include the correction constants $C_\kappa\approx 1.2$ and $C_\eta\approx 1.05$, as discussed at the end of Sec.~\ref{S:core_boltz} \cite{Shternin2013PhRvC}. These corrections are of course irrelevant for most applications in neutron star  physics. 

The main ingredients  for the calculation of the nucleon transport coefficients are the effective masses of the nucleons $m^*$ on the Fermi surface and the quasiparticle scattering rates $W_{ci}$. Both quantities are strongly affected by in-medium effects and should be calculated using a microscopic many-body approach. Thus, the results for the nucleon transport coefficients are model-dependent and, in principle, their calculation should be based on the same microscopic model as the calculation of the equation of state. 
From Eqs.~(\ref{eq:kappaeta_n_num}) we see that the effective mass enters the expressions for the transport coefficients in fourth power (because it describes the density of states and four quasiparticle states are involved in binary collisions). A moderate modification of the effective masses thus results in a strong modification of the transport coefficients. Therefore, the simplest way to include in-medium effects is to compute the effective mass modification, but use the free-space scattering rate which is well-known from experiment. This approach is particularly appealing because of its universality. The resulting expressions can be used for any equation of state of dense nuclear matter. Convenient fitting expressions for effective collision frequencies within this approach can be found in Refs.~\cite{Baiko2001AA,ShterninYakovlev2008}. Typical values of transport cross-sections at $n_{\rm B}=n_0$ in Eqs.~~(\ref{eq:kappaeta_n_num}) are $S_{\kappa nn}\sim S_{\eta nn}\sim S_{\eta np}\approx 0.2 m_\pi^{-2}$, while $S_{\kappa np}\approx 0.4 m_\pi^{-2}$. Note that older results by  \citet{FlowersItoh1979ApJ}, obtained via the same approach, turned out to be incorrect. Unfortunately, the in-medium modifications of the scattering rates themselves can be substantial. At present, the theoretical uncertainties are rather large and can result in order of magnitude differences in the final results.

Several many-body approaches have been employed in the calculation of the transport coefficients. The main problem is to properly take into account the particle correlations appearing in the strongly interacting liquid. Additional complications arise from the need to include  three-body nucleon forces, which are necessary to reproduce the empirical saturation point of symmetric nuclear matter (see for instance Ref.~\cite{Baldo1999Book}). Results have been obtained within the Brueckner-Hartree-Fock (BHF) scheme, where the in-medium $G$-matrix is used in place of the quasiparticle interaction \cite{Benhar2010PhRvC,Zhang2010PhRvC,Shternin2013PhRvC,Shternin2017JPhCS}, 
within the effective quasiparticle interaction constructed on top of the $G$-matrix \cite{Wambach1993NuPhA} (for neutron matter only), within the in-medium $T$-matrix approach (also for pure neutron matter) \cite{Sedrakian1994PhLB}, and using the Correlated Basis Function and the cluster expansion technique \cite{Benhar2010PhRvC,BenharValli2007PhRvL,CarboneBenhar2011JPhCS}, which employs the variationally constructed effective interaction. All these approaches start from  `realistic' nuclear potentials, which are designed to fit the data on the free-space scattering phase shifts and properties of bound few-body systems.

A somewhat different approach is based on the Landau-Migdal Fermi-liquid theory for nuclear matter. In this approach, the long-range pion-exchange part of the nucleon-nucleon interaction is considered explicitly, while the short-range part of the potential is absorbed into a number of phenomenological constants. The key point of the theory is an in-medium modification of the pion propagator \cite{Migdal1990PhR, Migdal1978RvMP} leading to the softening of the pion mode. This softening is strongly density-dependent and becomes important at $n_{\rm B}\gtrsim n_0$. At larger densities, this can lead to pion condensation. It is assumed that above the saturation density $n_0$, the nucleon quasiparticle scattering is fully determined by the medium-modified one-pion exchange (MOPE), where also the interaction vertices are modified due to short-range nuclear correlations. Since the pion mode is soft, the effective range of the nucleon interaction increases, which leads to a strong enhancement of the scattering rates, especially at high densities. As a consequence, the effective collision frequencies in Eq.~(\ref{eq:collfreq_system}) become larger and the transport coefficients reduce substantially. The calculations in this model (for pure neutron mater) 
were performed by \citet{Blaschke2013PhRvC} for  thermal conductivity and by \citet{Kolomeitsev2015PhRvC} for  shear viscosity.

Let us compare the lepton and nucleon contributions to the thermal conductivity and shear viscosity of non-superfluid nuclear matter.
The strong non-Fermi-liquid behavior of the lepton thermal conductivity (remember that $\kappa_{ e\mu}$ is constant in $T$) makes it smaller than the baryon contribution $\kappa_n\propto T^{-1}$ regardless of the microscopic model used to calculate the latter quantity. However, the key result is that the thermal conductivity is large such that the neutron star core is isothermal (more precisely, accounting for effects of general relativity, the redshifted temperature is spatially constant), and the precise value of $\kappa$ is not important. This value is only interesting for the cooling of young neutron stars, as it regulates the duration of the thermal relaxation in the newly-born star \cite{Gnedin2001MNRAS,ShterninYakovlev2008AstL,Blaschke2013PhRvC}. 
For instance, delaying the thermal relaxation due to the decrease of $\kappa_{n}$ in the MOPE model allowed \citet{Blaschke2013PhRvC} to fit the cooling data of the Cas~A neutron star.
The situation is different for the shear viscosity: here, the leptonic contribution is proportional to $T^{-5/3}$ to leading order and is not damped at low temperatures. The calculations reported in the literature show that $\eta_{e\mu}$ can be either larger or smaller than the nucleonic contribution $\eta_{n}$, see, however, the discussion in Ref.~\cite{Shternin2013PhRvC}.

All considerations above assume a uniform Fermi liquid. Let us briefly address the possibility of proton localization, originally proposed by \citet{Kutschera1989PhLB,Kutschera1990ActPolB}. In this scenario, for small proton fractions, the protons in neutron star cores can be localized in a potential well produced by neutron density fluctuations induced by the protons themselves. The protons occupy some bound ground state and do not form a Fermi sea. This model has its analogy in the polaron problem in solids \cite{Kutschera1993PhRvC}. Recently, proton localization was reconsidered for some realistic equations of state \cite{Szmaglinski2006AcPPB,Kubis2015PhRvC}. The authors find that protons can localize at densities $n_{\rm B} \gtrsim (0.5-1)\, {\rm fm}^{-3}$, i.e., well in the range that can occur in the interior of neutron stars. The transport properties of nuclear matter with localized protons were studied by \citet{BaikoHaensel1999AcPPB}. They considered $npe$ matter, where the localized protons are uncorrelated. Then the problem has much in common with transport properties of the neutron star crust with charged impurities, see Sec.~\ref{S:kincoeff_imp}. The electrons and neutrons now scatter off themselves and off the localized protons. Unless the protons can be excited in their sites, the scattering is elastic and the collision frequencies to be used in Eq.~(\ref{eq:collfreq_system}) become temperature-independent and dominate over temperature-dependent collision frequencies for other  scatterings (cf. Sec.~\ref{S:kincoeff_imp}). Clearly, this leads to a strong decrease of the transport coefficients at low temperatures compared to the results without localization \cite{BaikoHaensel1999AcPPB}. The consequences of proton localization on neutron star cooling was investigated in Ref.~\cite{BaikoHaensel2000A&A}, but other possible astrophysical implications are largely unexplored. \citet{BaikoHaensel1999AcPPB} considered a completely disordered system of localized protons, and it was proposed that these impurities can form a lattice \cite{Kutschera1995NuPhA}.

\subsubsection{Effects of Cooper pairing}\label{S:core_nuc_pairing}

So far we have neglected the effects of neutron and proton pairing on the transport coefficients. 
Pairing directly affects the dissipation in the neutron and proton subsystem 
since the structure of the excitations in superfluid matter is changed. However, as we will see immediately, proton pairing also affects 
the leptonic transport coefficients that were discussed in Sec.~\ref{S:core_nuc_lepton} without pairing.

We start by considering the electron and muon transport in the presence of proton pairing in the $^1S_0$ channel. The effect of proton superconductivity is twofold. First, it modifies the scattering rates of leptons off the protons (the protonic excitations)  and second, it modifies the screening properties of the plasma which regulates the electromagnetic interaction in Eq.~(\ref{eq:CoulombMatel}). In the static limit, the longitudinal part of the polarization operator $\Pi_l$, which describes longitudinal plasmon screening, remains unaffected (e.g.~\cite{Arseev2006PhyU,Gusakov2010PhRvC}). In contrast, the character of the transverse screening changes dramatically. The most important difference from the non-superconducting case is that now the transverse screening is predominantly static ($\Pi_t\neq 0$ for $\omega\to 0$). In this case, the collision probability becomes $\omega$-independent, which restores the standard Fermi-liquid behavior of the collision frequencies between the electrons and muons, $\nu_{ci}\propto T^2$. 

The long-wavelength ($q\to 0$) transverse plasmon (photon) in a superconductor acquires a Meissner mass,
\begin{equation}\label{eq:Meissner}
 q_M^2=\frac{4\alpha
_f}{3\pi} p_{\mathrm{F}p}^2 v_{\mathrm{F}p}.
\end{equation}
However, at larger $q$ the screening mass gradually drops and in the Pippard limit, $q \xi \gg 1$, 
it becomes inversely proportional to $q$. Here, $\xi\sim v_{Fp}/\Delta_{p}$ is the coherence length, with the energy gap $\Delta_{p}$ in the proton quasiparticle spectrum from Cooper pairing. It turns out that both the long wavelength (London) and small wavelength (Pippard) limits may be applicable to neutron star cores \cite{Shternin2018PRD}. 
At low temperatures $T\ll T_{cp}$, the protons give the dominant contribution to the photon polarization function $\Pi_t$, and one obtains [instead of Eq.\ (\ref{eq:Pi_T_nonsf})]
\begin{subequations}\label{eq:Pi_T}
\begin{eqnarray}
 \Pi_t&\approx&  q_M^2  \qquad \qquad\quad \;\; (\mbox{for}\;\; q\xi\ll 1) \, ,\label{eq:Pi_T_sup_london}\\[2ex]
\Pi_t&\approx& \pi \alpha_f p_{Fp}^2 \frac{\Delta_{p}}{q} \qquad (\mbox{for}\;\; q\xi\gg 1) \, .\label{eq:Pi_T_sup_pippard}
\end{eqnarray}
\end{subequations}
Thus the characteristic screening wavenumber in the London case is $q_M$ and does not depend on $\Delta_p$, while in the Pippard case it is $\propto \Delta_{p}^{1/3}$ instead of $\propto \omega^{1/3}$ in the unpaired case, Sec.~\ref{S:core_nuc_lepton}. It can be shown that the `transverse' part of the collision frequencies dominates in the superconducting  case as well \cite{Shternin2018PRD,ShterninYakovlev2007,ShterninYakovlev2008}.

Taking into account lepton collisions with protons is more involved. The main low-energy excitations of the proton system are the single-particle excitations, namely the Bogoliubov quasiparticles. In addition to the presence of the energy gap in the quasiparticle spectrum, one needs to take into account that the number of quasiparticles is not conserved. They can be excited from the Cooper pair condensate, or coalesce into it. As a consequence, the collision integral describing electron-proton scattering is more complicated than in Eq.~(\ref{boltz_coll_lin}).
However, at low temperatures the main effect of pairing is the exponential reduction of the number of quasiparticles and thus the exponential reduction in the collision frequencies, $\nu_{cp}\propto {\rm exp}(-\Delta_{p}/T)$ \cite{ShterninYakovlev2007, ShterninYakovlev2008}. Therefore, for temperatures much lower than the critical temperature for proton pairing, $T\ll T_{cp}$, the details of the lepton-proton collisions are not important, since they are suppressed. The transport coefficients are dominated in this case  by collisions in the lepton subsystem, and, taking into account the screening modification, one derives the following compact leading-order expressions for the thermal conductivity instead of Eq.~(\ref{eq:kappa_emu_trans_num}) \cite{Shternin2018PRD},
\begin{subequations}\label{eq:kappa_emu_SF}
\begin{eqnarray}
\kappa_{e\mu}^\mathrm{Lon} &=& \frac{5 q_M^3 c^2}{18\pi^2\alpha_f^2\hbar^2  T}=8.57\times 10^{23} \left(\frac{n_{\rm B}}{n_0}\right)^{3/2}\left(\frac{m_N}{m_p^*}\right)^{3/2} x_{p}^{3/2}\left(\frac{T}{10^8\,{\rm K}}\right)^{-1}\ \frac{\rm erg}{\mbox{cm s K}}  \qquad (\mbox{for}\;\; q_M\xi\ll 1) ,\label{eq:kappa_emu_Lon}\\[2ex]
\kappa_{e\mu}^{\rm Pip} &=& \frac{5}{24} \frac{k_B c p_{Fp}^2}{\alpha_f \hbar^2} \frac{\Delta_{p}}{k_B T}= 3.87\times 10^{24} \left(\frac{n_{\rm B}}{n_0}\right)^{2/3} x_{p}^{2/3} \frac{\Delta_{p}}{1\,{\rm MeV}} \left(\frac{T}{10^8\,{\rm K}}\right)^{-1}\ \frac{\rm erg}{\mbox{cm s K}}  \qquad (\mbox{for}\;\; q_M\xi\gg 1).\label{eq:kappa_emu_Pip}
\end{eqnarray}
\end{subequations}
Equation~(\ref{eq:kappa_emu_Lon}) corresponds to the screening in the London limit (\ref{eq:Pi_T_sup_london}) and is valid when $q_M\xi\ll 1$. Notice that the thermal conductivity is independent on the proton gap $\Delta_p$ in this case. Similarly, Eq.~(\ref{eq:kappa_emu_Pip}) corresponds to the Pippard limit (\ref{eq:Pi_T_sup_pippard}) and is valid for $q_M\xi \gg 1$. 
The interpretation of the two limits can also be understood  by noting that $q_M\xi=\pi/\varkappa$, where $\varkappa$ is the Ginsburg-Landau parameter in the theory of superconductivity. This parameter distinguishes between 
type-I and type-II superconductivity, and thus we conclude that 
 the London-limit expression (\ref{eq:kappa_emu_Lon}) is applicable in the type-II regime while Eq.~(\ref{eq:kappa_emu_Pip}) is appropriate in the type-I regime, which  
is realized at sufficiently high densities in the inner core of the star, or at low $\Delta_p$ \cite{Shternin2018PRD}. The general expression for the thermal conductivity in the intermediate case can be found in Ref.~\cite{Shternin2018PRD}; it is smaller than the 
expressions (\ref{eq:kappa_emu_SF}) for all densities $n_B$ and approaches them from below at small $n_B$ (\ref{eq:kappa_emu_Lon}) and large $n_B$ (\ref{eq:kappa_emu_Pip}). 
The result (\ref{eq:kappa_emu_SF})  shows the standard Fermi-liquid dependence $\kappa\propto T^{-1}$ and is several orders of magnitude larger than the non-superfluid result (\ref{eq:kappa_emu_trans_num}). This is not really important in practice since the star becomes isothermal in a short time in both cases. 

Similarly, one can compute the low-temperature expressions for the leptonic shear viscosity in the two different screening limits \cite{Shternin2018PRD}, 
\begin{subequations}\label{eq:eta_emu_SF}
\begin{eqnarray}
\eta_{e\mu}^{\rm Lon}&=&6.28\times 10^{21}\ \left(\frac{n_{\rm B}}{n_0}\right)^{11/6}\frac{\left(x_{e}^2+x_\mu^2\right) x_{p}^{1/2}}{x_{e}^{2/3}+x_\mu^{2/3}}\left(\frac{m_N}{m^*_p}\right)^{1/2}\left(\frac{T}{10^8\,{\rm K}}\right)^{-2}\ \frac{\rm g}{\mbox{cm s}} \qquad \;\;\;\;(\mbox{for}\;\; q_M\xi\ll 1), \label{eq:eta_emu_Lon}\\[2ex]
\eta_{e\mu}^{\rm Pip}&=&7.60\times 10^{21}\ \left(\frac{n_{\rm B}}{n_0}\right)^{14/9}\frac{\left(x_{e}^2+x_\mu^2\right) x_{p}^{2/9}}{x_{e}^{2/3}+x_\mu^{2/3}}\left(\frac{\Delta_{p}}{1\,{\rm MeV}}\right)^{1/3}\left(\frac{T}{10^8\,{\rm K}}\right)^{-2}\ \frac{\rm g}{\mbox{cm s}} \qquad (\mbox{for}\;\; q_M\xi\gg 1).\label{eq:eta_emu_Pip}
\end{eqnarray}
\end{subequations}
Comparing Eq.~(\ref{eq:eta_emu_SF}) with Eq.~(\ref{eq:eta_emu_trans}), we see  that the effect of proton Cooper pairing on the lepton shear viscosity is less dramatic than in the case of thermal conductivity. 
This is due to a weaker dependence of $\eta$ on the screening momentum than $\kappa$. 
In turns out that Eq.~(\ref{eq:eta_emu_SF}) is a better approximation to the full result for $\eta_{e\mu}^{\rm SF}$ than the non-superfluid expression (\ref{eq:eta_emu_trans}).
The electrical conductivity in the presence of proton pairing cannot be treated in similar 
simple way, see Sec.~\ref{S:core_nuc_mag}. The detailed behavior of screening at intermediate temperatures and a crossover from static to dynamical screening was discussed by \citet{ShterninYakovlev2007, ShterninYakovlev2008}, although only in the Pippard limit. 

Recently, an effective lepton-neutron interaction was proposed by \citet{Bertoni2015PhRvC}. The idea is that the neutron quasiparticle in the neutron star core is in fact a neutron dressed by a neutron-proton cloud. Thus it possesses an effective electric charge and interacts with charged leptons on the same ground as the protons. Within field-theoretical language, this lepton-neutron interaction is induced by a proton particle-hole excitation which is coupled to neutrons \cite{Bertoni2015PhRvC}. Estimates show that this effective interaction can be relevant when the protons are in the superconducting state. Moreover, at $T\ll T_{cp}$, the effective lepton-neutron collisions can dominate over the inter-lepton collisions, thus providing a dominant contribution to lepton transport coefficients. A detailed rigorous treatment of this interaction is yet to be done and would be highly desired. Notice that such a coupling can also modify the screening properties of the photons in the nuclear medium and therefore other collisions mediated by electromagnetic interactions. This is currently under investigation \cite{Stetina2017PRC}. 

Let us turn now to the nuclear (hadronic) sector in the presence of pairing. Recall that the main carriers are neutrons. If they are unpaired, but the protons are gapped, only neutron-proton collisions are affected. This situation can be treated in a similar way as the lepton-proton collisions above.
The result has not yet been computed in detail for $T\lesssim T_{cp}$, but at low temperatures the main effect is the exponential suppression of the collision frequency \cite{Baiko2001AA,ShterninYakovlev2008}. The damping of neutron-proton scattering leads to increase of the neutron effective relaxation times and, as  a consequence, of neutron transport coefficients, which are now governed by the neutron-neutron scattering only. 

A more interesting situation occurs if neutrons pair. In general, the transport equations in the superfluid are complicated, as one needs to account for anomalous contributions (the response of the condensate), just like in terrestrial fermionic superfluids such as liquid $^3$He \cite{Vollhardt1990}. However, the situation simplifies greatly when the temporal and spatial scales of the external perturbation are large compared to $\hbar\Delta^{-1}\sim 10^{-22}$~s and $\xi\sim 10^{-11}$~cm,
which is the appropriate limit for the transport coefficients. In this case, the response of the condensate is instantaneous, such that it can be considered to be in local equilibrium. Then, the kinetics of the system is described by the transport equation for Bogoliubov quasiparticles, whose streaming (left-hand side) term reduces to a standard streaming term of the Boltzmann equation \cite{Vollhardt1990}, see also Ref.~\cite{Gusakov2010PhRvC} for the case of superfluid mixtures. The transport coefficients can then, in principle, be calculated along the lines laid out in Sec.~\ref{S:core_boltz}, provided the collision integral is specified. The latter can be derived from the normal-state collision integral by applying the Bogoliubov transformations. The resulting expression takes into account non-conservation of quasiparticles. Since  neutron pairing in the core of a neutron star is expected to occur in the anisotropic $^3P_{j}$ state, additional complications arise. This is analogous to 
certain phases of superfluid $^3$He, which break rotational symmetry as well, resulting in anisotropic transport properties \cite{Vollhardt1990}. 
It is also comparable to anisotropic transport in nuclear pasta phases discussed in Sec.~\ref{S:kincoeff_pasta}. In the context of neutron stars, this was not studied in detail (see however Refs.~\cite{Shahzamanian2005IJMPD,JuriHernandez1988PhRvC}). A thorough investigation of the Bogoliubov quasiparticle contribution to transport coefficients remains an open problem, but the key features at low temperatures can be worked out, neglecting all the modifications to the 
quasiparticle collision integral except the spectrum modification by the gap (for the neutron thermal conductivity this was done by \citet{Baiko2001AA}). 

Naively, one might think that since the number of available quasiparticles is exponentially suppressed at low temperatures, their contribution to transport coefficients is exponentially suppressed as well. This is true if there exists a  scattering mechanism  which effectively limits the quasiparticle mean free path. If, however, the main contribution to the collision probability comes from collisions between the Bogoliubov quasiparticles, it is suppressed roughly by the same factor. As a result, the exponential factors cancel each other, and one is left with the standard Fermi-liquid dependence of the thermal conductivity, $\kappa\propto T^{-1}$, as shown in the context of the so-called B phase of superfluid $^3$He \cite{Pethick1977PhRvB_lowT}. Similar arguments show that the shear viscosity tends to a constant value which is not far from its value at $T_c$. Moreover, it can be shown that a relation analogous to the Wiedemann-Franz rule  (\ref{eq:Wiedemann}) applies \cite{Pethick1977PhRvB_lowT}, 
\begin{equation}\label{eq:Wiedemann-sf}
\frac{\kappa T}{\eta} = \frac{5\Delta^2}{p_F^2}\, .
\end{equation}
In neutron star cores, this situation can be realized when neutron pairing occurs at lower temperatures than proton pairing, such that $T<T_{cn}<T_{cp}$. Then, neutron-proton collisions are suppressed exponentially stronger than neutron-neutron collisions (there are much less proton excitations than neutron ones) and do not participate in neutron transport \cite{Baiko2001AA}. Nevertheless, with lowering temperature, the transport coefficients stay large until other neutron relaxation mechanisms start to dominate over neutron-neutron scattering, for instance neutron-lepton scattering due to the neutron magnetic moment or interactions with collective excitations. This will 
lead to a strong suppression of the transport coefficients compared to the limiting values discussed by \citet{Pethick1977PhRvB_lowT}.
Evidently, a similar analysis in the opposite case, $T_{cn}>T_{cp}$, leads to analogous results.

Let us note that the quasiparticle mean free path increases exponentially with
decreasing temperature. Eventually, at about $0.1\,T_c$, it becomes of the order of the size of the superfluid region. In this case, the bulk hydrodynamical picture is inappropriate to describe the transport since 
the quasiparticles move `ballistically'. Moreover, one needs to take into account the spatial structure of the superfluid region (baryon density, the gap value $\Delta$, and the gravitational potential change on the mean free path scale) and the interaction of the quasiparticles with the boundaries of the superfluid region (more precisely, with the edges of the critical temperature profile, where $T_{cn}(n_{\rm B})$ is lower and the macroscopic hydrodynamical picture is restored). If the spatial scale of the external perturbations is smaller than the quasiparticle mean free path (or the frequency of the perturbation is larger than the collision frequency), the response of the quasiparticle  system cannot be considered in the local equilibrium approximation. To our knowledge, these effects have not been studied yet. Nevertheless, one expects that leptons dominate the transport in this regime.

At low temperatures, 
since the number of single-particle excitations is suppressed exponentially, low-energy collective modes become the relevant degrees of freedom, if they exist. As the superfluid condensate spontaneously breaks the $U(1)$ internal symmetry related to baryon number conservation, at least one gapless collective mode must exist in the superfluid system according to the Goldstone theorem (see Sec.~\ref{S:cooper}). This fundamental mode is called superfluid phonon because of its acoustic dispersion relation. When this is the only low-lying collective excitation, it fully defines the 
transport properties of the system. This situation is realized, for instance, in superfluid $^4$He or in cold atomic gases. In this case, an effective theory can be constructed on general grounds that describes the phonon dispersion and the interactions between phonons (e.g., \cite{Son2002hep.ph,SonWingate2006AnPhy,Greiter1989MPLB}). In the context of neutron star cores, the phase that comes closest to this scenario (in the sense that there are no additional low-energy excitations such as leptons
or other unpaired fermions) is the color-flavor locked phase of quark matter, and we give some details on the field-theoretical description of superfluid phonons in Sec.~\ref{S:quark_colsup}. The phonon transport coefficients are calculated from the solution of the appropriate kinetic equation that includes phonon scatterings in the collision term \cite{Khalatbook1965}. In the case of neutron pairing, shear viscosity and thermal conductivity mediated by phonon-phonon interactions were investigated in Refs.~\cite{ManuelTolos2011PhRvD,ManuelTolos2013PhRvD,ManuelTolos2014PhRvC, Tolos2016AIPC}. These studies suggest that the phonon contribution is important in a narrow range of temperatures, $10^9~{\rm K}\lesssim T<T_{cn}$, where, in fact, the validity of the effective theory is questionable. 
In reality, the excitation spectra and in neutron star cores is richer and various scattering mechanisms can be important \cite{Bedaque:2013fja, Kolomeitsev2015PhRvC}. Superfluid phonons of the neutron component couple to leptons indirectly via the neutron-proton interaction; this provides an efficient scattering mechanism for phonons, decreasing their mean free path. According to \citet{Bedaque:2013fja}, this makes the superfluid phonon contribution to transport coefficients negligible (cf.\ discussion in Sec.~\ref{S:crust_neutrons}). The phonon coupling with the Bogoliubov quasiparticles was investigated by \citet{Kolomeitsev2015PhRvC} with a similar conclusion. Other low-energy excitations which can exist in neutron star cores in the presence of nucleon Cooper pairing are as follows. In metallic superconductors, the collective mode is massive due to presence of Coulomb interaction. However, it was proposed that the efficient plasma screening in nuclear matter `resurrects' the Goldstone mode of the proton condensate \cite{Baldo2011PhRvC} (it corresponds to the oscillation of a charge-neutral mixture of proton pair condensate and leptons). This mode was found to effectively scatter on leptons and does not contribute to transport \cite{Bedaque:2013fja}. Since neutrons in the core form Cooper pairs in the anisotropic $^3P_2$ state, the condensate spontaneously breaks rotational symmetry, and one expects the appearance of corresponding Goldstone modes. These modes were termed `angulons'\footnote{Note that the same term was recently used in a different context \cite{Schmidt2015PhRvL}.} by \citet{Bedaque2003PhRvC}. The properties of angulons were studied by \citet{Bedaque:2012bs}, and their contribution to transport properties of neutron star cores by \citet{Bedaque:2013fja}. 
\citet{Leinson2012PhRvC} analyzed the collective modes of the order parameter on a microscopic level for all temperatures and did not find a gapless mode. Instead, he found modes similar to `normal-flapping' modes of the A-phase of superfluid $^3$He \cite{Vollhardt1990}, which are not massless at finite temperatures. This, however, contradicts the recent study by \citet{Bedaque2015PhRvC}, who find that angulons have zero mass at any temperature. The reason for the contradiction is unknown, and a consistent picture of the low-lying excitations in the superfluid phases of neutron star cores is yet to be developed. Nevertheless, even the massive mode can contribute to the transport properties provided its mass is sufficiently small (of the order of $T$). Note also that Fermi-liquid effects can strongly modify the properties of the collective modes at nonzero temperatures. It is possible that such modes become purely diffusive, not being well-defined quasiparticle excitations at finite $q$. These issues were discussed in detail for a general Fermi liquid that becomes superfluid  by \citet{Leggett1966PhRv,Leggett1966PThPh}. 
Finally, we note that the warnings stated above regarding  transport in the `ballistic' regime apply also for the superfluid phonons (and other collective modes). The phonon mean free path grows by powers of $T$ in comparison to the exponential growth of the mean free path for Bogoliubov quasiparticles, but it can still easily become of the order of the size of the superfluid region in the star. The dissipation in this situation must be
treated with caution, see also remarks and references below Eq.\ (\ref{etaphiCFL}) in the context of quark matter.

\subsubsection{Magnetic field effects}\label{S:core_nuc_mag}
Transport properties of the magnetized neutron star core can be addressed using similar methods as for the crust, discussed in Sec.~\ref{S:crust_mag}, but generalized to the case of multi-component mixtures. As in the crust, the anisotropy induced by the magnetic field renders the transport coefficients anisotropic.  In contrast to the crust, the fermionic excitations are not expected to become strongly quantized by the magnetic field due to the larger effective masses\footnote{Remember that the effective mass of electrons becomes larger as we increase the density. For a rough estimate, let us assume $n_{\rm B}=n_0$, an electron density $n_{e}=n_{\rm B}/10$, and $v_{Fe}\approx c$. Then, the effective mass $m^*_{e}=p_{Fe}/v_{Fe}$ squared can be translated into a magnetic field $B\sim 4\times 10^{18}\, {\rm G}$, at which quantizing effects would become important. This field is larger than what is typically expected for neutron star cores, given that the largest measured surface magnetic fields are of the order of $10^{15}\, {\rm G}$ (according to the data in the ATNF pulsar catalogue on the time of writing (version 1.58) \url{http://www.atnf.csiro.au/research/pulsar/psrcat} \cite{Manchester2005AJ}). 
}.
Still, the collision probabilities can depend on the magnetic field, for instance  due to a modification of the plasma screening,
although this effect has never been investigated, to the best of our knowledge. The anisotropic thermal conductivity has never been considered since it is assumed to be very large anyhow. Similarly, the shear viscosity in magnetized star cores has not been calculated (except for the attempt to study the collisionless problem by \citet{Banik2013}).

However, the problem of electrical conductivity has gained considerable attention because this quantity is one of the key ingredients in the magnetic field evolution in neutron stars, see for instance Ref.\ \cite{Graber2015MNRAS} and references therein. The response of multi-component mixtures to an external electromagnetic field differs qualitatively  from the case of the electron conductivity described in Sec.~\ref{S:crust_mag},  because of the relative motion between all components in the plasma. This is especially pronounced if neutral species are present in the mixture, like in the case of a partially ionized plasma or in neutron star matter. We briefly review the details of the calculations following Ref.~\cite{Yakovlev1991Ap&SS}.

Let us assume that the plasma as a whole moves with a non-relativistic velocity $\bm{v}$ under the influence of an external force (electric field, but we will be a bit more general at this point). This force induces an $l=1$ perturbation to the distribution function that is given by $\Phi_i=-\bm{w}_i\cdot \bm{\vartheta}_i$, where $i=1,\ldots , N$ labels the constituents  of the multi-component system. The unknown vectors $\bm{\vartheta}_i$ are energy-dependent, but, as a first approximation,  can be assumed to be constant (as we saw above, this is a fairly good approximation in degenerate matter). Then, these vectors are exactly the diffusion velocities of the constituents of the mixture (relative to the total velocity $\bm{v}$). If the co-moving frame is defined to ensure zero total momentum of the fluid element, the diffusion velocities obey the linear constraints \cite{Yakovlev1991Ap&SS}
\begin{equation}\label{eq:diff_vel_constr}
\sum_i \frac{\mu_i n_i}{c^2} \bm{\vartheta}_i=0 \, .
\end{equation}
Here, in a generalization of the center-of-mass velocity to a degenerate fluid of relativistic particles, the mass density $m_in_i$ has been replaced by $\mu_i n_i/c^2$, where the chemical potentials $\mu_i$ include the rest mass. 

Instead of the single Eq.~(\ref{boltzlin}), one then obtains a system of linearized kinetic equations. In order to determine the vectors $\bm{\vartheta}_i$,  one multiplies these equations by $\bm{p}_i$ and integrates them over momenta to arrive at 
(neglecting temperature gradients for simplicity, i.e., there are no thermo-diffusion and thermo-electric effects)	
\begin{equation}\label{eq:momentum_system}
-\frac{\mu_i n_i}{c^2} \dot{\bm{v}} +\bm{F}_i+\frac{q_i n_i }{c} \bm{\vartheta}_i \times \bm{B} =\sum_j J_{ij} (\bm{\vartheta}_i-\bm{\vartheta}_j),
\end{equation}
where the friction term on the right-hand side is given by the symmetric matrix $J_{ij}$, which is  related to the effective collision frequencies [see Eq.~(\ref{eq:collfreq_system})] by $J_{ij}=n_i m^*_i \nu_{ij}$. 
The driving term on the left-hand side contains the body forces $\bm{F}_i=n_i \bm{R}_i$,  which do not depend on $\bm{\vartheta}_i$, and the magnetic part of the Lorentz forces. The system (\ref{eq:momentum_system}) contains more unknowns than equations since it also determines the plasma acceleration $\dot{\bm{v}}$, but is closed by Eq.~(\ref{eq:diff_vel_constr}). After $\dot{\bm{v}}$  is eliminated, the solution can be written as \cite{YakShal1991ApSS}
\begin{equation}\label{eq:drift_mobility_mix}
\bm{\vartheta}_i=-\hat{{\cal D}}_{ij} \left(\bm{F}_j-X_j \bm{\bm F}\right),
\end{equation}
where $\bm{F}$ is the total force and $X_j=n_j \mu_j \left(\sum_k n_k \mu_k\right)^{-1}$ is the mass fraction of species $j$. The auxiliary tensor $\hat{{\cal D}}_{ij}$ has rank $N-1$ and can be expressed in different ways \cite{Lam2006PhFl}. The resulting drift velocities are of course the same for any representation of $\hat{{\cal D}}_{ij}$. Due to the linear constraint (\ref{eq:diff_vel_constr}), one can use any $N-1$ independent linear combinations of diffusion velocities for forming the current terms in the final hydrodynamical expressions. A natural choice for one of these combinations is the electric (charge) current $\bm{j}=\sum_i q_i n_i \bm{\vartheta}_i$.  In the three-component $npe$ plasma containing a neutral ($n$)
component and a charged ($pe$) fluid, the natural choice of the second independent velocity is the `ambipolar drift' velocity, i.e., the relative velocity of the charged fluid with respect to neutrons. This description has become standard, starting from the work by \citet{Goldreich1992ApJ}. However, in the more general case where several charged and neutral species coexist (for instance when muons and/or hyperons are included), this description becomes less convenient. We thus prefer to keep the more symmetric choice, that is to work with the drift velocities $\bm{\vartheta}_i$ subject to (\ref{eq:diff_vel_constr}) and use the full system of multi-fluid  hydrodynamical equations. Of course any representation leads to the same physical results. For the generalized Ohm law and the induction equation in $npe$ matter this was discussed in detail by \citet{Shalybkov1995MNRAS}. It is instructive to write the expression for the entropy generation rate in collisions with the help of the general equation (\ref{eq:entr_rate}),    
\begin{equation}\label{eq:cond_core_coll_entropy}
\left.T\varsigma\right|_{\rm coll}=\frac{1}{2} \sum_{ij}  J_{ij}(\bm{\vartheta}_i-\bm{\vartheta}_j)^2\, .
\end{equation}
Now consider the electrical conductivity problem, where the force terms in Eq.~(\ref{eq:momentum_system}) are solely given by the electromagnetic field \cite{Yakovlev1991Ap&SS},
\begin{equation}\label{eq:Lorentz_multi}
\bm{F}_i=q_i n_i \bm{E}'\equiv q_i n_i \left(\bm{E}+\frac{1}{c}\bm{v}\times\bm{B}\right)\, .
\end{equation}
 Then, the electric current and electric field $\bm{E}'$  in the co-moving system are related via the (generalized) Ohm law
$\bm{j}=\hat{\sigma} \bm{E}',$
where the electrical conductivity tensor $\hat{\sigma}$ is expressed through the tensor $\hat{\cal D}_{ij}$ as 
\begin{equation}\label{eq:sigma_mobility_mix}
\hat{\sigma}=-\sum_{i,j} q_i q_j n_i n_j \hat{{\cal D}}_{ij} \, .
\end{equation}
It is convenient to introduce also the resistivity tensor $\hat{\cal R}=\hat{\sigma}^{-1}$. The final result for  non-superfluid $npe$  matter assuming charge neutrality and neglecting electron-neutron collisions reads
\begin{equation}\label{eq:resistivity_mag_core}
{\cal R}_\parallel = \left[\sigma(B=0)\right]^{-1}\, ,\qquad {\cal R}_\perp={\cal R}_\parallel + \frac{X_n^2}{J_{pn} c^2} B^2 \, ,\qquad {\cal R}_\Lambda=\frac{1-2X_{e}}{n_{e} e c }B
\, ,
\end{equation}
where ${\cal R}_\parallel$ and ${\cal R}_\perp$ are the resistivities parallel and perpendicular to the magnetic field, respectively, and ${\cal R}_\Lambda$ is the Hall resistivity. 
The second term in the expression for ${\cal R}_\perp$ is proportional to $B^2$ and thus is responsible for a considerable increase of the resistivity in strong magnetic fields. As we see from its form, it originates from the friction of the neutron fluid with the charged components, governed by the strong forces. In this sense, it can be viewed as the result of  ambipolar diffusion \cite{Goldreich1992ApJ}. Note that the increase of the transverse resistivity in a magnetized plasma containing neutral species is a well-known effect in physics of space plasmas \cite{Bykov2007PhyU}. This has important consequence on the dissipation of the magnetic field energy. Indeed, the field energy dissipation rate per unit volume is
\begin{equation}\label{eq:WB_dissipation}
\dot{W}_B = -\bm{j}\cdot \bm{E}'=-j^2_\parallel {\cal R}_\parallel-j_\perp^2 {\cal R}_\perp \, ,
\end{equation}
where $j_\parallel$ and $j_\perp$ are the components of the electric current along and transverse to $\bm{B}$, respectively. Therefore, as first noted by \citet{HUI1990A&A}, the increase of ${\cal R}_\perp$ in Eq.~(\ref{eq:resistivity_mag_core}) can lead to accelerated dissipation. Since the neutron-proton collision frequencies scale as $J_{np}\propto T^{2}$, the increase in $\cal{R}_\perp$ becomes pronounced at lower temperatures.
The microscopic calculation of the friction coefficients $J_{ij}$ (or effective collision frequencies) were performed by \citet{YakShal1991ApSS} and updated in Ref.~\cite{Shternin2008JETP} (as described in Sec.~\ref{S:core_nuc_lepton} and \ref{S:core_nuc_baryon}). Recall that the recent results by \citet{Bertoni2015PhRvC} indicate the 
possible importance of the electron-neutron collisions. The influence of this effect on the electrical conductivity in neutron star cores has not 
been studied yet.

This simple picture is modified if other driving forces (in addition to the  electromagnetic field) are present on the left-hand side of Eq.~(\ref{eq:momentum_system}). If gradients of temperature or chemical potentials are present, one deals with thermo-electric and electro-diffusion effects. The equations (\ref{eq:diff_vel_constr})--(\ref{eq:cond_core_coll_entropy}) still hold, but the expressions for the hydrodynamical currents are different \cite{Landau1987Fluid, Landau1984Electrodynamics}. For instance, the electric current is not proportional to $\bm{E}'$, but also depends on thermal and chemical gradients. 
In the 
following we do not consider any thermal gradient and focus on diffusion effects, which were found to be important in the problem of the magnetic field evolution in neutron stars \cite{Goldreich1992ApJ,Pethick1992sens}. The reason is that the timescale of the field evolution can be comparable to the typical times of the reactions that are responsible for chemical equilibration (see Sec.~\ref{sec:urca}). The driving forces for diffusion are the gradients of the chemical potentials, which are added to the force term in the kinetic equation\footnote{The driving force terms should also contain gravitational acceleration to ensure the proper equilibrium state \cite{Shalybkov1995MNRAS,Lam2006PhFl}; we omit the corresponding terms for brevity.} 
\begin{equation}\label{eq:force_diff}
\bm{F}_i=q_i n_i \bm{E}'-n_i\nabla \mu_i.
\end{equation}
In a one-component fluid (Sec.~\ref{S:crust_boltz}), the chemical potential gradient can be absorbed into an effective electric field, see Eq.~(\ref{eq:elfield_eff}). Multiple $\nabla \mu_i$'s do not allow this simple prescription. The diffusion velocities are still found by  Eq.~(\ref{eq:drift_mobility_mix}). They now receive contributions from the electric field and from the gradients. The microscopic equation for the entropy generated in the collisions still has the form given in Eq.~(\ref{eq:cond_core_coll_entropy}), but the macroscopic expression now reads
\begin{equation}\label{eq:entropy_gen_macro_diff}
\left.T \varsigma\right|_{\rm coll} = \sum_i \bm{F}_i\cdot \bm{\vartheta}_i = \bm{j}\cdot \bm{E}'-\sum_i n_i \bm{\vartheta}_i\cdot \nabla\mu_i \, .
\end{equation}
The first term on the right-hand side corresponds to Ohmic heating, while the second term describes entropy generation due to the irreversible diffusion process. The heat source for the latter is the particle (chemical) energy $ \mu_i d n_i$. 
If chemical reactions are taken into account, $d n_i = -\nu_i d \xi$, with the stoichiometric coefficients $\nu_i$ and  
the reaction extent $\xi$, such that $\Gamma \equiv \dot{\xi}$ is the reaction rate. The thermodynamic force that drives the relaxation to  chemical equilibrium is\footnote{In the chemistry literature, (the negative of) $\delta\mu$ is called reaction affinity and usually denoted by ${\cal A}$. All reactions with nonzero reaction affinity we discuss later (Secs.\ \ref{sec:urca}, \ref{S:bulk}, \ref{sec:bulkquark}) have stoichiometric coefficients $\nu_i=\pm 1$, and the notation $\delta\mu$ is used mostly (but not exclusively) in the  neutron star literature we refer to.} 
\begin{equation}\label{eq:affinity_def}
\delta\mu =\sum_i \nu_i \mu_i \, .
\end{equation}
In equilibrium, $\delta\mu=0$, and the reactions tend to move to this point. The thermodynamic flux conjugate  to $\delta\mu$ is $\Gamma$. In the linear regime, $\Gamma= \lambda\,  \delta\mu$, where $\lambda$ is the corresponding transport coefficient, which, to a first approximation, only depends on the equilibrium state. The entropy generation from the chemical reactions is 
\begin{equation}\label{eq:entr_react}
\left.T\varsigma\right|_{\rm react} = \Gamma \, \delta\mu = \lambda \, \delta\mu^2
\, ,
\end{equation}
thus the second law of thermodynamics requires $\lambda > 0$.

If we allow for chemical reactions, the conservation laws are modified:
source terms appear in the continuity equations for the constituent species, recoil terms emerge in the momentum conservation equations (usually neglected as second-order), and the energy conservation law is also  modified. For instance, using the continuity equation including the source term, 
\begin{equation}\label{eq:num_dens_cons_react}
\frac{\partial n_i}{\partial t} +\nabla\cdot (n_i \bm{v})+\nabla \cdot(n_i \bm{\vartheta}_i) = -\Gamma \nu_i \, , 
\end{equation}
 one rewrites Eq.~(\ref{eq:entropy_gen_macro_diff}) as 
\begin{equation}\label{eq:entropy_electrodiff_react}
-\dot{W}_{ B}-\sum_i \mu_i \left(\frac{\partial}{\partial t}+\bm{v}\cdot \nabla \right) n_i =\left.T\varsigma\right|_{\rm react}+ \left.T\varsigma\right|_{\rm coll}= \lambda\, \delta\mu^2 + \frac{1}{2}\sum\limits_{ij} J_{ij}(\bm{\vartheta}_i-\bm{\vartheta}_j)^2,
\end{equation}
where we have used Eq.\ (\ref{eq:diff_vel_constr}) and assumed ${\rm div}\,\bm{v}=0$. 
(The interplay between fluid compression, i.e., ${\rm div}\,\bm{v}\neq 0$, and the chemical reactions is discussed in detail in Sec.~\ref{S:bulk}.) 
It is frequently assumed that the magnetic field evolution is quasistationary,
such that the second (`chemical') term on the left-hand side of Eq.~(\ref{eq:entropy_electrodiff_react}) can be neglected. Then, Eq.~(\ref{eq:entropy_electrodiff_react}) describes the transfer of energy of the magnetic field to heat via binary collisions and reactions \cite{Gusakov2017arXiv, Castillo2017MNRAS}.

The problem of the magnetic field evolution in neutron star cores attracts persistent attention \cite{Castillo2017MNRAS,Gusakov2017arXiv,Passamonti2017,Glampedakis2011MNRAS, Beloborodov2016ApJ,Graber2015MNRAS,Hoyos2008A&A, Elfritz2016MNRAS, Bransgrove2017MNRAS}, and a more complete list of references can be found in the cited works. We did not discuss general relativistic 
effects and the influence of Cooper pairing, which 
can both be important. A full dynamical analysis that includes magnetic, thermal, and chemical evolution is highly demanded but has not been performed yet.

\subsection{Reaction rates from the weak interaction}
\label{sec:urca}
\subsubsection{General treatment}\label{S:urca_general}

Neutrino emissivity is a key ingredient in the study of the neutron star  evolution. 
Minutes after the birth of a neutron star, neutrinos escape the star freely, taking away energy and thus cooling the star. The core is the main source of neutrinos, which are produced in various reactions involving the weak interaction. 
In the exhaustive review by \citet{Yakovlev2001physrep} the wealth of neutrino-producing reactions possible in neutron star cores are described systematically. Naturally, the most important processes among them involve baryons. In this section we mainly focus on the recent results for these reactions, and mention others -- less efficient ones -- only briefly. 
As for the case of the transport coefficients discussed in Sec.~\ref{S:core_nuc_baryon}, the main recent efforts have been focused on improving the treatment of in-medium effects. 

The problem of calculating the neutrino emissivity is closely related to the more general problem of neutrino transport in dense matter, which is of utmost importance in supernovae and proto-neutron star studies. The emissivity can be calculated from the gain term in the corresponding neutrino transport equation. If this is done in the framework of non-equilibrium transport theory, one is in principle able to study the weak response of dense matter in a systematic way, including situations far from equilibrium. For a pedagogical discussion of the real-time Green's function approach see Ref.~\cite{Sedrakian2008PPN, Sedrakian2007PrPNP} and references therein. Alternatively, emissivities can be found using the optical theorem without employing the neutrino transport equation (see, e.g., Ref.~\cite{Voskresensky2001LNP}). In any case, the rates can be expressed through the contraction of the weak currents with the polarization of the medium. The latter accounts for all many-body processes 
which exist in dense matter, and its microscopic calculation is not straightforward. Fortunately, in neutron star  cores the quasiparticle approximation is well-justified. 
In this approximation, the reaction rates can be equivalently calculated using Fermi's Golden Rule based on the squared matrix element of the process. Due to its transparency, this approach is most commonly used for the weak reactions in hadronic cores of neutron stars. In this section we will follow this prescription and discuss the results beyond the quasiparticle picture at the end. This also allows us to make a close connection to the previous section. We note, however, that the approach based on Green's functions is particularly advantageous in the case of pairing, where some difficulties can arise with the use of Fermi's Golden Rule. We illustrate this approach in Sec.~\ref{sec:quarkneutrino}, where neutrino emission from quark matter is considered. 

The weak reactions are naturally classified by the number of quasiparticles involved, since each fermion generally adds a phase factor $T/\mu$ (cf. Sec.~\ref{S:core_boltz}), and by the type of the weak current (neutral or charged) responsible for the process\footnote{In this section we set $k_B=1$ for brevity, such that $T/\mu$ is dimensionless.}. 
Both types contribute to the neutrino emissivity, but only the flavor-changing reactions (that go via the charged weak current) are responsible for establishing beta-equilibrium.
The first kinematically allowed processes with the lowest number of involved quasiparticles  give the dominant contribution to the reaction rates. 
Therefore, the most powerful neutrino emission mechanism is the so-called baryon direct Urca process, which consists of a pair of reactions going via the charged weak current
\begin{subequations}\label{eq:Core_Urca_forrev}
\begin{eqnarray}
{\cal B}_1&\to& {\cal B}_2 + \ell + \bar{\nu}_\ell\label{eq:Core_Urca_forward} \, , \\[2ex]
{\cal B}_2 +\ell &\to& {\cal B}_1 + \nu_\ell \, ,\label{eq:Core_Urca_reverse}
\end{eqnarray}
\end{subequations}
where ${\cal B}_{1,2}$ stand for baryons [for instance $({\cal B}_1,{\cal B}_2)=(n,p)$ in the case of nuclear matter], $\ell$ for leptons, and $\nu_\ell$ for the corresponding neutrino. In equilibrium, the rates of the reactions (\ref{eq:Core_Urca_forward}) and (\ref{eq:Core_Urca_reverse}) are equal. They scale as $T^5$ and the neutrino emissivity as $T^6$, see below.  It is essential that direct Urca processes in strongly degenerate matter have a threshold: in the quasiparticle approximation all fermions in Eqs.~(\ref{eq:Core_Urca_forrev}) are placed on their Fermi surfaces [see Eq.~(\ref{eq:phase_space_decomp})]. Momentum conservation implies that the Fermi momenta of the triple $({\cal B}_1,{\cal B}_2,\ell)$ satisfy the triangle condition (if one neglects the small neutrino momentum which is of the order $T$). In the highly isospin-asymmetric cores of neutron stars [$({\cal B}_1,{\cal B}_2)=(n,p)$] this strongly limits both the electron ($\ell=e$) and muon ($\ell=\mu$) direct Urca processes. Only in matter with 
a sufficiently large proton fraction (and as a consequence of the electric neutrality, lepton fraction) the triangle condition for the Fermi momenta of the triple $(n,p,\ell)$ can be satisfied. In electrically neutral $npe$ matter, the threshold for the proton fraction $x_p$ is $11\%$. Therefore, not all equations of state allow for the direct Urca process to operate. Depending on the $x_p(n_{\rm B})$ profile, some equations of state can allow direct Urca processes for massive stars, and for some equations of state the proton fraction never exceeds the direct Urca threshold. 

The counterpart to (\ref{eq:Core_Urca_forrev}) via the neutral weak current is
\begin{equation}
{\cal B}_1 \to {\cal B}_1 + \bar{\nu}_\ell+\nu_\ell \, .\label{eq:Core_weak_neutral}
\end{equation}
In the quasiparticle approximation, this reaction is kinematically forbidden, unless the baryons ${\cal B}_1$ form a Cooper pair condensate,
see Sec.\ \ref{sec:nu_Cooper}. 

When the direct Urca processes are not allowed, the next-order processes in the number of quasiparticles take the lead. They include an additional baryon ${\cal C}$
which couples to the emitting baryons via the strong force\footnote{The spectator particle ${\cal C}$ can also be a lepton coupling via the electromagnetic forces, but this process is negligible.}. The presence of a spectator relieves the triangle condition, but the price to pay is the reduced phase space of the process by a factor of $(T/\mu)^2$. In the presence of the spectator, the neutral-current reaction 
\begin{equation}
{\cal B}_1+{\cal C} \to {\cal B}_1+{\cal C} + \bar{\nu}_\ell+\nu_\ell \label{eq:Core_Murca_Bremms}
\end{equation}
becomes possible, and it is the familiar bremsstrahlung emission. The most important processes are, however, the  charged-current reactions
\begin{subequations}\label{eq:Core_Murca_forrev}
\begin{eqnarray}
{\cal B}_1 + {\cal C}&\to& {\cal B}_2+{\cal C} + \ell + \bar{\nu}_\ell \, , \label{eq:Core_Murca_forward}\\[2ex]
{\cal B}_2 +{\cal C}+\ell &\to& {\cal B}_1 + {\cal C} + \nu_\ell \, , \label{eq:Core_Murca_reverse}
\end{eqnarray}
\end{subequations}
called modified Urca reactions (see Ref.~\cite{Yakovlev2001physrep} for the origin of the nomenclature). The reactions (\ref{eq:Core_Murca_Bremms})--(\ref{eq:Core_Murca_forrev}) are sometimes jointly called the electroweak bremsstrahlung of the lepton pairs. Depending on the relations between the Fermi momenta of the five degenerate fermions involved in the modified Urca reactions (\ref{eq:Core_Murca_forrev}), these reactions can also have thresholds. However, in practice, this is never really important in neutron star conditions. For the complete classification of all phase-space restrictions for reactions (\ref{eq:Core_Urca_forrev})--(\ref{eq:Core_Murca_forrev}) see Ref.~\cite{Kaminker2016Ap&SS}.

The expression for the rate $\Gamma$ and the neutrino emissivity $\epsilon_\nu$ of any of the reactions (\ref{eq:Core_Urca_forrev})--(\ref{eq:Core_Murca_forrev})  can be expressed via Fermi's Golden Rule as follows \cite[e.g.,][]{Yakovlev2001physrep,ShapiroTeukolsky1983book}
\begin{equation}\label{eq:weak_rate_gen_matel}
\left(\begin{array}{l}
\Gamma \\
\epsilon_\nu
\end{array} \right) = \int \prod\limits_{j=i,f} \frac{d^3 \bm{p}_j}{(2\pi)^3}\, {\cal F}_{fi}(2\pi)^4 \delta^{(4)}\left(P_f-P_i\right)\left(\begin{array}{l} 
1 \\
\omega_\nu
\end{array}\right) s\left|M_{fi}\right|^2,
\end{equation}
where $M_{fi}$ is the transition amplitude for the reaction (summed over initial and averaged over final polarizations), $P_i$ and $P_f$ are total four-momenta of initial and final particles, respectively, integration is done over the whole phase-space of reacting quasiparticles (including neutrinos)\footnote{We keep the same normalizations of the baryon and lepton wave functions for brevity, while traditionally one puts $2\varepsilon$ in the denominator for relativistic particles.}, and the symmetry factor $s$ corrects the phase volume in case of indistinguishable collisions. The quantity ${\cal F}_{fi}$ is the Pauli blocking factor
\begin{equation}\label{eq:Fermi_blocking_product}
{\cal F}_{fi}= \prod\limits_{i} f_F(\varepsilon_i-\mu_i) \prod\limits_{f\neq\nu} \left[1-f_F(\varepsilon_f-\mu_f)\right]\, ,
\end{equation}
where $f_F(y)\equiv (e^{y/T}+1)^{-1}$.
It contains products of the distribution functions for all fermions except neutrinos (which escape the star). At small temperatures,  it effectively puts all the quasiparticles on the respective Fermi surfaces when the matrix element is calculated [cf.~Eq.~(\ref{boltz_coll_lin})]. Finally, the  energy $\omega_\nu$, 
which enters the expression for the emissivity $\epsilon_\nu$, is the neutrino/antineutrino energy in the reactions (\ref{eq:Core_Urca_forrev}), (\ref{eq:Core_Murca_forrev}) and the total energy of the neutrino pair in the case of the bremsstrahlung reaction (\ref{eq:Core_Murca_Bremms}).  

The calculation of $\Gamma$ and $\epsilon_\nu$ is similar to the calculation of collision frequencies described in Sec.~\ref{S:core_boltz}. First, we assume that the matrix element $M_{fi}$ does not depend on the neutrino energy. Then, 
the integrals over the absolute values of the momenta are rewritten as energy integrals -- for the degenerate particles according to Eq.~(\ref{eq:phase_space_decomp}) and for the neutrino $d^3 \bm{p}_\nu=c^{-3}\varepsilon_\nu^2\, d\varepsilon_\nu \,d\Omega_\nu$. Introducing  the dimensionless energy variables $x=(\varepsilon-\mu)/T$ as in Sec.~\ref{S:core_boltz}, one can express the energy-conserving delta-function as
\begin{equation}
\delta(E_f-E_i)=T^{-1} \delta\left(\sum_f x_f - \sum_i x_i + \frac{\delta\mu}{T} \right),\label{eq:encons_delta}
\end{equation}
where $\delta \mu$ is the chemical potential difference between the final and initial particles in the reaction, as introduced in Eq.~(\ref{eq:affinity_def}). Recall that the freely escaping  neutrinos have zero chemical potential. Finally, we
can write both $\Gamma$ and $\epsilon_\nu$ in the following generic form,
\begin{equation}\label{eq:weak_rate_decomposed}
\left(\begin{array}{l}
\Gamma \\
\epsilon_\nu
\end{array} \right) =\frac{s}{(2\pi)^{3n-4}}\, \hat{\Omega}\, T^k\, I_\epsilon \left(\frac{\delta \mu}{T}\right)\langle\left|M_{fi}\right|^2\rangle \prod\limits_{j\neq \nu} p_{Fj} m^*_j\, ,
\end{equation}
 where $\hat{\Omega}$ is the angular integral over quasiparticle momenta orientations for fixed absolute values of momenta, so that the possible relative orientations are restricted by  momentum conservation \cite{Kaminker2016Ap&SS}, $\langle\left|M_{fi}\right|^2\rangle$ stands for the angular-averaged matrix element, $n$ is the number of reacting quasiparticles, and the last product comes from the quasiparticle densities of states on the respective Fermi surfaces. The factor $I_\epsilon(\delta \mu/T)$ in Eq.~(\ref{eq:weak_rate_decomposed}) is the energy integral over the dimensionless variables $x_{i,f}$. Due to Eq.~(\ref{eq:encons_delta}), it depends on the ratio $\delta\mu/T$.
In beta-equilibrium and for bremsstrahlung  reactions, $\delta\mu = 0$. The temperature dependence is given by the factor $T^k$ in Eq.~(\ref{eq:weak_rate_decomposed}), where the exponent $k$ depends on the specific reaction and on whether we compute $\Gamma$ or $\epsilon_\nu$: each degenerate fermion on either side of the reaction gives one power of $T$, the neutrino contributes a factor $T^3$, and one power of $T$ is subtracted due to the energy-conserving delta function. Therefore, the 
rate for the direct Urca process (three fermions) is proportional to $T^5$ and the rate for the modified Urca process (five fermions) is proportional to $T^7$. The corresponding emissivities have an extra 
factor $T$ due to the neutrino energy $\omega_\nu$ \cite{ShapiroTeukolsky1983book}. Slightly different considerations should be carried out for the bremsstrahlung reactions, where the assumption of an energy-independent matrix element no longer holds. Instead, as we discuss below, the leading energy-dependence of the bremsstrahlung matrix elements is $M_{fi}\propto \omega_\nu^{-1}$ and can be factored out from the angular integration. After the factorization, the decomposition (\ref{eq:weak_rate_decomposed}) still holds where the energy-independent part of the angular-averaged squared matrix element stays in place of $\langle\left|M_{fi}\right|^2\rangle$. The temperature dependence for the bremsstrahlung reaction (\ref{eq:Core_Murca_Bremms}) is the same as for the modified Urca reactions (\ref{eq:Core_Murca_forrev}) since the appearance of the squared neutrino pair energy in the denominator is compensated by the fact that the integration is now 
performed over the momenta of both neutrino and antineutrino in the outgoing channel of the reaction (see for instance Ref.\ \cite{Yakovlev2001physrep}). 

Let us return to the Urca reactions, which are responsible for the processes of beta-equilibration. We denote quantities related to the forward and backward reactions by superscripts `$+$' and `$-$', respectively. Then, the net neutrino emissivity from a pair of forward and backward reactions is $\epsilon_{\nu}(\delta\mu)=\epsilon_\nu^++\epsilon_\nu^-$ and the rate of the composition change is $\Delta \Gamma(\delta\mu) = \Gamma^+-\Gamma^-$. In equilibrium, $\delta\mu =0$ and $\Delta \Gamma =0$, while $\epsilon_{\nu}=2 \epsilon_\nu^+$.
Beta-equilibration processes increase the entropy of the system, while the neutrino emission takes away energy. Thus the total heat release of the Urca reactions in non-equilibrium is
\begin{equation}\label{eq:weak_total_heat}
T\varsigma = \Delta\Gamma \,\delta\mu - \epsilon_{\nu}.
\end{equation}
This means that the pair of the Urca reactions  can either cool or heat the star depending on the relation of two terms in Eq~(\ref{eq:weak_total_heat}), which in turn depends on the degree of departure from equilibrium. 

Both $\Delta \Gamma$ and $\epsilon_{\nu}$ can be expressed as a product of the equilibrium reaction rate times 
a function which depends solely on $\delta\mu/T$. The analytical expressions for these functions in non-superfluid matter can be found, for example, in Ref.~\cite{Reisenegger1995ApJ}. When the deviation from beta-equilibrium is small ($|\delta\mu| \ll T$), the response to this deviation is linear, $\Delta\Gamma \propto \delta\mu$, while $\epsilon_{\nu}\approx \epsilon_{\nu}^{\rm eq}$. In the suprathermal regime, when $|\delta\mu| \gg T$, the phase space available for the reacting particles is determined by $\delta \mu$ instead of $T$. Then, $I_\epsilon \propto (\delta\mu/T)^k$, such that the Urca reaction rates become temperature-independent, and $\delta \mu$ enters the final expressions in place of $T$, see for instance Ref.~\cite{Yakovlev2001physrep} for details. Moreover, \citet{FloresReisenegger2006MNRAS} proved the general expression
\begin{equation}\label{eq:Reiss_formula}
\frac{\partial \epsilon_{\nu}}{\partial \delta \mu} = 3\Delta \Gamma \, .
\end{equation}
Thus the reactions generate heat if $\epsilon_{\nu}(\delta \mu)$ is steeper than $\delta\mu^3$ and cool the star via neutrino emission otherwise. The advantage of the relation (\ref{eq:Reiss_formula}) is that it also works in case of pairing \cite{FloresReisenegger2006MNRAS}.

\subsubsection{Neutrino emission of nuclear matter}\label{S:urca_nucleon}

We now use the general formalism outlined above to quantitatively discuss the main neutrino reactions relevant in neutron star cores. We start with the most efficient one, the direct Urca reaction (\ref{eq:Core_Urca_forrev}). The possibility of its occurrence and importance for neutron stars was first pointed out by \citet{Boguta1981PhLB}, who found that the threshold conditions for the direct Urca process to operate can be fulfilled in some relativistic mean field models. This paper was unnoticed for a decade until \citet{Lattimer1991PhRvL} rediscovered this possibility and argued that the sufficiently large proton fraction can be achieved for many
realistic density dependencies of the symmetry energy of nuclear matter.	
In the limit of small momentum $q$ transferred from leptons to nucleons, the weak charged current of nucleons contains vector ($V$) and axial-vector ($A$) contributions.
For non-relativistic nucleons, when nucleon recoil can be neglected, the direct Urca matrix element averaged over the directions of neutrino momenta is \cite{Lattimer1991PhRvL, Yakovlev2001physrep}
\begin{equation}\label{eq:DU_matel}
|M_{\rm DU}|^2 = 2G_F^2 \cos^2 \theta_C (g_V^2+3g_A^2) \, ,
\end{equation}
where $G_F=1.17\times 10^{-5}$~GeV$^{-2}$ is the Fermi coupling constant, $\theta_C$ is the Cabibbo angle with $\sin\theta_C\approx 0.22$, and $g_V= 1$ and $g_A\approx 1.26$ are the nucleon weak vector and axial vector coupling constants. 

Inserting the expression (\ref{eq:DU_matel}) into Eq.~(\ref{eq:weak_rate_decomposed}), one obtains for the rate of the $np\ell$ reaction (in equilibrium) in physical units
\begin{equation}\label{eq:QDU_equil_phys}
\epsilon^{\rm DU}_{\nu} = 4\times10^{21} \left(\frac{n_{\rm B}}{n_0}\right)^{1/3} x_\ell^{1/3} \frac{m^*_p m_n^*}{m_N^2} \left(\frac{T}{10^8 K}\right)^6 \Theta_{np\ell}\ {\rm erg}~{\rm cm}^{-3}~{\rm s}^{-1},
\end{equation}
where $\Theta_{np\ell}$ is the step function accounting for the threshold. The reaction rates $\Gamma=\Gamma^+=\Gamma^-$  in equilibrium and the composition change rate in the subthermal regime are \cite{Yakovlev2001physrep}
\begin{equation}\label{eq:GammaDU_phys}
\Gamma^{\rm DU} = \frac{0.118}{T} \epsilon^{\rm DU}_{\nu}\, , \qquad \Delta \Gamma^{\rm DU} = \frac{0.158}{T} \frac {\delta \mu}{T} \epsilon^{\rm DU}_{\nu} \qquad (\mbox{for} \; \delta\mu \ll T) \, .
\end{equation}
We note that the characteristic time for a nucleon to participate in a direct Urca reaction is quite large, $\tau^{\rm DU} \sim n_{\rm B}/\Gamma^{\rm DU}\approx 500$~yr (for $n_{\rm B}=n_0$, $T=10^8$~K, and $x_\ell = 0.1$). Nevertheless, the direct Urca reaction is the strongest neutrino emission process and -- if it operates -- cools the star very fast.

The classical result (\ref{eq:QDU_equil_phys}) changes when relativistic corrections are taken into account. Apart from the nuclear recoil effect \cite{LeinsonPerez2001PhLBDu}, such corrections arise from additional terms in the nucleon charged weak current, corresponding to  a non-trivial spatial structure of nucleons. 
In addition to the vector and axial vector contributions, the weak currents contain tensor ($T$) and induced pseudoscalar ($P$) terms (see, for instance, Ref.~\cite{Timmermans2002PhRvC} for the discussion of the weak hadron currents). The $T$-terms are of the order $q/m_p$ and describe weak magnetism. According to Ref.~\cite{Leinson2002NuPhA} the weak magnetism contribution can be as large as 50\% of the rate (\ref{eq:QDU_equil_phys}). In the closely related context of the neutrino opacity due to charged weak current reactions, these corrections were taken into account in Ref.~\cite{HorowitzPerez2003PhRvC}, see also Ref.~\cite{RobertsReddy2017PhRvC} and references therein for a recent discussion on this subject. The contribution from the $P$-terms turns out to be proportional to the lepton mass and thus is found to be unimportant for the direct Urca processes with electrons \cite{Leinson2012PhRvC}. However, for the muonic direct Urca ($\ell=\mu$), this contribution can be substantial  \cite{Timmermans2002PhRvC}. For modifications of the direct Urca rate in various parameterizations in  the framework of relativistic mean field theories see also Ref.~\cite{Ding2016CoTPh}.

Another modification of the rate (\ref{eq:QDU_equil_phys}) results from in-medium effects (cf. Sec.~\ref{S:core_nuclear_kincoeff}). The simplest manifestation of the in-medium effects is through the modifications of the effective nucleon masses $m_{p}^*$ and $m_{n}^*$ (see, e.g., \citet{Baldo2014PhRvC} for an illustration of the range of uncertainty). \citet{Dong2016ApJ} investigated the suppression of the emissivities due to Fermi surface depletion. The depletion is quantified in terms of the `quasiparticle strength' $z_F<1$, which appears in the numerator of the single-particle propagator in the interacting system \cite{Dickhoff2008}. The reaction rates are basically multiplied by powers of $z_F$ depending of the number of quasiparticle involved. In this case it is important to use the effective masses calculated in the same order of the theory. Corrections leading to $z_F<1$ are counterbalanced by an increase of the effective mass \cite{Schwenk2003NuPhA, Zuo1998PhLB,Baldo2000PhLB,Schwenk2004PhLB}, thus the overall corrections are not dramatic.
Also, the couplings $g_V$ and $g_A$ (and, in principle, the tensor and pseudoscalar couplings $g_T$ and $g_P$) are renormalized in a dense medium due to nucleon correlations \cite{Voskresensky2001LNP}. This effect is assumed to be not very important for direct Urca processes \cite{Kolomeitsev2015PhRvC}.

In any case, all these corrections (relativistic, weak magnetism, in-medium)  modify the estimate (\ref{eq:QDU_equil_phys})  at most by a factor of a few, leaving all the principal astrophysical consequences based on its high rate intact.  

\begin{figure}[t]
\begin{center}
\includegraphics[width=0.75\textwidth]{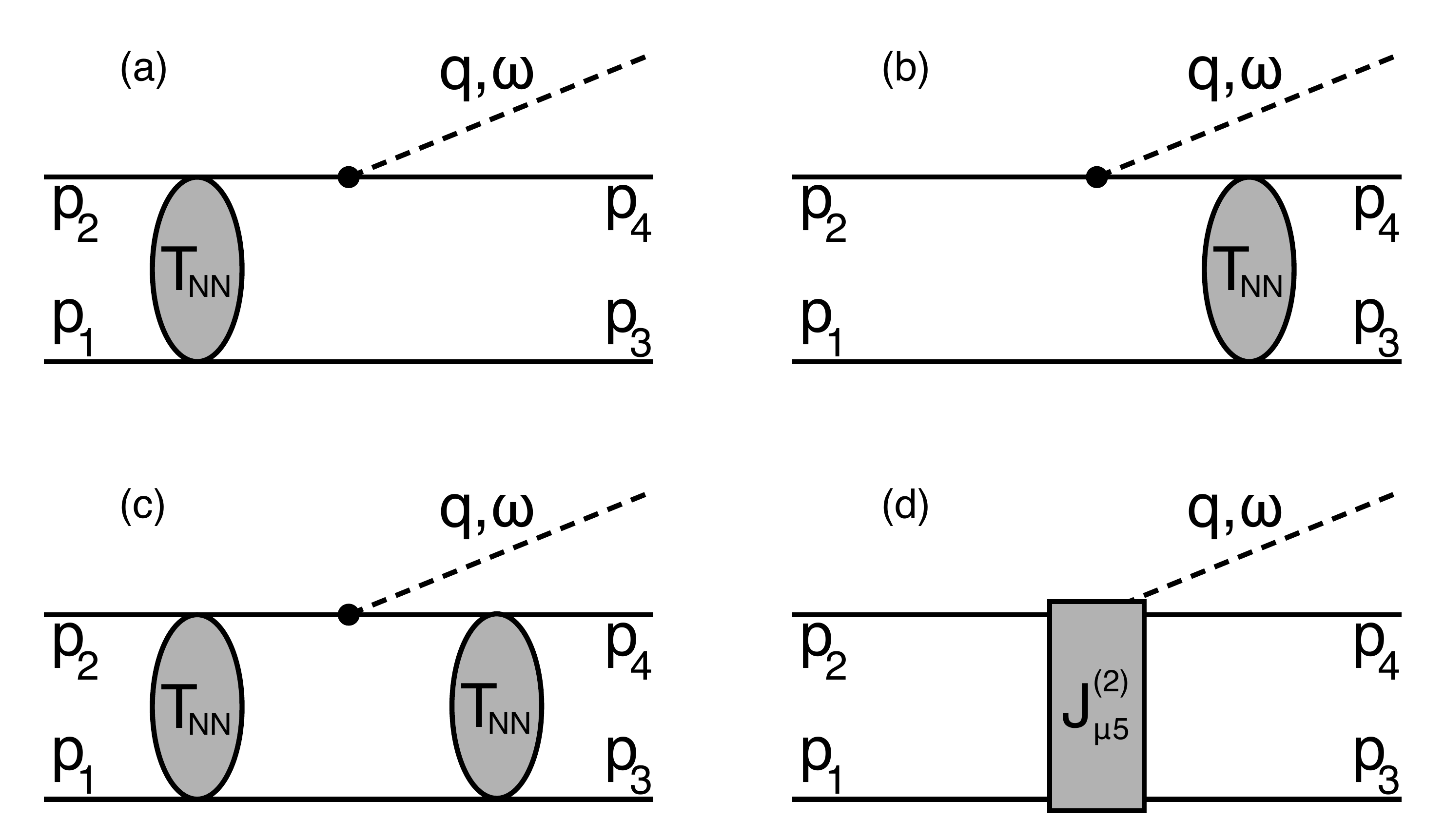}
\caption{Feynman diagrams contributing to the NN electroweak bremsstrahlung reactions. Here $p_1\dots p_4$ label initial and final nucleon momenta, $q,\omega$  are the emitted lepton pair [$\nu \bar{\nu}$ for bremsstrahlung reactions (\ref{eq:Core_Murca_Bremms}) and $\bar{\nu}_\ell \ell$ for modified Urca reactions (\ref{eq:Core_Murca_forrev})] momentum and energy, respectively. $T_{NN}$ is the quasiparticle scattering amplitude. In diagram (d), the $J_{\mu 5}^{(2)}$ block represents the two-body axial vector current. The figure is redrawn from Ref.~\cite{Hanhart2001PhLB}.}
\label{fig:NNdiagrams}
\end{center}
\end{figure}

Now we turn to the situation where the proton fraction is small and the triangle condition forbids the direct Urca process. Then the next order processes come into play, which all involve the strong interaction. The corresponding reaction rates are subject to uncertainties in the description of nucleon-nucleon interactions in medium. In this regard, the situation is similar to the problem of kinetic coefficients discussed in Sec.~\ref{S:core_nuclear_kincoeff} but is more complicated even if in-medium effects are not considered. This can be understood by looking at the relevant diagrams for the bremsstrahlung reaction (\ref{eq:Core_Murca_Bremms}) presented in Fig.~\ref{fig:NNdiagrams}. The dashed line represents the emitted lepton pair, while the block $T_{NN}$ represents the nucleon strong interaction amplitude. With respect to the strong interaction vertex, the emission of the lepton pair occurs from `external legs' in 
diagrams (a) and (b),
but there are also rescattering contributions (c) and emission from internal meson exchange lines (d) \cite{Hanhart2001PhLB}. Moreover, the amplitude $T_{NN}$ is half on-shell in contrast to the on-shell amplitude involved in the kinetic coefficients calculations. 

Despite the considerable progress achieved in recent decades in the treatment of the nucleon interaction, in practice the standard benchmark for the electroweak bremsstrahlung reactions follows the work of \citet{FrimanMaxwell1979ApJ}, who used the lowest-order one-pion exchange (OPE) model to describe the nucleon interaction. The neutron star cooling simulations which employ neutrino emissivities calculated in the \citet{FrimanMaxwell1979ApJ} model are called 'standard cooling scenarios' (e.g., \cite{YakovlevPethick2004ARA}). However, it is well-known that at energies relevant to neutron stars, OPE in the Born approximation overpredicts the cross-section by a factor of a few. 
\citet{FrimanMaxwell1979ApJ} also estimated the in-medium effects of long-range and short-range correlations by considering a special form of the correlated potential, utilizing the set of the Landau-Migdal parameters, and investigating the role of the one-$\rho$ exchange. In their calculations, Friman and Maxwell considered diagrams (a) and (b)
and used the non-relativistic V-A model for the weak vertices. They found that in the limit of small $q$, the vector current contributions of diagrams (a) and (b) cancels exactly for both OPE and Landau interactions. This is true for both neutral current and charged current (modified Urca) reactions (although the cancellation in latter case is nontrivial and involves exchange contributions). Therefore they concluded that the neutrino emission is dominated by the axial vector current. We will see below that this result survives in a more elaborate treatment.  

One may use a more universal approach based on the soft electroweak bremsstrahlung theorem. This method is similar to the soft-photon theorems for the electromagnetic emission \cite{Low1958PhRv,BurnettKroll1968PhRvL} which relates the cross-sections of the radiative processes in the leading and sub-leading orders to the corresponding cross-sections of the non-radiative processes. The soft photon theorem was extended to the axial vector currents by \citet{AdlerDothan1966PhRv} and first applied to neutrino emission in Refs.~\cite{Hanhart2001PhLB,Timmermans2002PhRvC}. \citet{Hanhart2001PhLB} employed the dominant term in the soft expansion, while \citet{Timmermans2002PhRvC} proved the general soft electroweak bremsstrahlung theorem. They also analyzed the full relativistic structure of the weak currents and the strong interaction amplitude. It was found that in  the extreme non-relativistic limit the vector current contribution vanishes irrespectively of the details of the structure of the amplitude $T_{NN}$, which generalizes the Friman \& Maxwell result. The basis of the soft emission theorem is the requirement of the vector current conservation and the partial axial vector current conservation.
In the soft limit, the dominant contribution comes from the diagrams (a)-(b) in Fig.~\ref{fig:NNdiagrams}, while the diagrams (c)-(d) are of higher  order in $q$, $\omega$. 

The hadronic part of the matrix element corresponding to diagram (b) in \ref{fig:NNdiagrams} can be written as
\begin{equation}\label{eq:weak_current_bremms}
M_{fi} \propto \langle f|\hat{\bm{\Gamma}}^{Z/W} {\cal G}_{N}(\bm{p}_2-\bm{q};\varepsilon_2-\omega) {T}_{NN} |i \rangle \, ,
\end{equation}
where $\hat{\bm{\Gamma}}^{Z/W}$ is the neutral or charged current weak vertex, and ${\cal G}_{N}$ is the nucleon propagator in the intermediate nucleon line. Non-relativistically, the quasiparticle propagator can be written in the form 
\begin{equation}\label{eq:nonrel_prop}
{\cal G}_{N}(\bm{p};\varepsilon)=\left[\frac{p^2}{2m_N^*}-\frac{p_{FN}^2}{2m_N^*} - (\varepsilon-\mu_N)\right]^{-1}.
\end{equation}
When the lepton pair energy and momentum are small, $\omega\ll \varepsilon_2$, $q\ll p_{2}$, one obtains 
${\cal G}_{N}(\bm{p}_2-\bm{q};\varepsilon_2-\omega)\approx \omega^{-1}$. Therefore, in the soft limit, $M_{fi}\propto \omega^{-1}$. 

For the bremsstrahlung reactions (\ref{eq:Core_Murca_Bremms}) the emitted energy is equal to the energy of the neutrino pair $\omega=\omega_\nu$, that is of the order of the temperature $T\lesssim 10$~MeV. Therefore, the soft limit is directly applicable. In contrast, for the  modified Urca processes (\ref{eq:Core_Urca_forrev}), the 
emitted energy is basically the degenerate lepton Fermi energy $\mu_\ell\sim 100$~MeV,  which is not small. We thus first discuss the reactions (\ref{eq:Core_Murca_Bremms}), although they are generally less important for neutron stars than the modified Urca processes.  For $nn$ (or $pp$) bremsstrahlung, the soft limit matrix element of the axial vector nucleon current in the non-relativistic limit can be expressed via the on-shell scattering amplitude as \cite{Hanhart2001PhLB, Timmermans2002PhRvC}
\begin{equation}\label{eq:MatelSoftnn}
\bm{J}^{A}_{fi}\propto \frac{g_A}{\omega_\nu}\left[T_{NN},\bm{S}\right]_{fi},
\end{equation}
where $\bm{S}$ is the operator of the total spin of the nucleon pair and the square brackets denote the commutator. 
The denominator $\omega_\nu$ comes from the virtual nucleon propagator in the dominant `external legs' diagrams [diagrams (a) or (b) in Fig.~\ref{fig:NNdiagrams}]. This allows us to calculate the reaction rates and emissivities in a model-independent way based on the experimentally measured phase shifts (in other words, $T_{NN}$). A similar approach was used in Ref.~\cite{Baiko2001AA} for transport coefficients (see Sec.~\ref{S:core_nuclear_kincoeff}). An analogous (but different) expression can be written for the $np$ bremsstrahlung \cite{Timmermans2002PhRvC}. 

It is customary to write the expression for the neutrino emissivity in the Friman \& Maxwell form \cite{FrimanMaxwell1979ApJ,Yakovlev2001physrep} 
\begin{equation}\label{eq:nn_brems_ope}
\epsilon^{nn}_\nu=7.5\times 10^{11} \left(\frac{m_n^*}{m_N}\right)^4 \left(\frac{n_n}{n_0}\right)^{1/3} \left(\frac{T}{10^8\,{\rm K}}\right)^8 \alpha^{\rm ex}_{nn}\beta_{nn}{\cal N}_\nu\ \rm erg~cm^{-3}~s^{-1},
\end{equation}
where  ${\cal N}_\nu=3$ is the number of neutrino flavors, and $\alpha^{\rm ex}_{nn}\approx 0.8$ is the dimensionless factor coming from the angular averaged OPE matrix element; its density dependence is very mild\footnote{The numerical prefactor in Eq.~(\ref{eq:nn_brems_ope}) is calculated assuming a charge-independent value of the pion-nucleon coupling constant $f^2_{NN\pi}=0.08$, as used in Refs.~\cite{FrimanMaxwell1979ApJ,Yakovlev2001physrep}. Using $f^2_{NN\pi}=0.075$, as in Ref.~\cite{Timmermans2002PhRvC}, results in a prefactor 6.6 instead of 7.5. This difference plays no role in practical applications.}. (The superscript `ex' indicates that the exchange contribution is included.) Note that \citet{FrimanMaxwell1979ApJ} overestimated the exchange contribution by a factor of two because they used incorrect symmetry factor $s=1/2$ [see Eq.~(\ref{eq:weak_rate_gen_matel})] instead of $s=1/4$ which should be used when working with the antisymmetrized amplitudes\footnote{This gives $1/2$, and another $1/2$ is needed to account for double counting of collisions when integrating over distributions of initial particles.} \cite{Lykasov2008PhRvC,  Hannestad1998ApJ, Yakovlev2001physrep,Hanhart2001PhLB}. 
The same incorrect factor seems to have been used in Refs.~\cite{Maxwell1987ApJ,Schwenk2004PhLB,VanDalen2003PhRvC}.
The factor $\beta_{nn}$ takes into account all other corrections beyond OPE, see below. Similar expressions can be written down for $np$ and $pp$ bremsstrahlung, resulting in $\epsilon^{pp}_\nu<\epsilon^{np}_\nu<\epsilon^{nn}_\nu$ (see for 
instance \cite{Yakovlev2001physrep}). 
The calculations show that the use of the realistic $T_{NN}$ matrix instead of the OPE leads to the suppression of the neutrino emissivity approximately by a factor of four [so $\alpha^{\rm ex}_{nn}(T_{NN})\approx 0.2$] \cite{Hanhart2001PhLB,Timmermans2002PhRvC,VanDalen2003PhRvC,Li2009PhRvC,LiLiou2015PhRvC}. Note that the inclusion of additional meson exchange terms \cite{FrimanMaxwell1979ApJ,VanDalen2003PhRvC} results in a 
better agreement with $T$-matrix calculations.  \citet{LiLiou2015PhRvC} quantified the limits of applicability for the soft bremsstrahlung theorem for a certain realistic model of the nucleon interactions. They found that the approximation (\ref{eq:MatelSoftnn}) is accurate within 10\% up to $\omega_\nu=60$~MeV. Relativistic corrections for the densities of interest are within 5-15\% \cite{VanDalen2003PhRvC}.

Up to now we have not considered in-medium effects. Since the bremsstrahlung reactions are based on nucleon-nucleon collisions, one deals with similar complications as in Sec.~\ref{S:core_nuc_baryon}, where we discussed transport coefficients. As usual, one medium effect is in the renormalization of the effective masses. Since $m^*$ enters the bremsstrahlung emissivities in the fourth power, bremsstrahlung 
is more sensitive to the effective mass than the direct Urca processes. Apart from the effective masses and possible renormalization of weak coupling constants, correlations  in the medium modify the quasiparticle scattering rates. It was found by \citet{VanDalen2003PhRvC} that the  use of the $G$-matrix of BHF theory (three-body forces not included) instead of the $T$-matrix results in some 30\% increase of the bremsstrahlung rate. This is in contrast to estimates by \citet{Blaschke1995MNRAS}, who found a decrease of the in-medium emissivity by a factor of 10--20. A possible source of this discrepancy may be the omission of the tensor $^3P_2 - {^3F}_2$ coupling in the latter work \cite{VanDalen2003PhRvC}. \citet{Schwenk2004PhLB} used quasiparticle scattering amplitudes which are constructed from renormalization group methods based on the low momentum universal potential $V_{\rm low-k}$ \cite{SchwenkFriman2004PhRvL}. Their results computed to second-order give an overall reduction of the emissivity by a density-dependent factor of $4-10$. This reduction is higher at lower densities and thus more relevant to supernova studies (remember that their values must be divided by two \cite{Lykasov2008PhRvC}). 

More general calculations can be performed in the framework of linear response theory by studying the response of nuclear matter to a weak probe. As it is clear from Eq.~(\ref{eq:MatelSoftnn}) and the discussion above, essential information is contained in the dynamical spin response function $S_\sigma(\omega,q)$, see for example Ref.\ \cite{Lykasov2008PhRvC,Hannestad1998ApJ} and references therein. For  results within Landau's Fermi liquid theory see \cite{Lykasov2008PhRvC} and a general overview of the correlated basis function approach to the weak response is given in Ref.~\cite{BenharLovato2015IJMPE}. In any case, at present the medium modifications of the bremsstrahlung rates deep in the interior of neutron stars are quite uncertain. One can easily imagine a modification  by a factor of two in any direction.

In the medium-modified OPE (MOPE) model  of Refs.~\cite{Migdal1990PhR,Voskresensky2001LNP}, the emissivity receives a strong density-dependent correction due to the softening of the pion mode. This correction results in a factor that can be written as \cite{Kolomeitsev2015PhRvC} 
\begin{equation}\label{eq:beta_Voskr_mope}
\beta^{\rm MOPE}_{nn}=3\left(\frac{n}{n_0}\right)^{4/3} \frac{\left[\Gamma(n)/\Gamma(n_0)\right]^6}{\left(\tilde{\omega}_{\pi}/m_\pi\right)^3},
\end{equation}
where $\Gamma(n)\approx \left[1+1.6(n/n_0)^{1/3}\right]^{-1}$ and  $\tilde{\omega}_\pi$ is the effective pion gap in the medium. The adopted density dependence of the pion gap results in suppression at $n\lesssim n_0$ and in significant enhancement at $n\gtrsim n_0$ up to a factor of $\beta^{\rm MOPE}_{nn}\approx 100$ depending on a model adopted for the pion gap. 

Finally, in the exotic case of proton localization, also discussed in Sec.~\ref{S:core_nuclear_kincoeff}, the neutrino emissivity due to scattering of neutrons off the localized protons was considered in Ref.~\cite{BaikoHaensel1999AcPPB}. Its interesting feature is the $T^6$ temperature dependence of the emissivity, compared to $T^8$ for the ordinary bremsstrahlung processes. This contribution could thus be 
very important, provided that the phenomenon of proton localization is realized in neutron stars. According to the results of Ref.~\cite{BaikoHaensel1999AcPPB}, the ratio $\epsilon^{n-{\rm loc.}p}_\nu/\epsilon^{\rm MU}_\nu$ can be as large as $2\times 10^{3}\ T_8^{-2}$, dominating the neutrino emission in the neutron star core. 

Unfortunately, the situation for the modified Urca reactions (\ref{eq:Core_Murca_forrev}) is even more cumbersome. As stated above, the applicability of soft electroweak theorems is not justified since the energy of the lepton pair is not small, $\omega\sim p_{F\ell}$. Therefore it is not immediately clear that the `leg' diagrams (a) and (b) in Fig.~\ref{fig:NNdiagrams} give the dominant contribution. Moreover, off-shell amplitudes should be used.
In $npe\mu$ matter one considers two branches of the modified Urca processes, namely neutron and proton branches [${\cal C}=n$ and ${\cal C}=p$ in Eq.\ (\ref{eq:Core_Murca_forrev})]. We focus on the neutron branch, for the proton branch see Ref.~\cite{Yakovlev2001physrep}\footnote{Note that the angular factor $\hat{\Omega}$ [see Eq.~(\ref{eq:weak_rate_decomposed})] for the proton branch is slightly incorrect in Ref.~\cite{Yakovlev2001physrep}, see Refs.~\cite{Gusakov2002A&A, Kaminker2016Ap&SS} for details. 
}. In the OPE approximation, the emissivity of the modified Urca process from the leg diagrams is \cite{FrimanMaxwell1979ApJ,Yakovlev2001physrep}
\begin{equation}\label{eq:n_MU_ope}
\epsilon^{\rm MU}_{\nu} = 8.1\times 10^{13} \left(\frac{m_n^*}{m_n}\right)^3 \left(\frac{m_p^*}{m_p}\right)^4 \left(\frac{p_{F\ell}c}{\mu_\ell}\right)\left(\frac{n_p}{n_0}\right)^{1/3} \left(\frac{T}{10^8~K}\right)^8 \alpha^{n\ell}_{\rm MU}~\rm erg~cm^{-3}~s^{-1},
\end{equation}
where $\alpha^{n\ell}_{\rm MU}\approx 1 $ comes from the averaged matrix element \cite{Yakovlev2001physrep}. Comparing with Eq.~(\ref{eq:nn_brems_ope}), one sees that the neutrino emissivity from the modified Urca process is more than 50 times stronger than that of the bremsstrahlung process.
The reaction rate is given by the equation analogous to Eq.~(\ref{eq:GammaMU_phys}), but with different numerical constants in the prefactors
\begin{equation}\label{eq:GammaMU_phys}
\Gamma^{\rm MU} = \frac{0.106}{T} \epsilon^{\rm MU}_{\nu}, \qquad \Delta \Gamma^{\rm MU} = \frac{0.129}{T} \frac {\delta \mu}{T} \epsilon^{\rm MU}_{\nu} \qquad (\mbox{for}\;\delta\mu \ll T) \, .
\end{equation}
The correlation effects considered by \citet{FrimanMaxwell1979ApJ} reduce the rate in Eq.~(\ref{eq:n_MU_ope}) approximately by a factor 1/2. From the above discussion of the $nn$ bremsstrahlung, one expects further reduction of the emissivity when going beyond the OPE approximation towards the full scattering amplitude. Indeed, according to estimates in Ref.~\cite{Blaschke1995MNRAS}, the expected reduction is about $1/4$ with respect to the OPE result and the use of the in-medium $T$-matrix leads to further reduction by an additional factor of $0.6-0.9$.

In the recent work by \citet{Niri2016PhRvC}, the in-medium modified Urca emissivity  was calculated starting from the correlated nucleon pair states (see also Refs.~\cite{SawyerSoni1979APJ,HaenselJerzak1987A&A}). 
The pair correlation functions are determined in the lowest-order constrained variational (LOCV) procedure. 
The LOCV functions turn out to be similar to those obtained in the BHF scheme \cite{BaldoMoshfegh2012PhRvC}. The modified Urca emissivity calculated in this way by construction effectively includes rescattering diagrams of type (c) in Fig.~\ref{fig:NNdiagrams} (as pointed out already in Ref.~\cite{FrimanMaxwell1979ApJ}).   
It was found that the LOCV result at two-body level shows a reduction of the emissivity from the Friman \& Maxwell result. 
The reduction becomes more pronounced with increasing density, reaching a factor of 4 at $n=3n_0$. However, the inclusion of  phenomenological three-body forces (in the Urbana IX model \cite{Carlson1983NuPhA}) eliminates this reduction, and the LOCV result with three-body forces turns out to be  close to that of Ref.~\cite{FrimanMaxwell1979ApJ}. 

One might expect that in the MOPE model of in-medium nuclear interactions \cite{Migdal1990PhR,Voskresensky2001LNP} the correction of the emissivity of the modified Urca process is similar to the bremsstrahlung correction given in Eq.~(\ref{eq:beta_Voskr_mope}). This is not the case. According to the analysis of Ref.\ \cite{Voskresensky2001LNP}, at $n\gtrsim n_0$, diagrams of type (d) dominate, which describe in-medium conversion of a virtual charged pion to a neutral pion with the emission of a real lepton pair. The modification factor for the medium modified Urca reaction with respect to the free one-pion exchange result is
\begin{equation}\label{eq:beta_Urca_Voskr_mope}
\beta_{\rm MMU}=3\left(\frac{n}{n_0}\right)^{10/3} \frac{\left[\Gamma(n)/\Gamma(n_0)\right]^6}{\left(\tilde{\omega}_{\pi}/m_\pi\right)^8} \, .
\end{equation}
Comparing with Eq.~(\ref{eq:beta_Voskr_mope}),  one finds a higher power of the pion gap $\tilde{\omega}_\pi$ in the denominator and a stronger density dependence in the prefactor. With typical parameters, one obtains enhancement by a factor of $\beta_{\rm MMU}\sim 3$ at $n\sim n_0$ and up to $\beta_{\rm MMU}\approx 5\times 10^{3}$ at $n=3n_0$.
Note, however, that this enhancement strongly depends on the uncertain values of the pion gap $\tilde{\omega}_\pi$ and the vertex correction $\Gamma$ entering Eqs.~(\ref{eq:beta_Voskr_mope}) and (\ref{eq:beta_Urca_Voskr_mope}).

The modified Urca process is dominant when the direct Urca process is forbidden. When the density is sufficiently close to (but still below) the direct Urca threshold, one needs to take into account the softening of the nucleon propagation in the virtual lines when examining the `leg' contributions (a) and (b) in Fig.~\ref{fig:NNdiagrams}. This can  increase the modified Urca rates significantly in comparison to the standard result \cite{Shternin2018PLB}. For example, consider diagram (b) for the neutron branch of 
the modified Urca reaction, i.e., $p_2$ corresponds to a neutron line. After emitting the lepton pair with momentum $q\approx p_{F\ell}$ and energy $\omega\approx \mu_\ell + \omega_\nu$, the neutron transforms to a virtual proton with energy $\varepsilon =\mu_n-\omega$ and momentum $\bm{k}=\bm{p}_2-\bm{q}$ well above the Fermi surface ($k>p_{Fp})$. 
The standard practice (e.g., \cite{FrimanMaxwell1979ApJ}) is to set the proton propagator to ${\cal G}^{-1}_0=-\omega\approx -\mu_\ell$ in Eq.~(\ref{eq:nonrel_prop}). However, in the case of backward emission $\bm{q}\uparrow\downarrow \bm{p}_2$ ($k=p_{Fn}-p_{F\ell}$),  the intermediate momentum $k$ can be close to $k_{Fp}$ and in beta-stable matter
 $\varepsilon\approx \mu_p$ (we neglect $\omega_\nu$ here). When $\rho > \rho_{\rm DU}$ this results in a pole on the real-axis (${\cal G}^{-1}=0$ for some values of $\bm{q}$), manifesting opening of the direct Urca process, while for $\rho\to\rho_{\rm DU}$, the intermediate proton line softens  in a certain (backward) part of the phase space. In other words, ${\cal G}^{-1}$ can be much smaller than ${\cal G}_0^{-1}$ when $k\to p_{Fp}$, leading to a strong enhancement of the neutrino emissivity. As a consequence, only the vicinity of $\bm{k}\approx \bm{p}_2-\bm{q}$ is important when calculating the matrix element in (\ref{eq:weak_current_bremms}), i.e., only weakly off-shell values of $T_{NN}$ are needed. In this sense, one reinstalls  the soft bremsstrahlung theorem in a certain way. A crude estimate of the effect of the nucleon softening can be obtained by neglecting all momentum dependence in (\ref{eq:weak_current_bremms}) except for ${\cal G}$. For the contribution of diagram (b) in Fig.~(\ref{fig:NNdiagrams}) one gets \cite{Shternin2018PLB}
\begin{equation}\label{eq:mUrca_soft_bmatel}
\frac{\left\langle \left|M_{fi}^{(b)}\right|^2\right\rangle_\Omega}{\left\langle \left|M_{fi}^{(b)0}\right
|^2\right\rangle_\Omega}\approx \frac{m_p^{*2}\mu_\ell}{2p_{Fp}^2p_{F\ell}\;\delta p}\, , \qquad \delta p \ll p_{Fn}\, ,
\end{equation}
where $M_{fi}^{(b)}$ and $M_{fi}^{(b)0}$ are calculated using ${\cal G}$ and ${\cal G}_0$, respectively, and 
$\delta p = p_{Fn}-p_{Fp}-p_{F\ell}$ measures the distance from the direct Urca threshold in terms of momenta. A slightly different result is found for the (a) diagram\footnote{One should substitute $m_p^*$ by $m_n^*$ and one of the factors $p_{Fp}$ in the denominator by $p_{Fn}$ in Eq.~(\ref{eq:mUrca_soft_bmatel}).} but the $\delta p^{-1}$ asymptotic behavior is the same. The exchange contributions somewhat complicate this picture, however they are of next order in $\delta p$. The correction (\ref{eq:mUrca_soft_bmatel})  to the modified Urca rates leads to a more pronounced density dependence  of $Q^{\rm MU}$ than given in Eq.~(\ref{eq:n_MU_ope}) and a significant rate enhancement at $\rho\to \rho_{\rm DU}$. Moreover, calculations show that the rates are enhanced by a factor of several for all relevant densities in neutron star cores (even far from $\rho_{\rm DU}$). Notice that this result is universal in a sense that it does not depend on the particular model employed for the strong interaction. A more  detailed study of this effect would be interesting.

Let us note that the softening of the intermediate nucleon, which results in the enhancement of the modified Urca rate given by Eq.~(\ref{eq:mUrca_soft_bmatel}), has similarity with the MOPE result (\ref{eq:beta_Urca_Voskr_mope}) where the enhancement is due to softening of the intermediate pion. We note, however, that the dominance of the diagram (d) contribution over diagrams (a)-(c) in MOPE calculations was found without taking into account enhancement of the latter by the effects described above. This can alter the MOPE result\footnote{Notice that it is not sufficient simply to compare Eqs.~(\ref{eq:beta_Urca_Voskr_mope}) and (\ref{eq:mUrca_soft_bmatel}), since the `leg' diagrams (a)-(b) also possess MOPE enhancement, c.f. Eq.~(\ref{eq:beta_Voskr_mope}).}.

At high temperatures, one needs to go beyond the quasiparticle approximation and  take into account coherence effects such as the Landau-Pomeranchuk-Migdal (LPM) effect. 
When the quasiparticle lifetime becomes small [the spectral width $\gamma(\omega)$ of the quasiparticle becomes large], it undergoes multiple scattering during the formation time of the radiation. In this case, the picture of well-defined quasiparticles fails and the nuclear medium basically plays the role of the spectator in reactions (\ref{eq:Core_Murca_Bremms})--(\ref{eq:Core_Murca_forrev}), such that the process (\ref{eq:Core_weak_neutral}) essentially becomes allowed. Calculations of the reaction rates and emissivities become more involved \cite{KnollVoskresensky1995PhLB}. The finite spectral width in the nucleon propagators regularize the infrared divergence (\ref{eq:MatelSoftnn}), leading to the LPM suppression of the reaction rates. This becomes important when $\omega\sim T\lesssim \gamma(\omega)$. According to various calculations, the threshold temperature is rather large, $T\gtrsim 5$~MeV \cite{VanDalen2003PhRvC,Sedrakian1999PhLB,Lykasov2008PhRvC,Shen2013PhRvC}. 

\subsubsection{Neutrino emission in the presence of Cooper pairing of nucleons} \label{sec:nu_Cooper}

The onset of the pairing instability has a strong effect on the reaction rates and the neutrino emission, as already mentioned in Sec.~\ref{S:cooper}. As discussed in Sec.~\ref{S:core_nuc_pairing}, the presence of the gap in the quasiparticle spectrum reduces the available phase space for the reaction to proceed and the reaction rates become strongly suppressed (in the close-to-equilibrium situation). In addition, the number of quasiparticles is not conserved now, which opens new reaction channels, namely those corresponding to Cooper pair breaking and formation. 

The modifications due to Cooper pairing are usually described by superfluid `reduction factors', 
\begin{subequations}\label{eq:QdG_sf}
\begin{eqnarray}
\epsilon_{\nu}^{\rm SF}&=&\epsilon_{\nu}^{N} R_\epsilon\left(\left\{\frac{\Delta_i}{T}\right\},\frac{\delta\mu}{T}\right), \label{eq:Q_sf}\\[2ex]
\Delta\Gamma^{\rm SF}&=&\Delta\Gamma^{N} R_\Gamma\left(\left\{\frac{\Delta_i}{T}\right\},\frac{\delta\mu}{T}\right),\label{eq:dR_sf}
\end{eqnarray}
\end{subequations}
where SF refers to superfluid and $N$ to non-superfluid. The factors $R_\epsilon$ and $R_\Gamma$ describe the superfluid modifications of the total emissivity and the equilibration rate (for the composition-changing reactions), respectively, and depend on the superfluid gaps $\Delta_i$, where $i$ labels the superfluid species\footnote{For simplicity, here we 
only use the term superfluidity, including proton Cooper pairing. The distinction between 
superfluidity and superconductivity is not important in the present context.}, and on the chemical potential imbalance $\delta\mu$. Calculating these factors accurately 
is a complicated task. Effects of superfluidity enter the original expressions for the rate and the emissivity (\ref{eq:weak_rate_gen_matel}) through the superfluid quasiparticle distribution functions in (\ref{eq:Fermi_blocking_product}) and the energy spectra in the delta-function (\ref{eq:encons_delta}). They also affect the matrix element $M_{fi}$, allowing for the quasiparticle number non-conservation, and, moreover, the weak interaction vertices can be affected by the response of the condensate [this has crucial consequences when the emission due to Cooper pairing (\ref{eq:Core_weak_neutral}) is considered, see below]. 
All effects of pairing can be taken into account consistently by starting from the full propagators in the so-called Nambu-Gorkov space. We shall briefly discuss this approach in Sec.~\ref{sec:quarkneutrino} for the direct Urca process in color-superconducting quark matter, see Eq.~(\ref{df1}) and discussion below that equation. In almost all calculations of the reduction factors $R_{\epsilon/\Gamma}$ we are aware of for nuclear matter, the modifications  of the reaction cross-sections are not considered, and only the phase-space modifications are included.
For the direct Urca processes (\ref{eq:Core_Urca_forrev}) this approach is well-justified, see section 4.3.1 in Ref.~\cite{Yakovlev2001physrep}.
The relative contribution of the number-conserving channels of the reaction and channels which include breaking and formation of Cooper pairs are considered in Refs.~\cite{Sedrakian2005PhLB,Sedrakian2007PrPNP}. At high temperatures, the scattering contribution dominates, while at $T\to 0$ its contribution decreases to one half of the total rate. In practice, there is no need to consider these contributions separately and one can use $M_{\rm DU}$ (\ref{eq:DU_matel})  without superfluid modifications \cite{Yakovlev2001physrep}. The effects of the superfluid coherence factors on the matrix element of the electroweak bremsstrahlung reactions have, to our knowledge, not been explored. Therefore, in the following, we briefly discuss the results for the superfluid reduction factor obtained without superfluid effects on $M_{fi}$. Such factors are universal since they do not depend on the details of the strong interaction and are assumed to reflect the main properties of the correct results.

Let us first consider beta-equilibrated matter, $\delta \mu =0$ \cite{Yakovlev1999PhyU, Yakovlev2001physrep,Gusakov2002A&A}.  Even with the above simplifications, the calculation of $R_{\epsilon/\Gamma}$ require laborious efforts because it has to account for the possible coexistence of proton pairing and (anisotropic) neutron pairing. Recall
that protons are assumed to pair in the isotropic $^1S_0$ state, while neutrons in the neutron star core are paired in the $^3P_2(m_J=0,\ \pm 1,\ \pm2)$ channel with a possible admixture from the $^3F_2$ channel. Only the cases $m_J=0$ and $|m_J|=2$ are considered in detail since they do not include integration over the azimuthal angle of the quasiparticle momentum about the quantization axis. The angular integration over the polar angle must be carried out, which precludes using the angular-energy decomposition in the form of Eq.~(\ref{eq:weak_rate_decomposed}).\footnote{In the case of  bremsstrahlung or modified Urca, which can include two neutrons with anisotropic pairing, the matrix element in principle cannot be taken out of the integration since it depends on the relative orientation of the scattered particles even without superfluid modifications. This is always neglected. Note, however, that in this case the region of the momenta orientation that imply lowest gaps will be extracted from $M_{fi}$. One can expect considerable modifications if the angular dependence of $M_{fi}$ is not flat (this is the case for $np$ scattering, which contributes to modified Urca and $np$ bremsstrahlung rates).} At very low temperatures, $T\ll T_c$, where $T_c$ is the 
critical temperature for pairing, participation in the reaction of a paired fermion species in the $^1S_0$ or $^3P_2(m_J=0)$ channels leads to an exponential suppression of the rates. The case $|m_J|=2$ is qualitatively different since the gap contains nodes on the Fermi surface. Then the suppression is given by a power law in $T/T_c$. At intermediate temperatures, most interesting in practice, the suppression factors show an approximate power-law dependence for any superfluidity type, even for a fully gapped spectrum. Numerical results and fitting expressions for direct Urca, modified Urca, and bremsstrahlung reactions for various combinations of pairing types, isotropic and anisotropic, can be found in Ref.~\cite{Gusakov2002A&A}, which also contains a review of other works. 

The beta-equilibrium conditions can be perturbed by various processes, for instance by compression. In the superfluid case, since the reaction rates are suppressed, the system cannot counterbalance a growing perturbation $\delta\mu$. If $\delta\mu$ becomes larger than the pairing gap $\Delta$ 
the reactions become unblocked. If the direct Urca process is allowed by momentum conservation, the threshold value is $\delta\mu_{\rm th}=\Delta_n+\Delta_p$.  Otherwise, if the modified Urca process is responsible for beta-equilibration, $\delta\mu_{\rm th}=\Delta_n+\Delta_p+2\min (\Delta_n,\, \Delta_p)$ \cite{Reisenegger1997ApJ}. When $\delta \mu >\delta\mu_{\rm th}$, the beta-equilibration reaction which decreases $\delta\mu$ is allowed and is no longer suppressed by the presence of gaps. The value of $\delta\mu$ determines the phase space for the reaction, like in case of normal matter in the supra-thermal regime. \citet{Reisenegger1997ApJ} first suggested this effect\footnote{This idea was re-discovered recently under the name of `gap-bridging' in Refs.~\cite{AlfordReddy2012PhRvL,AlfordPangeni2017PhRvC}.} and qualitatively described it by introducing step-like suppression factors $R_{\epsilon,\Gamma}=\Theta(\delta\mu-\delta\mu_{\rm th})$ [it is understood that the `$N$' quantities in Eqs.~(\ref{eq:QdG_sf}) are calculated including $\delta\mu$]. Later these results were improved in Refs.~\cite{VillainHaensel2005A&A,Pi2010PhRvC,PetrovichReisenegger2010,Gonzalez2015MNRAS}, where discussions of the behavior of the $R$-factors and complications of the numerical scheme can be found. In the recent Ref.~\cite{Gonzalez2015MNRAS}, the most general case of anisotropic pairing in considered, but unfortunately no analytical approximation for the reduction factors are given. It would be nice (but not easy) to obtain approximations similar to those presented in Ref.~\cite{Gusakov2002A&A},  but for the non-equilibrium case. In fact, according to Eq.~(\ref{eq:Reiss_formula}), it is enough to find one of the factors $R_\epsilon$ or $R_\Gamma$ \cite{FloresReisenegger2006MNRAS}.

Now let us turn to the neutral weak current emission associated with the Cooper pair breaking and formation (CPF) processes in the reaction given in 
Eq.\ (\ref{eq:Core_weak_neutral}). These processes were already mentioned in Sec.~\ref{S:cooper}. 
The process (\ref{eq:Core_weak_neutral}) is a first-order process in the number of quasiparticles and therefore does not explicitly depend on the strong interaction details (although strong interactions determine, for instance, the value of the gap). This process is kinematically forbidden in the normal matter but becomes allowed if the nucleons pair. It was proposed by \citet{FlowersRuderman1976ApJ} and later rediscovered by \citet{VoskresenskySenatorov1987SJNP} 
The expression for the emissivity  can be written as \cite{Yakovlev2001physrep}
\begin{equation}\label{eq:PBF_gen}
\epsilon^{\rm CP}_\nu = 1.17\times 10^{14} \left(\frac{m_N^*}{m_N}\right)\left(\frac{p_{FN}}{m_N c}\right) \left(\frac{T}{10^8~{\rm K}}\right)^7\alpha^{\rm CP} F(\Delta_N/T) {\cal N}_\nu\ {\rm erg~cm^{-3}~s^{-1}} \, ,
\end{equation}
where, as usual, $\alpha^{\rm CP}=\alpha^{\rm CP}_V+\alpha^{\rm CP}_A$ is a dimensionless number that arises from the matrix element of the process containing vector $\alpha^{\rm CP}_V$ and axial-vector $\alpha^{\rm CP}_A$ contributions and $F(\Delta_N/T)$ comes from the energy integration (and angular integration in the anisotropic case).
Near the critical temperature $T\to T_{cN}$, the function $F$ approaches zero linearly, $F\propto (1-T/T_{cN})$, and for low temperatures $F$
behaves qualitatively like the reduction factors $R_{\epsilon/\Gamma}$, i.e., the emissivity is exponentially suppressed, unless there are nodes of the gap on the Fermi surface, in which case $F$ behaves according to a power-law in temperature \cite{Yakovlev2001physrep}. Thus, at low temperatures, the CPF emission is strongly suppressed. However, the function $F$ has a maximum at  $T\sim 0.8 \,T_{cN}$ and in the vicinity of this temperature the CPF emission can be the dominant neutrino emission mechanism in the superfluid neutron star core. Therefore, the CPF process is an important ingredient in the so-called minimal cooling scenario of the thermal evolution of isolated neutron stars \cite{Page2004ApJS,Gusakov2004A&A, Page2009ApJ}. 

During the last decade, significant improvements in CPF emission studies were made. 
Crucially, one has to take into account consistently the response of the condensate (collective modes), which enters the emissivity through the anomalous part of the weak vertices. This is achieved by a proper renormalization of the vertices, which ensures vector current conservation \cite{KunduReddy2004PhRvC}. As a consequence, the CPF emission for singlet-paired matter is strongly suppressed in the non-relativistic limit, as pointed out by  \citet{LeinsonPerez2006PhLB}, and later elaborated in Refs.~\cite{Sedrakian2007PhRvC, Leinson2008PhRvC, Kolomeitsev2008PhRvC, SteinerReddy2009PhRvC, Leinson2009PhRvC,KolomeitsevVoskresensky2010PhRvC,Kolomeitsev2011PAN,Sedrakian2012PhRvC}. Although some controversy about the results from different approaches still exists, the main conclusion is that $\alpha_V^{\rm CP}\propto g_V^2 v_{FN}^4$, being small in the non-relativistic limit. Recall that the singlet nucleon pairing is present in the low-density, hence non-relativistic ($v_{FN}\ll 1$), domain. The axial-vector contribution does not receive any vertex correction because there is no spin response from the condensate in the case of singlet pairing. However, this contribution itself is a relativistic effect, $\alpha^{\rm CP}\approx\alpha_A^{\rm CP}=\tilde{\alpha}_A^{\rm CP} g_A^2 v_{FN}^2$, 
where $\tilde{\alpha}_A^{\rm CP}$ is a numerical factor of order unity whose precise value is subject to debates. The final conclusion is that the CPF emission from the singlet pairing is not important, being much smaller than the bremsstrahlung contribution even when the latter is suppressed by the superfluid reduction factor $R_\epsilon$.

The situation is different for the case of triplet pairing of neutrons, which is thought to occur in a large fraction of the neutron star core. Without taking into account the condensate response, the CPF emission from the triplet superfluid is given by Eq.~(\ref{eq:PBF_gen}) with $\alpha^{\rm CP}=g_V^2+2g_A^2\approx 4.17$ \cite{Yakovlev2001physrep}. Up to very high densities, where the triplet pairing is usually found to vanish, the neutrons can still be considered non-relativistic. By  analogy with the case of singlet pairing, one expects that the vector current conservation would suppress the vector contribution to the emissivity also in the triplet case. This leads to the approximation  $\alpha^{\rm CP}=2g_A^2$, by a factor of about $0.78$ less than the initial result \cite{Page2009ApJ}. However, in contrast to the singlet case, the order parameter of the triplet superfluid varies under the action of the axial-vector field \cite{Leinson2010PhRvC}. As a consequence, this modifies the axial-vector contribution to the emissivity. This was considered  by \citet{Leinson2010PhRvC}, using an angular-averaged gap as an approximation. He indeed found the suppression of the vector contribution in the non-relativistic limit, while for the axial-vector contribution the result is $
\alpha^{\rm CP}\approx\alpha_A^{\rm CP}=\frac{1}{2}g_A^2\approx 0.8$. 
Thus taking into account the condensate response reduces the emissivity by a factor of 0.19 compared to Ref.~\cite{Yakovlev2001physrep}. This quenching has observational consequences if the real-time thermal evolution of the superfluid neutron star can be observed, see for instance Ref.~\cite{ShterninYakovlev2015MNRAS}. The actual angular dependence of the gap and Fermi-liquid effects modify this result only slightly (within 10\% according to Ref.~\cite{Leinson2013PhRvC}). Thus, even with the more elaborate treatment of the superfluid response to weak perturbations, the neutrino emissivity due to CPF processes from triplet neutron superfluidity can be the dominant neutrino emission mechanism at $T\sim 0.8 \, T_{cn}$.

Collective modes in the superfluid (see Sec.~\ref{S:core_nuc_pairing}) can also contribute to the neutrino emissivity. The emission related to the collisions of the Goldstone modes -- angulons -- in the triplet superfluid was considered in Ref.~\cite{Bedaque2003PhRvC} and found to be always negligible. 
However, \citet{Bedaque2014PhRvC} recently considered the case 
of a strong magnetic field, to which the neutron fluid couples through the neutron magnetic moment. Since the magnetic field breaks rotational symmetry explicitly, one of the angulon modes acquires a gap of the order of $eB/(m^*_nc)$ and its decay to a neutrino pair becomes kinematically allowed. The resulting neutrino emissivity can be written in a form similar to Eq.~(\ref{eq:PBF_gen}), where the function $F(\Delta_N/T)$ is replaced by the $B$-dependent function $h(g_nB/(aT))$, where $g_n$ is the neutron magnetic moment, and $a=4.81$.  This function $h(x)$ peaks at $x\sim 7$ and is exponentially suppressed at large $x$ (small $T$).  According to the numerical estimates in Ref.~\cite{Bedaque2014PhRvC}, the neutrino emissivity due to the `magnetized angulon' decay can be larger than that of the CPF 
process at $T\approx 10^7$~K provided the interior magnetic field is as large as $B\sim 10^{15}$~G (the situation where the magnetic field is  confined in flux tubes of the proton superconductor is also discussed).

\subsubsection{Electromagnetic bremsstrahlung}\label{Sec:ee_brems}

The preceding sections do not contain all neutrino emission processes in the cores of neutron stars. Other possibly 
relevant processes are discussed in Ref.\ \cite{Yakovlev2001physrep}, see also Ref.\ \cite{Potekhin:2015qsa}. 
Here we briefly discuss new results for the electromagnetic bremsstrahlung emission, obtained after, and thus not included in, Ref.\ \cite{Yakovlev2001physrep}.
The emission from the electromagnetic bremsstrahlung
\begin{equation}
\ell+{\cal C}\to \ell + {\cal C} +\nu +\bar{\nu} \, ,
\end{equation}
where $\ell=e,\, \mu$ is a lepton and ${\cal C}$ is some electrically charged   particle, is thought to be several orders of magnitude smaller than those from collisions mediated by strong interactions \cite{Yakovlev2001physrep}. Still, the lepton-lepton bremsstrahlung may be the dominant process for low-temperature superfluid matter (with both neutrons and protons in the paired state), where the neutrino emission processes involving baryons are suppressed. The studies reviewed in  \cite{Yakovlev2001physrep} underestimated the significance of the bremsstrahlung in electromagnetic collisions. The reason is the same as discussed in Sec.~\ref{S:core_nuc_lepton} -- correctly taking into account screening of the transverse part of the interaction makes these collisions much more efficient. The proper transverse screening was considered for the electron-electron bremsstrahlung in Ref.~\cite{Jaikumar2005PhRvD}, with the result 
\begin{equation}\label{eq:ee_Br_Jaikumar}
Q^{ee}=1.7\times 10^{12} \left(\frac{T}{10^8~{\rm K}}\right)^7 \left(\frac{n_{e}}{n_0}\right)^{2/3} \tilde{N}\left(\frac{m_{D}c^2}{2  T}\right){\cal N}_\nu~{\rm erg}~{\rm cm}^{-3}~{\rm s}^{-1},
\end{equation}
where $m_{D}$ is the electric (Debye) screening mass (corresponding to $\hbar q_l/c$ in the notation of Sec.~\ref{S:core_nuc_lepton}) and $\tilde{N}\leqslant 1 $ is a slowly varying function, approaching unity in the strongly degenerate limit.
Like in the case of the thermal conductivity  (Sec.~\ref{S:core_nuc_lepton}), dynamical screening borrows one power of the temperature from the expression for emissivity, so it behaves like $T^7$ instead of $T^8$ for standard bremsstrahlung reactions (here we neglect the temperature dependence of $\tilde{N}$, which becomes important if the temperature approaches the plasma temperature). According to  Eq.~(\ref{eq:ee_Br_Jaikumar}), the emissivity in $ee$ collisions becomes increasingly important with lowering the temperature and was underestimated by more than five orders of magnitude before that work. The bremsstrahlung emission from electron-proton (or other charged baryons) collisions should obey a similar enhancement, although the transverse channel is suppressed by the relativistic factor $v_{Fp}^2$ (see Sec.~\ref{S:core_nuc_lepton}). 

The domain of immediate importance of Eq.~(\ref{eq:ee_Br_Jaikumar}) is in the possible region of the inner core where the singlet proton pairing is absent, but the neutron triplet pairing exists. Then the neutrino emission due to the process in question can compete with the proton-proton bremsstrahlung due to strong forces. In the case of  proton pairing, the expression (\ref{eq:ee_Br_Jaikumar}) is expected to modify in the same way as the transport coefficients discussed in Sec.~\ref{S:core_nuc_pairing}. Detailed studies of these effects  for  realistic conditions in neutron star cores are desirable but not performed yet.

\subsection{Bulk viscosity}\label{S:bulk}

As we can see from Eq.\ (\ref{dotS}), the bulk viscosity $\zeta$ is responsible 
for dissipation in the presence of a nonzero divergence $\nabla\cdot {\bm v}$. Via the continuity equation (\ref{contin}), this divergence is identical to 
compression and expansion of a fluid element. In a neutron star, certain 
oscillations lead to local, periodic compression and expansion. Therefore, bulk viscosity is an important transport property of the matter inside the star if we are 
interested in the damping of these oscillations. The dominant contribution to bulk viscosity is given by electroweak reactions because their time scale becomes comparable
to the period of the oscillations of the star, which are typically of the order of the rotation period. Since rotation periods are of the order of a millisecond or larger, re-equilibration processes from the strong interaction play no role for bulk viscosity. The `resonance' between the weak interaction and the oscillation frequency occurs in a certain temperature 
regime, usually for relatively high temperatures of about 1 MeV or higher. Bulk viscosity is thus particularly important for young neutron stars or in neutron star mergers. 

To explain the interplay between the reaction rates of the weak processes and an externally given volume oscillation, let us briefly review how the bulk viscosity of dense 
hadronic matter is computed \cite{Sawyer:1989dp,Haensel:2000vz}. 
We denote the angular frequency of the volume oscillation by $\omega$, such that we can write the volume as $V(t) = V_0[1+\delta v(t)]$, 
with a (dimensionless) volume perturbation $\delta v(t) = \delta v_0\cos\omega t \ll 1$. Then, on the one hand, we can write the dissipated energy density,
averaged over one oscillation period $\tau=2\pi/\omega$, as
\be \label{Epsdot}
\langle\dot{{\cal E}}\rangle_\tau = -\frac{\zeta}{\tau}\int_0^\tau dt \, (\nabla\cdot{\bm v})^2 \approx  -\frac{\zeta \omega^2\delta v_0^2}{2} \, ,
\ee
where we have used the continuity equation (at zero velocity ${\bm v}=0$) to relate the divergence of the velocity field to the change in the total 
particle number density, which, in turn, is directly related to the change in volume if the total particle number is fixed. On the other hand, the dissipated energy density
can be expressed in terms of the mechanical work done by the induced pressure oscillations,  
\be \label{Epsdot1}
\langle\dot{{\cal E}}\rangle_\tau = \frac{1}{\tau}\int_0^\tau dt \, P(t)\frac{d\delta v}{d t} \, ,
\ee
where the pressure is
\be \label{pt}
P(t) = P_0 + \frac{\partial P}{\partial V}V_0\delta v + \sum_{x=n,p,e} \frac{\partial P}{\partial n_x}\delta n_x \, .
\ee
The oscillation in the pressure is in general out of phase compared to the volume oscillation because of the microscopic re-equilibration processes which induce changes 
in the number densities of the particle species $\delta n_x$. For this derivation, we consider the simplest form of hadronic matter, made of neutrons, protons and electrons. We 
discuss extensions to more complicated forms of matter and their bulk viscosity below.  

From Eqs.\ (\ref{Epsdot}) and (\ref{Epsdot1}) we compute the bulk viscosity. 
Let us for now assume the electroweak re-equilibration process is the direct Urca process, given by  
\bea \label{dUbulk}
p+e \to n + \nu_e \, , \qquad n &\to& p + e +\bar{\nu}_e  \, . 
\eea
In chemical equilibrium, the reactions (\ref{dUbulk}) do not change the various densities because they occur with the same rate, and the sum of the chemical potentials of the 
ingoing particles is the same as the sum of the chemical potentials of the outgoing particles, $\delta\mu\equiv \mu_p+\mu_e-\mu_n=0$. We assume that 
neutrinos and anti-neutrinos leave the system once they are produced. They can thus only be outgoing particles and we set their chemical potential to zero, $\mu_{\nu}\approx 0$. This assumption has to be dropped 
for very young (proto-)neutron stars where the temperature is large and the mean free path of neutrinos becomes much smaller than the size of the star. 
Then, neutrino absorption processes need to be taken into account in the calculation of the bulk viscosity, as discussed by \citet{Lai:2001jt}. A non-equilibrium situation occurs if the equality of 
chemical potentials is disrupted, $\delta\mu\neq 0$. Such a disruption can be induced by the volume oscillation if the various particle species respond differently to 
compression and expansion. 
The situation considered here is particularly simple because there is a single process and a single $\delta\mu$. In general, there can be multiple processes related to the 
same $\delta\mu$, for instance if we include modified Urca processes (whose contribution, if the more efficient direct Urca process is allowed, can be neglected).
A more complicated situation occurs if multiple processes are related to multiple $\delta\mu$'s, for instance if we include 
strangeness in the form of hyperons. We shall sketch the derivation of the bulk viscosity for such a case in the context of quark matter, see Sec.\ \ref{sec:bulkunpaired}. 
Here we proceed with the single process (\ref{dUbulk}). In this case, the changes in the densities are all locked together, 
\be\label{dnnep}
\frac{dn_n}{dt} =-\frac{dn_e}{dt} =-\frac{dn_p}{dt} =  \Gamma[\delta\mu(t)] \approx \lambda\,\delta\mu(t) \, , 
\ee
where $\Gamma[\delta\mu(t)]$ is the number of neutrons produced per unit time and volume in the process  $p+e \to n + \nu_e$. Using the general terminology employed at the end of Sec.\ \ref{S:core_nuc_mag}, the stoichiometric coefficients of the reaction $p+e\to n+\nu_e$ are $-1$ for $n$ and $+1$ for $e$ and $p$, counting how many particles of a given species are created and annihilated in the given process. These numbers (in this case simply plus or minus signs) appear in Eq.\ (\ref{dnnep}). On the right-hand side of Eq.\ (\ref{dnnep}) we have applied 
the linear approximation for small $\delta \mu$, such that now all information about the reaction rate is included in $\lambda$, using the same notation as in Eq.\ (\ref{eq:entr_react}).  According to our definition of $\delta\mu$, a net production 
of neutrons sets in for $\delta\mu>0$, from which we conclude that $\lambda >0$.  The difference in chemical potentials $\delta\mu$ oscillates due to the volume oscillation and due to the weak reactions,
\bea \label{dmu0}
\frac{d\delta\mu}{dt} &=& \frac{\partial\delta\mu}{\partial V}\frac{dV}{dt} + \sum_{x=n,p,e} \frac{\partial \delta\mu }{\partial n_x}\frac{dn_x}{dt} \non[2ex]
&=& - B \frac{d\delta v}{dt}  - \lambda C  \delta\mu(t) \, , 
\eea
where we have used Eq.\ (\ref{dnnep}) and abbreviated
\be
B \equiv \frac{\partial P}{\partial n_p} + \frac{\partial P}{\partial n_e} - \frac{\partial P}{\partial n_n} \, , \qquad C\equiv \frac{\partial \delta\mu }{\partial n_p}
+\frac{\partial \delta\mu }{\partial n_e}-\frac{\partial \delta\mu}{\partial n_n} \, .
\ee
These quantities are evaluated in equilibrium, i.e., they only depend on the equation of state, not on the electroweak reaction rate. We can also express the pressure (\ref{pt}) 
with the help of $B$,
\be \label{ptB}
P(t) = P_0 + \frac{\partial P}{\partial V}V_0\delta v + B \delta n_e \, .
\ee
In general, $\delta\mu(t)$ oscillates out of phase with the volume $\delta v(t)$, and we make the ansatz $\delta\mu(t) = {\rm Re}[\delta\mu_{0}e^{i\omega t}]$, with 
the complex amplitude $\delta\mu_{0}$. The differential equation (\ref{dmu0}) then yields algebraic equations for real and imaginary parts of $\delta\mu_{0}$, which can easily be solved. We compute $\delta n_e$ by integrating Eq.\ (\ref{dnnep}), then insert the result into the pressure (\ref{ptB}) and the result into the expression for the bulk viscosity, which 
is obtained from Eqs.\ (\ref{Epsdot}) and (\ref{Epsdot1}). This yields 
\be\label{zetanep}
\zeta = -\frac{\lambda B\,{\rm Re}(\delta\mu_0)}{\omega^2\delta v_0} = \frac{\lambda B^2}{(\lambda C)^2+\omega^2} \, . 
\ee
This is the basic form of the bulk viscosity of nuclear matter, as a function of the 
thermodynamic quantities $B$ and $C$, the reaction rate $\lambda$, and the external angular frequency $\omega$. It shows that bulk viscosity is a resonance phenomenon: the viscosity is maximal when the time scales set by the external oscillation frequency and the microscopic
reaction rate match. Since the microscopic reaction rate typically increases with temperature $T$, the bulk viscosity as a function of $T$ at fixed $\omega$ is a function 
with a maximum at $T$ given by $C\lambda(T)=\omega$. It is now obvious that the strong processes, which are responsible for {\it thermal} equilibrium, do not 
contribute to the bulk viscosity because they operate on much shorter time scales. Bulk viscosity in a neutron star is utterly dominated by weak processes, which 
are responsible for {\it chemical} re-equilibration. It has been pointed out by \citet{Alford:2006gy} that the dissipation due to the out-of-phase oscillations of volume
(externally given) and chemical potentials (response of the system) is completely analogous to an electric circuit with alternating voltage (externally given) and 
electric current (response of the system). In this analogy, which is mathematically exact and physically very plausible, the analogue of the resistance is $B^{-1}$ and the analogue of the capacitance 
is $B/(C\lambda)$,  while the inductance is zero. 

\begin{figure}[t]
\begin{center}
\includegraphics[width=0.5\textwidth]{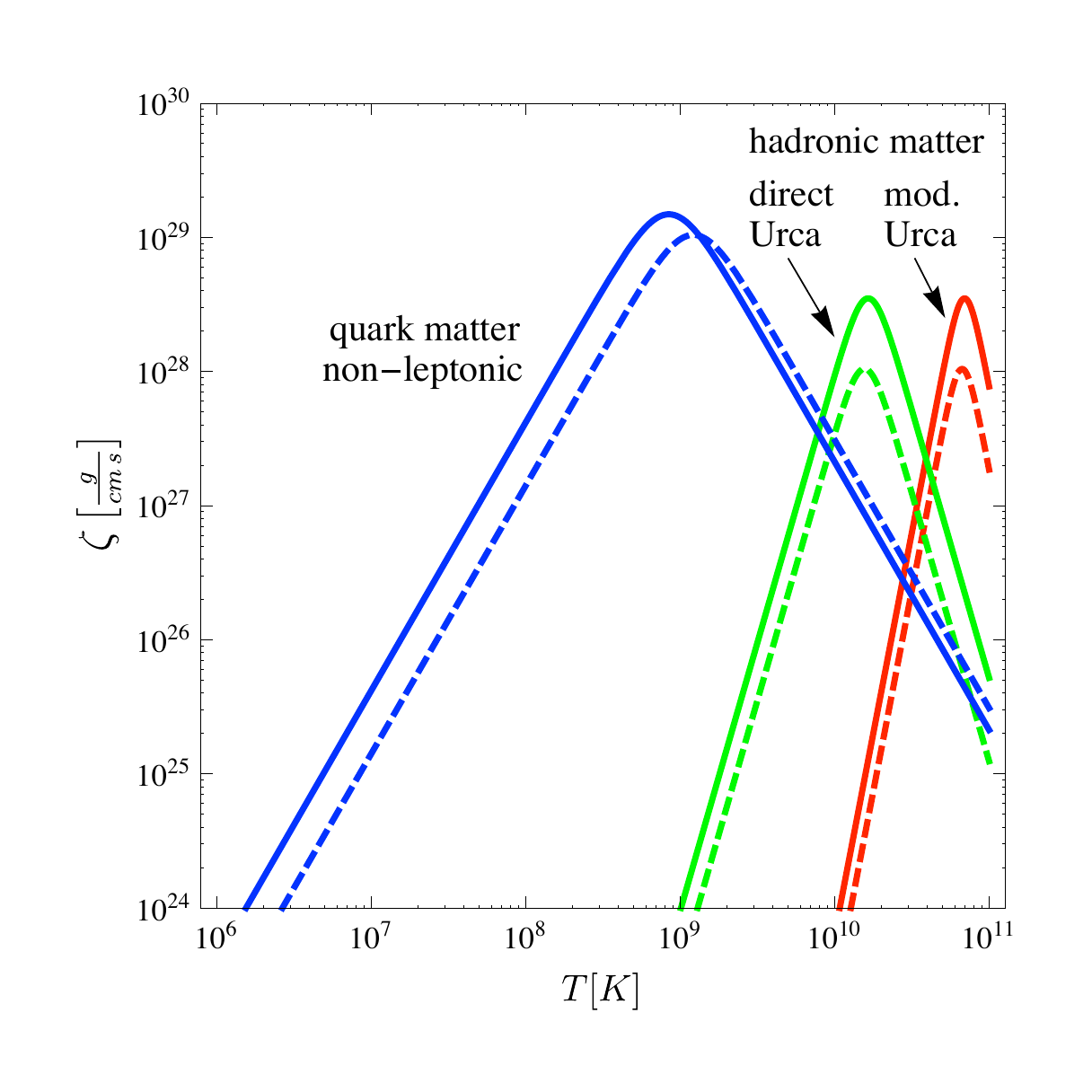}
\caption{Bulk viscosity of unpaired nuclear matter from Urca processes with angular frequency $\omega=8.4\,{\rm kHz}$ and baryon density $n=2n_0$, taken from 
Ref.\ \cite{Alford:2010gw}. If the direct Urca process is allowed, the reaction rate is faster and the maximum of the bulk viscosity is at a lower temperature. The dashed lines are obtained by using 
non-interacting matter for the susceptibilities, while the solid curves use the equation of state of \citet{Akmal:1998cf}. The plot also shows the result for unpaired, strange quark matter from the non-leptonic process
$u+d\leftrightarrow u+s$, to be discussed in Sec.\ \ref{sec:bulkunpaired}, see also Fig.\ \ref{fig:bulkquarks}. Again, the dashed curve represents non-interacting quark matter, while the solid curve includes effects of the interaction on the thermodynamics. }
\label{fig:bulkhadron}
\end{center}
\end{figure}

We show the bulk viscosity of hadronic matter in Fig.\ \ref{fig:bulkhadron}. If the direct Urca process is allowed, the conversion of neutrons into protons and vice versa is faster
and thus the maximum of the bulk viscosity occurs at a smaller temperature compared to the case where only the modified Urca process is at work. 
Since the strong interaction is needed for the modified Urca process, the corresponding rates are prone to large uncertainties, as discussed in the previous section. The bulk viscosity  
due to the modified Urca process used by \citet{Alford:2010gw}, from which Fig.\ \ref{fig:bulkhadron} is taken, is based on free one-pion exchange  interaction 
\cite{FrimanMaxwell1979ApJ,1992A&A...262..131H,Reisenegger1995ApJ}.
\citet{Kolomeitsev2015PhRvC} showed that medium modifications in the MOPE model can enhance the rate of the modified Urca process  and thus may shift the maximum of the bulk viscosity to lower temperatures. 
Fig.\ \ref{fig:bulkhadron}
also includes a comparison with quark matter, whose bulk viscosity we discuss in Sec.\ \ref{sec:bulkquark}. We see that the bulk viscosity peaks at even lower temperatures than 
that of hadronic matter with the direct Urca process. The reason is that in (unpaired) quark matter the more efficient non-leptonic, strangeness-changing, process $u+d\leftrightarrow u+s$ is the dominant chemical re-equilibration process. The figure also shows that the equation of state, through the susceptibilities $B$ and $C$, has a sizable effect on the bulk viscosity. This 
effect has also been studied by \citet{Vidana:2012ex}, with an emphasis on the role of the symmetry energy for the bulk viscosity.

The bulk viscosity also receives contribution from muons. Muons appear in the direct (or modified) Urca processes 
(\ref{dUbulk}) with electron and electron neutrino replaced by muon and muon neutrino \cite{Haensel:2000vz,Haensel:2001mw}. 
One can also consider the purely leptonic processes 
that convert an electron into a muon and vice versa, 
\be
e+e\leftrightarrow\mu+e+\nu+\bar{\nu} \, , \qquad e+\mu\leftrightarrow\mu+\mu+\nu+\bar{\nu} \, .
\ee
These processes are the dominant contribution to the bulk viscosity for temperatures well below the critical temperature for hadronic superfluidity \cite{Alford:2010jf}.

As the result for quark matter in Fig.\ \ref{fig:bulkhadron} suggests, the presence of strangeness has a significant effect on the bulk viscosity. The reason is that 
the phase space for a non-leptonic (strangeness-changing) process is typically much larger than that for a semi-leptonic process because the leptons have a negligibly small 
Fermi momentum. In hadronic matter, the presence of hyperons thus leads to a very different result for the bulk viscosity, with a maximum typically at smaller 
temperatures that for ordinary nuclear 
matter. The bulk viscosity based on the strangeness-changing processes
\begin{subequations}
\bea
n+n&\leftrightarrow& p+\Sigma^- \, , \\[2ex]
n+p&\leftrightarrow& p+\Lambda \, , \\[2ex]
n+n&\leftrightarrow& n+\Lambda \, , 
\eea
\end{subequations}
has been computed by \citet{PhysRevD.64.084003,Haensel:2001em,Lindblom:2001hd,vanDalen:2003uy,Chatterjee:2006hy}. 
Effects of a large magnetic field were taken into account by \citet{Sinha:2008wb}, 
and the bulk viscosity in quark/hadron mixed phases was computed by 
\citet{Drago:2003wg}. 

The curves in Fig.\ \ref{fig:bulkhadron} show the result for unpaired matter. Cooper pairing of nucleons and/or hyperons change the underlying reaction rates dramatically and 
(to a much smaller extent) the susceptibilities that enter the bulk viscosity. Therefore, the energy gap in the nucleon dispersions has to be taken into account, leading to a suppression of the reaction rates. This suppression is exponential for temperatures much smaller than the critical temperature if at least one of the participating particles 
[say the neutron or the proton for the direct Urca process (\ref{dUbulk})] is gapped with an isotropic gap. A power law suppression occurs if the pairing leaves a node at the 
Fermi sphere where excitations with infinitesimally small energy are possible. This is conceivable for certain phases of $^3P_2$ pairing of neutrons (the milder suppression 
is of course only possible if at the same time there are unpaired protons). As a consequence of  
the suppression of the reaction rate, the bulk viscosity is suppressed for small temperatures  $T\ll T_c$. The effect of pairing on the bulk viscosity of hadronic matter was calculated by 
\citet{Haensel:2000vz,Haensel:2001mw,Haensel:2001em}. If neutrons form a superfluid, the corresponding Goldstone mode 
may contribute to the bulk viscosity and, depending on the equation of state, there may be a temperature regime where its contribution is dominant \cite{Manuel:2013bwa}.
Superfluidity also has an effect on the hydrodynamics of the system. Since a superfluid at finite temperature is effectively a two-fluid system, there is more than a single 
bulk viscosity coefficient. The additional coefficients have been computed for superfluid nuclear matter from Urca processes by \citet{Gusakov:2007px}, from 
phonons by \citet{Manuel:2013bwa}, and for superfluid nucleon-hyperon matter by \citet{Gusakov:2008hv}. The effect of these additional coefficients on the
instability window for $r$-mode oscillations appears to be small \cite{Haskell:2010ab}.

\section{Transport in the core: quark matter}
\label{sec:quarks}

\subsection{General remarks} 
\label{sec:quarkremarks}

Matter at sufficiently large baryon density is deconfined and quarks and gluons rather than baryons and mesons become the relevant degrees of freedom. 
This phase of matter is called quark matter or, especially at large temperatures where the gluons contribute to the thermodynamics, quark-gluon plasma. In the 
context of neutron stars, by quark matter we always mean three-flavor quark matter (or `strange quark matter') made of up, down, and strange quarks.
The reason is that the charm, bottom, and top quarks are too heavy to exist at the densities and temperatures typical for a neutron star. Therefore, even when we use perturbative methods which can only be trusted at extremely large densities, we ignore the heavy flavors because eventually we are interested in extrapolating our results down to neutron star densities.  
It is uncertain whether quark matter exists in the interior of neutron stars because we do not precisely know the central density of the star and, more importantly, 
we do not know the critical density at which nuclear matter turns into quark matter. It is conceivable that this transition is a crossover \cite{Schafer:1998ef,Alford:1999pa,Hatsuda:2006ps,Schmitt:2010pf}, such that there is no well-defined
transition density, just like the transition from the hadronic phase to the quark-gluon plasma at large temperatures and small baryon densities \cite{Aoki:2006we}.
Astrophysical data may provide important clues for the question of the location and nature of the deconfinement transition at large densities. Connecting observations 
from neutron stars to properties of ultra-dense matter is an intriguing example of probing our understanding of fundamental theories such as QCD 
with the help of astrophysics. In the case of quark matter (and also ultra-dense nuclear matter), the interplay between astrophysics and theory is particularly important because
currently there is no rigorous first-principle calculation of dense QCD, unless we go to even larger densities where perturbative methods become reliable \cite{Kurkela:2009gj,Kurkela:2010yk,Kurkela:2014vha}. The reason is that even brute force methods on the lattice fail due to the so-called sign problem, although there has been recent 
progress towards evading and/or mitigating the sign problem \cite{Aarts:2015tyj,Glesaaen:2015vtp,Gattringer:2016kco}. Ideally, we would like a given phase of dense
matter to be identifiable in an unambiguous way from a set of astrophysical observations. Of course, in reality, several distinct phases may show very similar behavior 
with respect to the observables that are accessible to us. For instance, many of the 
quark matter phases that we discuss in the following are basically 
indistinguishable from each other through bulk properties such as the equation of state and thus mass and radius of the star. But they do differ from each other in their low-energy properties, for instance 
because of different Cooper pairing patterns. Therefore, it is mostly the transport, 
less the thermodynamics, that differs from phase to phase. 

When we compute transport properties of quark matter, many aspects are similar to what we have discussed for hadronic matter in the previous sections: we are
obviously interested in the same quantities, i.e., neutrino emissivity, viscosity coefficients, etc., and the methods we use are often the same, even though the formulations
in the literature may sometimes look different. Nevertheless, there are some general differences which are useful to keep in mind before we go into more details. 
Firstly, quark matter is relativistic because the quark masses are small compared to the quark chemical potential and thus compared to the Fermi momentum\footnote{In this section, we work in natural units, $c=\hbar=k_B=1$, such that mass, energy, momentum, and temperature have the same units.}. (For the up and down quarks, the Fermi velocity $v_F$ introduced in Eq.\ (\ref{eq:qp_spectrum}) is very close to the speed of light, 
while the strange quark is heavy enough to induce sizable corrections to this ultra-relativistic limit.) Therefore, all microscopic calculations are performed in a relativistic framework, which for nuclear matter is only necessary at very large densities where
the nucleon rest mass becomes comparable to the Fermi energy. 
Secondly, when we want to treat quark matter rigorously with currently available methods, we need to approach neutron star densities `from above', i.e., we often assume 
quarks to be weakly interacting to be able to apply perturbation theory and then extrapolate the results down in density. This becomes 
relevant for some of the results discussed here, but not for all since, as we know from the previous section, not all transport properties rely on a precise knowledge of the strong interaction and are rather dominated by the electroweak interaction. Thirdly, quark matter has
a larger variety of candidate phases for neutron star cores than nuclear matter because there are 9 fermion species in three-flavor, three-color quark matter. As a result, there is a multitude of different possible patterns of Cooper pairing \cite{Alford:2007xm}, which is particularly interesting with regard to transport properties. 

We will summarize the current state of the art of reaction rates and transport properties in quark matter that are relevant for neutron stars. We attempt to give a comprehensive account of the current knowledge, which is possible because there are considerably 
fewer studies about quark matter transport than about nuclear matter transport. The results about quark matter we present here 
were obtained starting from a few works in the early eighties through a peak period around 2005 -- 2008 and including very recent progress that is still ongoing, with 
interesting ideas and prospectives for future work. 

\subsection{Phases of quark matter: overview}
\label{sec:quarkover}

As a preparation, especially for readers unfamiliar with dense QCD, it is 
useful to start with a brief overview about the relevant quark matter phases and their basic properties. In many cases, these basic properties already give us a rough idea about 
the behavior of the transport properties which we shall then discuss in more detail.

Just as we know the properties of low-density `ordinary' nuclear matter, we have solid, albeit only theoretical, knowledge about quark matter at extremely 
high densities. If the density is sufficiently large to apply weak-coupling methods and to neglect all three quark masses compared to the quark chemical potential, 
the ground state is the color-flavor locked (CFL) phase \cite{Alford:1997zt,Alford:1998mk}. While in nuclear matter more complicated phases including meson condensates and hyperons may occur as we move away from ordinary nuclear matter to {\it larger} densities (towards the center of the neutron star), more complicated phases of quark matter occur as we move towards {\it lower} densities (starting from the asymptotically dense regime, which is beyond neutron star densities). 

In the CFL phase, all quarks participate in Cooper pairing\footnote{Cooper pairing in quark matter always implies color superconductivity because at least some of the gluons acquire a Meissner mass, in CFL all eight of them. Whether a color superconductor is also an electromagnetic superconductor and a superfluid is more subtle and will not be discussed in full detail here.}, and as a consequence the dispersions of all fermionic quasiparticles are gapped. At zero temperature, the number densities of all quark species are identical, and therefore the CFL phase is `automatically' neutral, without any electrons or muons
(recall that the electric charges of up, down and strange quarks happen to add up to zero). This makes CFL very special with respect to transport because at low 
energies all fermionic degrees of freedom are suppressed and can be neglected. The CFL phase breaks chiral symmetry spontaneously, and thus there is a set of 
(pseudo-)Goldstone modes, very similar to the mesons that arise from `usual' chiral symmetry breaking through a chiral condensate. At low temperatures, 
these light bosons dominate (some of) the transport properties of the CFL phase. The CFL mesons appear with the same quantum numbers as the mesons from quark-antiquark condensation, which is a consequence of the identical symmetry breaking pattern. However, their masses 
are different: in CFL, the kaons, not the pions, are the lightest mesons. 

The CFL phase is a superfluid because it spontaneously breaks the $U(1)$ symmetry associated with baryon number conservation, and thus, as discussed in Sec.\ \ref{S:cooper}, the usual complications of superfluid transport arise, such as the two-fluid picture at nonzero temperature, or the existence of quantized 
vortices in rotating CFL. Moreover, superfluidity implies the existence of an exactly massless Goldstone mode, which yields the dominant contribution for instance to 
the shear viscosity of CFL. The transport properties of CFL are determined by an effective theory for the pseudo-Goldstone modes and the superfluid mode. The form of 
this effective theory, in turn, is entirely given by the symmetry breaking pattern of CFL, just like usual chiral perturbation theory. Therefore, if CFL 
persists down to neutron star densities, we have a very solid knowledge of the low-energy physics of quark matter, although the numerical coefficients of the 
effective theory can only be determined reliably at weak coupling and become uncertain as we move towards lower densities. 

The opposite of CFL, in a way, is unpaired quark matter, where none of the quark species forms Cooper pairs. 
Unpaired quark matter probably exists only at high temperatures
$T\gtrsim 10\, {\rm MeV}$, because at lower temperatures Cooper pairing in some form seems unavoidable \cite{Alford:2007xm}. Nevertheless, unpaired quark matter is an 
important concept and its transport properties, even at low temperature, are relevant. The reason is that almost all quark matter phases except for CFL have some unpaired quarks 
or quarks with a very small pairing gap, which dominate transport. Thus, up to numerical prefactors, the result for unpaired quark matter is a good approximation for these 
phases in many instances. The calculation of transport properties for unpaired quark matter is, in a sense, more difficult than for CFL because 
we need to know the interaction via gluons in a strongly coupled regime (unless the transport property of interest is dominated by the electroweak interactions). Therefore, 
shear viscosity, electrical and thermal conductivities of unpaired quark matter are typically based on perturbative calculations, assuming the strong coupling constant $\alpha_s$ to be small.  

Between the two extreme cases of CFL (present at asymptotically large densities and sufficiently small temperatures) and completely unpaired quark matter
(strictly speaking only present at temperatures higher than in a neutron star but important conceptually), there are many possible quark matter phases where quarks `partially' pair.
These phases are likely to be relevant for neutron stars and thus their transport properties have been studied extensively. `Partial' pairing means that certain 
quark colors or flavors remain unpaired and/or that Cooper pairing does not occur in all directions in momentum space and even may vary spatially. Such phases 
necessarily appear at moderate densities because going down in density means a decrease in quark chemical potential and an increase in the strange quark mass, somewhere between 
the current mass of about $100\, {\rm MeV}$ and the vacuum constituent mass of about 
$500\, {\rm MeV}$.
As a consequence, the strange quark mass cannot be neglected 
at densities in the cores of neutron stars, and the particularly symmetric situation at asymptotically large densities is disrupted. Why does the less symmetric situation of different quark masses eventually lead to a breakdown of CFL? The reason is that the gain in free energy 
from Cooper pairing is maximized if the two participating fermion species have the same Fermi surface and pairing occurs over the entire surface in momentum space. 
If the fermions that `want' to pair have different Fermi surfaces, an energy cost is involved in Cooper pairing, and this cost may be too large to create a paired state. 
Different masses, together with the conditions of beta-equilibrium and charge neutrality, provide such a difference in Fermi surfaces because, at least for the most favorable 
spin-0 channel, pairing takes place between quarks of different color and flavor. Therefore, CFL is under stress if we move away from asymptotically large densities. 
The system is expected to react in a series of phase transitions, producing more complicated quark matter phases. The exact sequence of these phases can be determined in a controlled way at very large densities and weak coupling, but as we move to lower densities, we have less rigorous knowledge of the phase structure and rely mostly 
on model calculations or bold extrapolations from ultra-high densities. In particular, it is not known where in this sequence of phases nuclear matter takes over. It is 
conceivable that CFL persists down to densities where the transition to hadronic matter occurs, possibly leading to a nuclear/CFL interface inside a neutron star. It is 
also possible that other color-superconducting phases exist in the core of neutron stars, possibly breaking rotational and/or translational invariance. 
Also, since the QCD coupling increases with lower energies, $\alpha_s \gtrsim 1$ at densities relevant for astrophysics, the color-superconducting 
phases may be replaced by something qualitatively different, possibly involving elements from the Bardeen-Cooper-Schrieffer--Bose-Einstein-condensation (BCS-BEC) crossover \cite{Deng:2006ed} seen in atomic gases or possibly 
showing features of the quarkyonic phase that is predicted in QCD in the limit of infinite 
number of colors $N_c$ \cite{McLerran:2007qj}.

\subsection{Neutrino emissivity}
\label{sec:quarkneutrino}

As for hadronic matter, neutrino emissivity is interesting in itself because it is the main cooling mechanism of the star, and 
the rates for the neutrino processes can be relevant for the bulk viscosity of quark matter inside a neutron star. We consider the processes 
\be \label{urcaquark}
d\to u + e +\bar{\nu}_e \, , \qquad u + e \to d+\nu_e \, .
\ee
These are the analogues of the direct Urca processes in nuclear matter (\ref{eq:Core_Urca_forrev}). In quark matter, the triangle inequality from momentum conservation does not pose a severe 
constraint on this process because the Fermi surfaces between up and down quark are not as different as the ones for neutrons and protons in nuclear matter.  Therefore, second-order neutrino processes such as bremsstrahlung are usually negligible in quark matter. 
We first discuss the processes (\ref{urcaquark}) and later summarize the results that involve strangeness. Generalizing the definition (\ref{eq:weak_rate_gen_matel}), where Fermi's Golden Rule was applied directly, the neutrino emissivity is
\be \label{defeps}
\epsilon_\nu \equiv 
 2\frac{\partial}{\partial t}\int\frac{d^3{\bm p}_\nu}{(2\pi)^3}\,p_\nu\,
f_\nu (t, {\bm p}_\nu) \, ,
\ee
where ${\bm p}_\nu$ is the neutrino momentum, and the factor 2 accounts for neutrinos and anti-neutrinos. The change of the neutrino distribution function $f_\nu$
is computed from the gain term in the neutrino transport equation, which takes the form
\bea \label{df}
\frac{\partial}{\partial t} f_\nu(t,{\bm p}_\nu) &=& -\frac{\cos^2\theta_CG_F^2}{8} \int\frac{d^3{\bm p}_e}{(2\pi)^3 p_\nu p_e} 
L_{\lambda\sigma}f_F(p_e-\mu_e)f_B(p_\nu+\mu_e-p_e){\rm Im}\,\Pi_R^{\lambda\sigma}(Q) \, , 
\eea
with the Fermi and Bose distribution functions $f_{F,B}(x)=(e^{x/T}\pm 1)^{-1}$ for the electron with energy $p_e$ and chemical potential $\mu_e$ and the $W$-boson with four-momentum  
$Q =(q_0,{\bm q})= (p_e-p_\nu-\mu_e, {\bm p}_e-{\bm p}_\nu)$. 
We have abbreviated
$L^{\lambda\sigma}\equiv\mbox{Tr}\left[(\gamma_0p_e-\vg\cdot{\bm p}_e)\, 
\gamma^\sigma (1-\gamma^5)(\gamma_0p_\nu-\vg\cdot{\bm p}_\nu)\, \gamma^\lambda
(1-\gamma^5)\right]$,  with the Dirac  matrices $\gamma^\sigma$ ($\sigma =0,1,2,3$) and 
$\gamma^5 = i\gamma^0\gamma^1\gamma^2\gamma^3$. (Note that the subscript $\nu$ stands for neutrino and is thus not used as a Lorentz index.) 
Finally, ${\rm Im}\,\Pi_R^{\lambda\sigma}$ is the imaginary part of the retarded $W$-boson self-energy
\be \label{PiQ}
\Pi^{\lambda\sigma}(Q) = \frac{T}{V}\sum_K\mbox{Tr}[\Gamma_-^\lambda S(K)\Gamma_+^\sigma S(P)] \, , 
\ee
with the quark propagator $S$,  which is a matrix in color, flavor, and Dirac space 
and -- in the case of Cooper pairing -- in Nambu-Gorkov space. The electroweak vertices are diagonal in Nambu-Gorkov space, $\Gamma_\pm^\lambda={\rm diag}[\gamma^\lambda(1-\gamma^5)\tau_\pm,- \gamma^\lambda(1+\gamma^5)\tau_\mp]$, where $\tau_\pm=(\tau_1\pm i\tau_2)/2$ with the Pauli matrices $\tau_1$, $\tau_2$ are matrices in flavor space which ensure that an up and a down quark interact at the vertex, i.e., $K$ and $P$ correspond to the up and down quark four-momenta, respectively. 

\begin{figure}[t]
\begin{center}
\includegraphics[width=\textwidth]{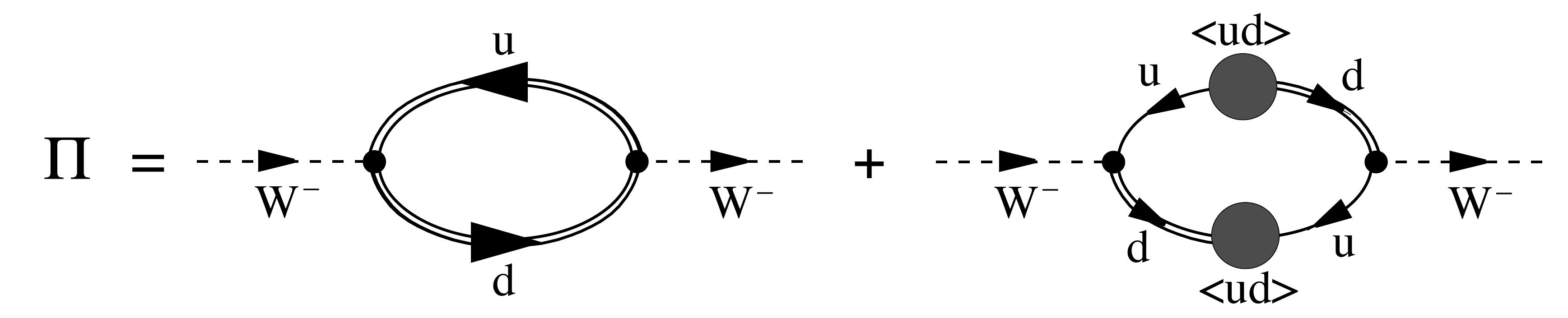}
\caption{Normal and anomalous contributions to the $W$-boson self-energy needed to compute the
neutrino emissivity in the presence of Cooper pairing, here shown for pairing between up and down quarks, for instance in the 2SC phase. The first diagram contains two `normal' propagators (diagonal elements in Nambu-Gorkov space), the double line indicating that they include the effect of pairing through a modified dispersion relation. The second diagram contains two `anomalous' propagators (off-diagonal elements in Nambu-Gorkov space), whose diagrammatic structure indicates that they describe a propagating fermion that is absorbed by the condensate $\langle ud\rangle$ and continues to propagate
as a charge-conjugate fermion. The two contributions are obtained naturally in the Nambu-Gorkov formalism by performing the trace over Nambu-Gorkov space in Eq.\ (\ref{PiQ}). 
}
\label{fig:Pianomalous}
\end{center}
\end{figure}

As an instructive example for the neutrino emissivity in quark matter due to the Urca processes, let us discuss the 2SC phase \cite{Bailin:1983bm}. In the 2SC phase all strange 
quarks and all quarks of one color, say blue, are ungapped. This phase is an example for the less symmetrically paired phases mentioned above. The up and down quarks participating in the processes (\ref{urcaquark}) can be either gapped or ungapped. Since the weak interaction does not 
change the color of the quarks, they are either both gapped (when they are red or green) or both ungapped (when they are blue). The contribution of the gapped sector,
here shown for a single color, is \cite{Schmitt:2010pn}
\bea \label{df1}
\frac{\partial}{\partial t} f_\nu(t,{\bm p}_\nu) &=& \frac{\pi \cos^2\theta_CG_F^2 }{4} \sum_{e_1,e_2=\pm}\int\frac{d^3{\bm p}_ed^3{\bm k}}{(2\pi)^3(2\pi)^3 p_\nu p_e} 
L_{\lambda\sigma}\left({\cal T}^{\lambda\sigma} B_k^{e_1}B_p^{e_2} + {\cal U}^{\lambda\sigma}\frac{\Delta^2}{4\varepsilon_k\varepsilon_p}\right)\non[2ex]
&&\times f_F(p_e-\mu_e)f_F(-e_1\varepsilon_k)f_F(e_2\varepsilon_p)\delta(q_0-e_1\varepsilon_k+e_2\varepsilon_p) \, , 
\eea
where  ${\bm k}$ and ${\bm p}$ are the up and down quark three-momenta, respectively. We have 
denoted the Bogoliubov coefficients by $B_k^{e}=(1-e\xi_k/\varepsilon_k)/2$ with $\xi_k=k-\mu_u$ and the quasiparticle dispersion $\varepsilon_k^2 = \xi^2_k+\Delta^2$, where $\mu_u$ is the up quark chemical potential, and analogously for the down quark with chemical potential $\mu_d$. Moreover, we have abbreviated ${\cal T}^{\lambda\sigma}\equiv \Tr\left[\gamma^\lambda(1-\gamma^5)\gamma^0\Lambda_k^-
\gamma^\sigma(1-\gamma^5)\gamma^0\Lambda_p^-\right]$, and ${\cal U}^{\lambda\sigma}\equiv \Tr\left[\gamma^\lambda(1-\gamma^5)\gamma^5\Lambda_k^+
\gamma^\sigma(1+\gamma^5)\gamma^5\Lambda_p^-\right]$, with the energy projectors $\Lambda_k^\pm = (1\pm\gamma^0\vg\cdot\hat{\bm k})/2$. 
The contribution 
proportional to $\Delta^2$ comes from the so-called anomalous propagators, the off-diagonal components of the quark propagator $S(K)$ in Nambu-Gorkov space. Their effect was discussed in detail and evaluated numerically for the 2SC phase by \citet{Jaikumar:2005hy}, see Fig.\ \ref{fig:Pianomalous} for a diagrammatic representation of normal and anomalous contributions to the $W$-boson self-energy.

The expression in Eq.\ (\ref{df1}) is instructive for the neutrino emissivity in the presence of Cooper pairing because it shows 
4 potential subprocesses that arise from summing over $e_1$ and $e_2$. Naively, one would expect the distribution functions for the process $u + e \to d+\nu_e$
to appear in the form $f_ef_u(1-f_d)$ (we have neglected $f_\nu$ since the neutrinos leave the star once they are created). We see that the combinations 
$f_ef_uf_d$, $f_e(1-f_u)(1-f_d)$, $f_e(1-f_u)f_d$ appear as well [note that for the Fermi distribution $f(-x)=1-f(x)$]. The reason is that quasiparticles in the superconductor are mixtures of particles and holes 
(this momentum-dependent mixture is quantified by the Bogoliubov coefficients) and are thus allowed to appear 
on either side of the reaction. 
Since we have started from a general form of the reaction rate that is based on the full structure of the propagator, all four reactions are included automatically
and we do not have to set up a separate calculation of these Cooper pair breaking and formation processes.  
This is analogous to nuclear matter with superfluid neutrons, see Sec.\ \ref{sec:nu_Cooper}. 
In that case, since the direct Urca process is usually suppressed, the Cooper pair breaking and formation processes are 
discussed for the neutral current process $n + n \to n+n+\nu+\bar{\nu}$, which, in the 
presence of Cooper pairing, allows for the processes $\{nn\} \to n+n+\nu+\bar{\nu}$ and $n+n \to \{nn\}+\nu+\bar{\nu}$, where $\{...\}$ denotes the Cooper pair condensate. 
The quark version of these processes is $\{uu\} \to u+u+\nu+\bar{\nu}$ and $u+u \to \{uu\}+\nu+\bar{\nu}$ (assuming 
single-flavor quark Cooper pairing), which yields a neutrino emissivity $\epsilon_\nu\propto T^7$ \cite{Jaikumar:2001hq}, just like in nuclear matter. 

If we are only interested in small temperatures compared to the energy gap $\Delta$, the Cooper pair breaking and formation processes are 
irrelevant, and the contribution from the 
gapped quarks is exponentially suppressed, $\epsilon_\nu\propto e^{-\Delta/T}$. As a consequence, the neutrino emissivity of the 2SC phase is, at small temperatures, utterly dominated by the unpaired blue quarks. At higher temperatures, as we approach the critical temperature $T_c$ (for the 2SC phase, $T_c \sim 10\, {\rm MeV}$), Eq.\ (\ref{df1}) has to be evaluated numerically.

To compute the neutrino emissivity for unpaired quarks, we may set $\Delta=0$ in Eq.\ (\ref{df1}). As a result, the dispersions $\varepsilon_k$ of the quarks (assumed to be massless) 
become dispersions of free fermions. However, it is crucial to include the effect of the 
strong interaction, i.e., to treat the system as a Fermi liquid rather than a non-interacting 
system of quarks. Otherwise, the phase space for the Urca process is zero and the neutrino 
emissivity vanishes. Fermi liquid corrections are included by writing the Fermi momenta of up and down quarks as $\mu_u[1-{\cal O}(\alpha_s)]$ and $\mu_d[1-{\cal O}(\alpha_s)]$. The result for 2-flavor unpaired quark matter is (reinstating all color degrees of freedom, $N_c=3$)
\be \label{emitfinal}
\epsilon_\nu^{\rm unp.} \approx \frac{457}{630}\cos^2\theta_CG_F^2\alpha_s\mu_e\mu_u\mu_d T^6 \left(1+\frac{4\alpha_s}{9\pi}\ln\frac{\Lambda}{T}\right)^2 \, ,
\ee
where the electron chemical potential is related to the quark chemical potentials via $\mu_u+\mu_e=\mu_d$ in $\beta$-equilibrium. 
The logarithmic correction of \citet{Schafer:2004jp} to the standard result by \citet{Iwamoto:1980eb,Iwamoto:1982zz} arises if non-Fermi liquid effects are included for the quarks, and
can lead to an enhancement of  the neutrino emissivity at low temperatures \cite{Schafer:2004jp}. The energy scale that appears in the logarithm is of the order of the 
screening scale, $\Lambda \propto g\mu$, where $g$ is the strong coupling constant related to $\alpha_s$ by $\alpha_s = g^2/(4\pi)$, see Ref.\ \cite{Gerhold:2005uu} for a calculation of $\Lambda$.
Higher order corrections to this result have been computed by \citet{Adhya:2012sq}.
The strange quark mass has to be included if the result is generalized to strange quark matter. A mass term can easily be added in the quark dispersion, 
but the result for the emissivity becomes more complicated and is best evaluated numerically. One effect of the mass is to open up the phase space such that the emissivity would be 
nonzero even if the Fermi liquid corrections were neglected, as discussed by \citet{Iwamoto:1980eb,Wang:2006tg}. A dynamical quark mass 
from a chiral density wave has a similar effect. The chiral density wave is an anisotropic phase in which the chiral condensate oscillates between 
between scalar and pseudoscalar components, and the neutrino emissivity depends on the dynamical mass and the wave vector that determines this oscillation 
\cite{Tatsumi:2014cea}. This phase, possibly in coexistence with quark Cooper pairing is a candidate phase in the vicinity of a potential first-order chiral phase transition 
between the hadronic matter and quark matter.

A similar calculation as outlined here for the unpaired and 2SC phases applies to the so-called Larkin-Ovchinnikov-Fulde-Ferrell (LOFF) phases  and 
to color superconductors where Cooper pairs have total spin one. These two classes of phases are 
further important examples of the less symmetric phases that are expected to arise for a large mismatch in Fermi surfaces. An estimate of this mismatch, based on an expansion for 
small strange quark masses $m_s$, is given by comparing $m_s^2/\mu$ to the energy gap $\Delta$, where $\mu$ is the quark chemical potential (baryon chemical potential divided by $N_c=3$). In neutron stars, exotic phases like LOFF or spin-one pairing thus occur 
if the attractive interaction (for which $\Delta$ is a measure) is not strong enough to overcome the mismatch $m_s^2/\mu$ (which increases with decreasing density because 
$m_s$ increases and $\mu$ decreases). In the LOFF phase, the system reacts to the mismatch in Fermi surfaces 
by forming Cooper pairs only in certain directions in momentum space, resulting in Cooper pairs with nonzero momentum \cite{Alford:2000ze,Anglani:2013gfu}. 
In general, a finite number of different 
Cooper pair momenta will be realized in a given phase, resulting in counter-propagating currents and in a crystalline structure with periodically varying gap function. Since there are 
directions in momentum space where the quasiparticle dispersion is ungapped, the neutrino emissivity of the LOFF phase is qualitatively very similar to unpaired quark matter, as shown by
\citet{Anglani:2006br}.
Spin-one color superconductors arise unavoidably in single-flavor Cooper pairing. This form of pairing is the only possible one if the mismatch in Fermi momenta of quarks of 
different flavor is sufficiently large to prevent any form of cross-flavor pairing. Spin-one color superconductors break rotational symmetry and typically 
exhibit ungapped directions in momentum space as well. Therefore, as for the LOFF phase, their neutrino emissivity has the same $T^6$ behavior as 
unpaired quark matter. A possible exception is the color-spin locked phase (CSL), where, in a certain variant, all quarks are gapped. However, weak coupling calculations suggest
that another variant of CSL, where there {\it are} unpaired quasifermions, is energetically preferred \cite{Schmitt:2004et} (although there are fewer paired quarks, the larger 
value of the gap function overcomes this lack of pairing). If we only consider the gapped branches, there are striking similarities of the neutrino emissivity in spin-one 
color superconductors, computed by \citet{Schmitt:2005wg,Wang:2006tg,Berdermann:2016mwt}, 
to the neutrino emissivity of $^3P_2$ phases in nuclear matter 
\cite{Yakovlev2001physrep}, 
which we have briefly discussed in Sec.\ \ref{sec:nu_Cooper}.

The neutrino emissivity of CFL is qualitatively different from the phases with ungapped fermions. In CFL, neutrino emissivity is dominated by the Goldstone modes,
and the relevant processes are 
\be
\pi^\pm, K^\pm \to e^\pm + \bar{\nu}_e \, , \qquad \pi^0\to \nu_e+\bar{\nu}_e \, ,\qquad \phi+\phi \to \phi + \nu_e+\bar{\nu}_e \, .
\ee
Here, $\pi^\pm$ and $K^\pm$ are the CFL mesons mentioned above, which have the same quantum numbers as, but different masses than, their counterparts from usual
chiral symmetry breaking. In particular, the kaons are the lightest mesons in CFL, with masses of a few MeV. Since these masses are larger than typical temperatures 
of neutron stars, the resulting neutrino emissivities are exponentially suppressed. The superfluid mode $\phi$ is massless and thus does not show this exponential suppression.
However, the emissivity is proportional to a large power of $T$, which makes this result very small as well \cite{Jaikumar:2002vg},
\be
\epsilon_\nu^{\rm CFL} \sim \frac{G_F^2 T^{15}}{f_\phi\mu^4} \, , 
\ee
where $f_\phi$ is the analogue of the pion decay constant for the spontaneous breaking of baryon number. We conclude that the CFL phase basically does not contribute to the 
neutrino emissivity. 

Neutrino emissivities of quark matter have been included in cooling calculations for hybrid stars \cite{Grigorian:2004jq,Popov:2005xa,Hess:2011qw,Noda:2011ag}, and 
quark matter may provide an explanation for the rapid cooling of the neutron star in Cassiopeia A \cite{Sedrakian:2013xgk,Sedrakian:2015qxa}. In this scenario, the 
star cools through a transition from the 2SC phase with very inefficient cooling to a crystalline color superconductor, where there are unpaired fermions. 
This explanation assumes that there is no contribution of the strange quarks and -- on purely phenomenological grounds -- that there is some residual pairing mechanism for the blue quarks in 2SC. While the explanation of the rapid cooling in nuclear matter 
is based on the transition from an unpaired phase to the superfluid phase, quark matter may thus potentially  show a similar behavior via a transition from one paired phase 
to another.

\subsection{Bulk viscosity}\label{sec:bulkquark}

\subsubsection{Unpaired quark matter}
\label{sec:bulkunpaired}

We have already discussed the definition and physical meaning of the bulk viscosity $\zeta$ in Sec.\ \ref{S:bulk} and can immediately start from the expression 
\be \label{zeta1}
\zeta = -\frac{2}{\omega^2\delta v_0^2}\frac{1}{\tau} \int_0^\tau dt \, P(t)\frac{d\delta v}{d t} \, , 
\ee
which follows from Eqs.\ (\ref{Epsdot}) and (\ref{Epsdot1}). The pressure $P(t)$ is given by Eq.\ (\ref{pt}), where now, for unpaired quark matter, 
the oscillations in density occur for the three quark flavors and the electron, $\delta n_u$, $\delta n_d$, $\delta n_s$, $\delta n_e$. We consider the processes
\begin{subequations} \label{uuu}
\bea
u+d &\leftrightarrow& u + s  \, ,  \label{uuu1}\\[2ex]
u+e &\to& d + \nu_e \, , \qquad d \to u+e+ \bar{\nu}_e\label{uuu2} \\[2ex]
u+e &\to& s + \nu_e \, , \qquad s \to u+e+ \bar{\nu}_e\, . \label{uuu3}
\eea
\end{subequations}
The non-leptonic process $u+d \leftrightarrow u + s$ will turn out to be the dominant one, but it is instructive to keep the leptonic processes. This allows us to sketch the calculation of the bulk viscosity for a more complicated scenario as outlined at the beginning of Sec.\ \ref{S:bulk}. Namely, we now have two out-of-equilibrium chemical potentials $\delta\mu_1 \equiv \mu_s-\mu_d$ and $\delta\mu_2 \equiv \mu_d-\mu_u-\mu_e$, relevant for the reactions (\ref{uuu1}) and (\ref{uuu2}). The relevant difference in chemical potentials for the reaction (\ref{uuu3}) is then $\delta\mu_1+\delta\mu_2$, and thus not an independent quantity. 
As independent changes in densities we keep $\delta n_d$, $\delta n_e$. The changes in 
up and strange quark densities then are $\delta n_u = \delta n_d - \delta n_e$ and $\delta n_s = -\delta n_d - \delta n_e$. The change in the electron density comes from the processes (\ref{uuu2}) and  (\ref{uuu3}), and the change in the down quark number density comes from the processes (\ref{uuu1}) 
and  (\ref{uuu2}), and in analogy to Eq.\ (\ref{dnnep}) we write in the linear approximation
\be \label{dned}
\frac{d n_e}{dt} \approx (\lambda_2+\lambda_3)\delta\mu_2(t)+\lambda_3\delta\mu_1(t) \, , \qquad  
\frac{d n_d}{dt}\approx \lambda_1\delta\mu_1(t)-\lambda_2\delta\mu_2(t) \, ,
\ee
where $\lambda_1$, $\lambda_2$, $\lambda_3$ have to be computed from the microscopic processes. The result for the non-leptonic process (\ref{uuu1}) is
\cite{Wang:1985tg,Sawyer:1989uy,Madsen:1993xx}
\be \label{lam1}
\lambda_1 \approx \frac{64\sin^2\theta_C\cos^2\theta_CG_F^2}{5\pi^3}\mu_d^5 T^2 \, , 
\ee
while the leptonic processes (\ref{uuu2}) and (\ref{uuu3}) yield \cite{Iwamoto:1980eb,Iwamoto:1982zz}
\begin{subequations} \label{lam23}
\bea
\lambda_2 &\approx&\frac{17\cos^2\theta_CG_F^2}{15\pi^2}\alpha_s\mu_u\mu_d\mu_e T^4 \, , \\[2ex]
\lambda_3 &\approx&\frac{17\sin^2\theta_CG_F^2}{40\pi^2}\mu_s m_s^2 T^4 \, ,
\eea
\end{subequations} 
where $\lambda_2$ is obtained from the same calculation that leads to the neutrino emissivity (\ref{emitfinal}), and the leptonic process including the strange quark is computed to lowest order in the strange quark mass $m_s$. Generalizing Eq.\ (\ref{dmu0}), we have two differential equations for $\delta\mu_1$ and $\delta\mu_2$,
\bea \label{dmu12}
\frac{d\delta\mu_i}{dt} &=& \frac{\partial \delta\mu_i}{\partial V} V_0 \frac{d\delta v}{dt} +  \sum_{x=u,d,s,e} \frac{\partial \delta\mu_i}{\partial n_x}
\frac{d n_x}{dt}  \non[2ex]
&=& -B_i \frac{d\delta v}{dt}  - \alpha_i \delta\mu_1(t) - \beta_i \delta\mu_2(t) \, , \qquad i=1,2 \, ,
\eea
where $B_i$ are combinations of thermodynamic functions in equilibrium, and $\alpha_i$, $\beta_i$ contain thermodynamic functions and the reaction rates $\lambda_1$, $\lambda_2$, $\lambda_3$.  As in Sec.\ \ref{S:bulk}, we use the ansatz $\delta\mu_i(t) = {\rm Re}(\delta\mu_{i0}e^{i\omega t})$ with complex amplitudes $\delta\mu_{i0}$, such that (\ref{dmu12}) can be solved for 
real and imaginary parts of $\delta\mu_{10}$ and $\delta\mu_{20}$. The bulk viscosity (\ref{zeta1}) then becomes 
\be
\zeta = \frac{a_1{\rm Re}(\delta\mu_{10})+a_2{\rm Re}(\delta\mu_{20})}{\omega^2\delta v_0} \, , 
\ee
where $a_1$ and $a_2$ are combinations of $B_1$, $B_2$, $\lambda_1$, $\lambda_2$, $\lambda_3$. Computing ${\rm Re}(\delta\mu_{10})$ and ${\rm Re}(\delta\mu_{10})$ from Eq.\ (\ref{dmu12}) yields the final expression in terms of thermodynamic functions in equilibrium, the reaction rates, and the externally given frequency $\omega$. This result is very lengthy and entangles all reaction rates in a complicated way with the thermodynamic functions \cite{Alford:2006gy,Sad:2007afd}. 
\begin{figure}[t]
\begin{center}
\includegraphics[width=0.6\textwidth]{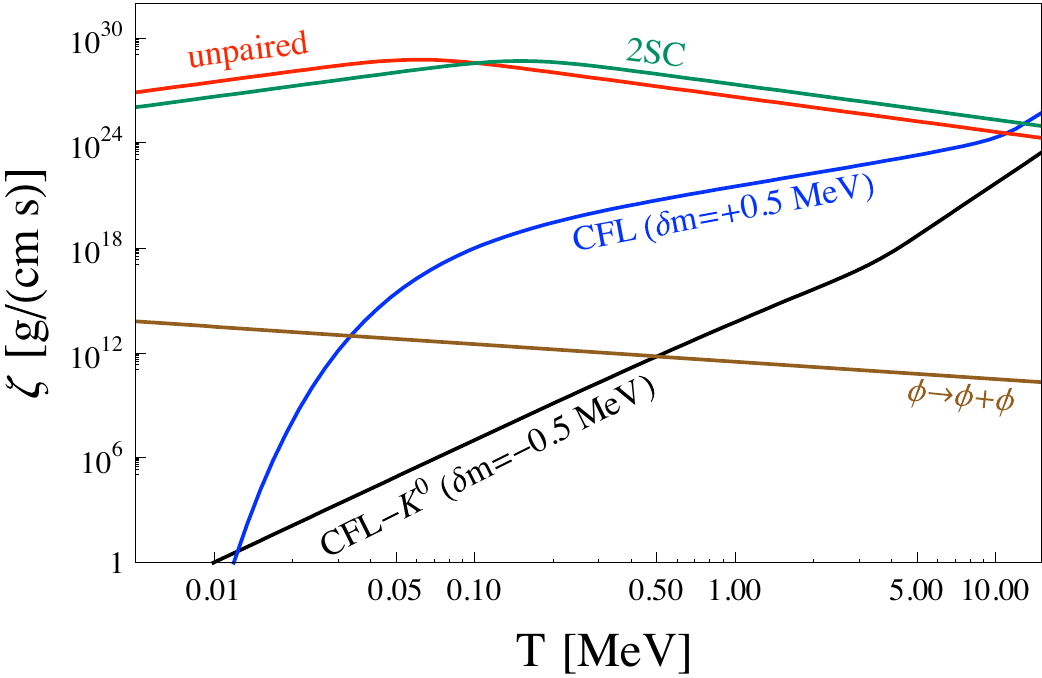}
\caption{Bulk viscosity for unpaired quark matter, the 2SC phase, and the CFL phase. For the CFL phase, the result is shown from the process that only involves the superfluid mode $\phi$ and
from processes that involve kaons and the superfluid mode, for thermal kaons $\delta m = m_{K^0}-\mu_{K^0} > 0$ and condensed kaons $\delta m <0$. The parameters
chosen here are the quark chemical potential $\mu = 400\,{\rm MeV}$, the frequency $\omega/(2\pi) = 1\,{\rm ms}^{-1}$, and the kaon mass $m_{K^0}= 10\, {\rm MeV}$. The figure is reproduced with modifications from Ref.\ \cite{Alford:2008pb}.}
\label{fig:bulkquarks}
\end{center}
\end{figure}
For a qualitative discussion we introduce the inverse time scales $\gamma_{\rm nl} = \lambda_1/\mu_s^2$ for the non-leptonic 
process (\ref{uuu1}) and $\gamma_{\rm l} = \lambda_2/\mu_s^2\approx \lambda_3/\mu_s^2$ for the leptonic processes (\ref{uuu2}) and (\ref{uuu3}) and assume 
$\gamma_{\rm nl} \gg \gamma_{\rm l}$. Then, with some simple estimates of the thermodynamic functions, and ignoring numerical prefactors, we 
find \cite{Alford:2006gy} 
\be
\zeta \propto  \gamma_{\rm nl}\frac{\gamma_{\rm nl}\gamma_{\rm l}+\omega^2}{\gamma_{\rm nl}^2\gamma_{\rm l}^2+\gamma_{\rm nl}^2\omega^2+\omega^4} \, .
\ee
From this result, various limit cases can be derived, depending on whether the external frequency $\omega$ is of the order of the leptonic rate, the nonleptonic rate, in between these 
rates etc. The most relevant case turns out to be $\omega \approx \gamma_{\rm nl} \gg \gamma_{\rm l}$, in which the slower leptonic processes can be completely 
neglected. Reinstating the thermodynamic functions, we obtain \cite{Madsen:1992sx}
\be
\zeta \approx  \frac{\lambda_1 B^2}{(\lambda_1 C)^2+\omega^2} \, , \label{zetaunp}
\ee
where
\be
B \equiv n_d \frac{\partial \mu_d}{\partial n_d} - n_s \frac{\partial \mu_s}{\partial n_s} \, , \qquad C \equiv \frac{\partial \mu_d}{\partial n_d} +\frac{\partial \mu_s}{\partial n_s} \, .
\ee  
We have recovered the result (\ref{zetanep}) derived in the context of nuclear matter for a single reaction rate. The result for unpaired quark matter as a function of temperature for a fixed frequency 
$\omega$ is plotted in 
Fig.\ \ref{fig:bulkquarks}.

\begin{figure}[t]
\begin{center}
\includegraphics[width=0.5\textwidth]{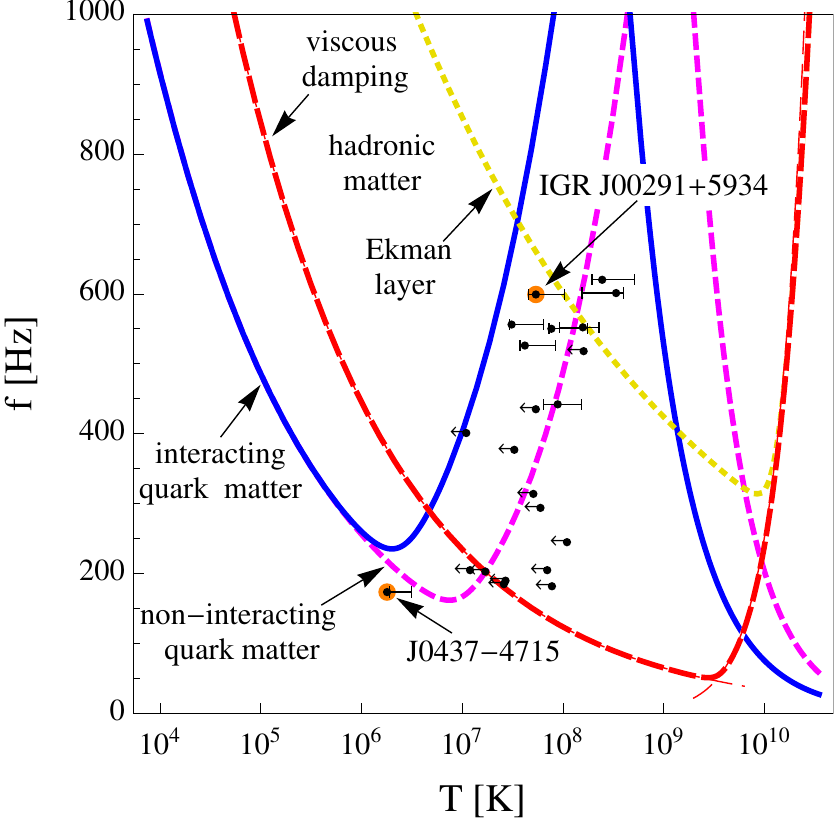}
\caption{$R$-mode instability window, computed from bulk viscosity (high $T$) and shear viscosity (low $T$) in nuclear and quark matter, taken from Ref.\ \cite{Alford:2013pma}.  Stars are unstable with respect to the emission of gravitational waves through the $r$-mode instability if they rotate with higher 
frequencies than given by the critical curves shown here. 
The observed pulsars shown as data points are in the stable region only for interacting quark matter with the non-Fermi-liquid corrections from Eq.\ (\ref{lam1nf}). These corrections enhance the 
reaction rate for the conversion of down into strange quarks and thus shift the maximum of the bulk viscosity towards lower temperatures, leading to a 
shifted stability region compared to noninteracting quark matter. Hadronic matter is consistent with the data only by including additional processes such as dissipation from boundary layer rubbing at the crust of the star.} 
\label{fig:rmode1}
\end{center}
\end{figure}

The calculation of the bulk viscosity in unpaired quark matter outlined here has been improved and extended in the literature in several ways. Firstly, Cooper pairing needs to be taken into account, and we shall discuss the results for various phases in the following subsection (Fig.\ \ref{fig:bulkquarks} collects most of these results). Secondly, the 
supra-thermal regime, where the amplitude of the oscillations in 
chemical potential become large compared to  the temperature, has been studied by \citet{Alford:2010gw}, who have generalized earlier numerical results for strange quark matter by \citet{Madsen:1992sx}.  \citet{Shovkovy:2010xk} studied this regime together with the interplay of leptonic and non-leptonic processes. The bulk viscosity 
in the presence of large amplitudes is important if the time evolution and in particular the saturation of unstable $r$-modes is studied \cite{Alford:2011pi}. Thirdly, 
as for the neutrino emissivity, see Eq.\ (\ref{emitfinal}), non-Fermi liquid effects can be included in the calculation of unpaired quark matter. Most importantly, they modify
the result for the dominant non-leptonic process $u+d\leftrightarrow u+s$ \cite{Schwenzer:2012ga}
\be \label{lam1nf}
\lambda_1 \approx \frac{64\sin^2\theta_C\cos^2\theta_CG_F^2}{5\pi^3}\mu_d^5 T^2 \left(1+\frac{4\alpha_s}{9\pi}\ln\frac{\Lambda}{T}\right)^4 \, . 
\ee
The correction factor has a higher power compared to the leptonic process that leads to the neutrino emissivity (\ref{emitfinal}) because now 4, not 2 quarks participate in the 
process. If this result is extrapolated to realistic values of the strong coupling, $\alpha_s \sim 1$, the enhancement due to the long-range interactions is larger than for the emissivity. As a consequence,
the maximum of the bulk viscosity -- at a fixed frequency $\omega$ -- is shifted to smaller temperatures. \citet{Alford:2013pma} pointed out that this may have interesting consequences for the $r$-mode instability,
see 
Fig.\ \ref{fig:rmode1}. 
We have to keep in mind, however, firstly, that completely unpaired quark matter is unlikely to exist in this (sufficiently cold) temperature regime because some form of Cooper pairing 
is expected to occur, and, secondly, that unpaired quark matter provides extremely efficient  cooling due to the presence of direct Urca processes (\ref{emitfinal}), which is difficult to reconcile with observations \cite{Chugunov2017MNRAS}. Finally, \citet{Huang:2009ue} have computed the bulk viscosity for unpaired strange quark matter in the presence of a magnetic field, which induces an anisotropic bulk viscosity and, for very large
fields, $B\gtrsim 10^{18}\, {\rm G}$, a hydrodynamical instability.

\subsubsection{Color-superconducting quark matter} \label{S:quark_colsup1}

If Cooper pairing between the quarks is taken into account, the derivation outlined above is still valid, but the thermodynamic functions $B$ and $C$ in Eq.\ (\ref{zetaunp}) become different, and of 
course the reaction rates (\ref{lam1}) and (\ref{lam23}) have to be recomputed. Let us first discuss the 2SC phase and its bulk viscosity from the process $u+d\leftrightarrow u+s$, which 
was computed by \citet{Alford:2006gy}.
Since the weak interaction does not change color, there are $N_c\times N_c=9$ subprocesses from the $N_c=3$ possible colors at each of the two 
electroweak vertices of the process $u+d\leftrightarrow u+s$. In the 2SC phase, only if 
the color at both vertices is blue, all participating quarks are ungapped. Therefore, at temperatures much smaller than the 
critical temperature $T_c$, where all processes with at least one gapped quark are exponentially suppressed, the reaction rate in the 2SC phase is 1/9 times the rate
of unpaired quark matter. This does not necessarily mean that the bulk viscosity in the 2SC phase is smaller. It rather means that the maximum of the bulk viscosity, 
where the rate is in resonance with the frequency $\omega$, is assumed at a larger temperature (if the approximation $T\ll T_c$ is still valid at this temperature). The peak 
value of the bulk viscosity in the 2SC phase is different from the unpaired phase because the peak frequency is different and the thermodynamic functions $B$ and $C$ are different. 
It turns out, however, that for typical values of the strange quark mass and the superconducting gap, the peak values are very similar, as we can see in Fig.\ \ref{fig:bulkquarks}.

The bulk viscosity has also been computed in spin-one color superconductors, from the process $u+e \leftrightarrow d + \nu_e$ \cite{Sad:2006egl,Berdermann:2016mwt}, which 
is the dominant one if strange quarks are ignored, and taking into account the non-leptonic process $u+d \leftrightarrow u + s$ \cite{Wang:2010ydb}, whose reaction rate for four different spin-one color superconductors was computed by \citet{Wang:2009if}. The conclusion is very similar as for the 2SC phase: since there are unpaired quarks
in all possible spin-one phases (with the exception mentioned above in the context of neutrino emissivity), one can, for a rough estimate, neglect the contributions of the 
gapped branches, and the reaction rates become, up to a numerical prefactor, the same as for unpaired quark matter. The bulk viscosity thus behaves qualitatively 
similar to 2SC quark matter. 

The CFL phase behaves differently because there are no ungapped quarks that can contribute to chemical re-equilibration processes, and the bulk viscosity is dominated from bosonic 
low-energy degrees of freedom, such as the kaon and the superfluid mode via the processes
\begin{subequations}
\bea
K^0 &\leftrightarrow& \phi + \phi \, , \label{K0phi} \\[2ex]
\phi &\leftrightarrow& \phi + \phi \, . \label{phiphi}
\eea
\end{subequations}
The process involving the neutral kaon has been computed for thermal kaons by \citet{Alford:2007rw}. As mentioned above, the kaons are the lightest 
pseudo-Goldstone modes from chiral symmetry breaking in CFL. If the strange quark mass is taken into account, kaon condensation occurs on top of the 
Cooper pair condensation, giving rise to the so-called CFL-$K^0$ phase \cite{Bedaque:2001je,Kaplan:2001qk}. This phase is the next phase down in density if we
start from the CFL phase at asymptotically large densities and include the effects of the strange quark mass in a systematic way. It is therefore a very important phase and a viable candidate for the interior of neutron stars. The bulk viscosity of the CFL-$K^0$ phase is also dominated by the process (\ref{K0phi}), where $K^0$ now denotes the 
Goldstone mode from kaon condensation \cite{Alford:2008pb,Schmitt:2008zzc}. This Goldstone mode would be exactly massless if strangeness was an exact symmetry. 
Taking into account the effect of the 
weak interactions, one finds a mass of about $50\,{\rm keV}$ for this mode \cite{Son:2001xd}, smaller than the temperatures at which the bulk viscosity of the CFL-$K^0$
phase becomes sizable. Since all the above arguments about the bulk viscosity remain valid, the bulk viscosity of CFL also peaks at a certain temperature. In 
Fig.\ \ref{fig:bulkquarks} we see that this maximum is reached only at temperatures larger than $10\, {\rm MeV}$. This is due to the less efficient reaction (\ref{K0phi})
compared to contributions from quarks.  Therefore, inside neutron stars, the CFL bulk viscosity, with or without kaon condensation, is very small compared to other quark matter phases. One might think that the 
potentially large result for the bulk viscosity at high temperatures is relevant for proto-neutron stars or neutron star mergers, where temperatures may well reach $10\, {\rm MeV}$ or more. However, we expect the 
critical temperatures for kaon condensation \cite{Alford:2007qa} and the critical temperature of CFL itself to be of the order of $10\, {\rm MeV}$, and thus the results beyond 
this temperature have to be taken with care. 

At much lower temperatures, the process  (\ref{phiphi}), which only involves the exactly massless superfluid mode $\phi$, 
is expected to be dominant \cite{Manuel:2007pz}. The result shown in Fig.\ \ref{fig:bulkquarks} for this process should be taken seriously only for temperatures larger than about
$50\,{\rm keV}$, because for smaller temperatures the mean free path is of the order of or larger than the size of the star, indicating that we are no longer in the hydrodynamic
regime.  
Since the CFL phase is a superfluid, there is more than one bulk viscosity coefficient because 
a superfluid at nonzero temperature can be viewed as a two-fluid system, as mentioned in 
Sec.\ \ref{S:cooper}. 
Let us denote the full relativistic stress-energy tensor of a superfluid by $T^{\mu\nu}_{\rm ideal} + T^{\mu\nu}_{\rm diss}$, as we did in Sec.\ \ref{sec:relativistic} for a normal fluid. Then, 
the dissipative terms in first-order hydrodynamics that are usually considered are 
\bea \label{dTmunu}
T^{\mu\nu}_{\rm diss}  &=& \eta\Delta^{\mu\gamma}\Delta^{\nu\delta}\left(\partial_\delta v_\gamma+\partial_\gamma v_\delta - \frac{2}{3}g_{\gamma\delta} \partial\cdot v\right) +\Delta^{\mu\nu}\left[\zeta_1\partial_\gamma\left(\frac{n_s}{\sigma} w^\gamma\right)+\zeta_2\partial\cdot v\right] \nonumber\\[2ex]
&&+\kappa(\Delta^{\mu\gamma} v^\nu +\Delta^{\nu\gamma} v^\mu)[\partial_\gamma T + T(v\cdot\partial)v_\gamma] \, . 
\eea
We have denoted $\Delta^{\mu\nu}=g^{\mu\nu}-v^\mu v^\nu$ with the metric tensor $g^{\mu\nu}={\rm diag}(1,-1,-1,-1)$ and the four-velocity of the normal fluid $v^\mu$. Moreover, $n_s$ is the superfluid density, $w^\mu \equiv \partial^\mu\psi-\mu v^\mu$ with the chemical potential $\mu$ measured in the rest frame of the normal fluid
and $\partial^\mu\psi/\sigma$ the four-velocity of the superfluid with the phase of the condensate $\psi$ and 
the chemical potential measured in the rest frame of the superfluid $\sigma=(\partial_\mu\psi \partial^\mu\psi)^{1/2}$. For a single fluid, $n_s=0$, Eq.\ (\ref{dTmunu})
reduces to the normal-fluid expression (\ref{Tmunudiss}) with $\zeta=\zeta_2$. The Josephson equation, which 
relates the chemical potential $\mu$ to the phase of the condensate, is also modified by dissipative corrections, 
\be
v\cdot \partial\psi = \mu + \zeta_3 \partial_\mu\left(\frac{n_s}{\sigma} w^\mu\right)+\zeta_4\partial\cdot v \, .
\ee
In general, there are even more possible dissipative terms and thus more coefficients in a two-fluid system \cite{Gusakov:2007px}, which are usually neglected. 
As mentioned in Sec.\ \ref{sec:relativistic}, the form given here corresponds to the Eckart frame, where, in contrast to the Landau frame, there is no explicit dissipative correction to the conserved current. 
With $\zeta_4=\zeta_1$ due to the Onsager symmetry principle, there are three independent bulk viscosity coefficients $\zeta_1$, $\zeta_2$, $\zeta_3$ (which have different units), and 
$\zeta_2$ corresponds to $\zeta$ discussed above for a single fluid. 
The bulk viscosity coefficients have been estimated for the process (\ref{phiphi}) in the zero-frequency limit by \citet{Mannarelli:2009ia} with the result
\be
\zeta_1 \sim \frac{m_s^2}{T\mu} \, , \qquad \zeta_2 \sim \frac{m_s^4}{T} \, ,\qquad \zeta_3 \sim \frac{1}{T\mu^2} \, ,
\ee
and for the process (\ref{K0phi}) by \citet{Bierkandt:2011zp}. The bulk viscosity coefficients of CFL have been applied
to the damping of $r$-modes by \citet{Andersson:2010sh}, but, as argued above, the dissipative effects from bulk viscosity in CFL are very small. 
This is not changed by the additional bulk viscosity coefficients from superfluidity.

\subsection{Shear viscosity, thermal and electrical conductivity}
\label{sec:shearquark}

\subsubsection{Unpaired quark matter}\label{S:kincoeff_uqm}

As we have already seen in Sec.\ \ref{S:core_nuclear}, the physics behind shear viscosity of dense matter in neutron stars is different from the physics behind bulk viscosity. 
In the case of shear viscosity, it is thermal, not chemical, re-equilibration and thus for the baryonic and the quark contributions the strong, not the electroweak, interaction becomes relevant. (Recall, however,  that even in the calculation of electroweak processes and chemical
re-equilibration the strong interaction plays a role because the participating quarks interact strongly with each other.) Electrically neutral unpaired quark matter must contain electrons because the strange quark mass induces an imbalance between the number densities of up, down, and strange quarks, and electrons are needed to neutralize the system. We shall discuss the electron contribution in the context of the 2SC phase below, but first focus on the contribution from quarks alone.
If we assume the QCD coupling to be weak, the quasiparticle picture is valid
and we can use a kinetic approach to compute the quark-quark scattering rate from 
one-gluon exchange. This calculation proceeds along the same lines as outlined in Sec.\ \ref{S:core_nuclear_kincoeff}. Here we simply give the final results for shear viscosity $\eta$, thermal conductivity $\kappa$, and electrical conductivity $\sigma$ for unpaired quark matter at low temperatures $T\ll \mu$, computed by \citet{Heiselberg:1993cr},  
\begin{subequations} \label{etakapsigQuarks}
\bea \label{etaunpaired}
\eta^{\rm unp.} &\approx& 4.4\times 10^{-3} \frac{\mu^4 m_D^{2/3}}{\alpha_s^2 T^{5/3}} = 5.5\times 10^{-3}\frac{\mu^4}{\alpha_s^{5/3}T(T/\mu)^{2/3}}\nonumber\\[2ex]
& \approx &2.97 \times 10^{15}
\left(\frac{\mu}{500\, {\rm MeV}}\right)^{14/3}\left(\frac{T}{1\, {\rm MeV}}\right)^{-5/3}\frac{\rm g}{{\rm cm}\,{\rm s}} \, , \\[2ex]
\kappa^{\rm unp.} &\approx& 0.5\frac{m_D^2}{\alpha_s^2} \approx 2.53\times 10^{21}\left(\frac{\mu}{500\, {\rm MeV}}\right)^2\frac{\rm erg}{{\rm cm}\,{\rm s}\, {\rm K}} \\[2ex] 
\sigma^{\rm unp.} &\approx& 0.01\frac{e^2\mu^2m_D^{2/3}}{\alpha_s T^{5/3}} \approx 2.72\times 10^{25}\left(\frac{\mu}{500\, {\rm MeV}}\right)^{8/3}
\left(\frac{T}{1\, {\rm MeV}}\right)^{-5/3}\,{\rm s}^{-1} \, ,
\eea
\end{subequations}
where $m_D^2 = N_f g^2\mu^2/(2\pi^2)$ is the gluon electric screening mass (squared). The results show a similar non-Fermi-liquid behavior as the  leptonic results discussed in Sec.\ \ref{S:core_nuc_lepton} for the same reason: the magnetic interaction that governs the quasiparticles collisions is screened dynamically. For the estimates given here, we have set $\alpha_s\approx 1$ and $N_f=3$. \citet{Jaccarino:2012zz} have performed a numerical comparison of the quark matter shear viscosity to that of nuclear matter.  

It is interesting to compare these results, in particular the shear viscosity, with other QCD calculations and general expectations from strongly coupled systems. The weak-coupling result for the QCD shear viscosity in the opposite limit, $T\gg\mu$, 
is $\eta\approx aT^3/[\alpha_s^2\ln(b/\alpha_s)]$, with numerical coefficients $a$ and $b$ (for massless quarks and $N_c=N_f=3$, $a\approx 1.35$, $b\approx 0.46$) \cite{Huot:2006ys}. 
This result, together with the entropy density $s\propto T^3$ yields a prediction for the 
dimensionless ratio $\eta/s$. This ratio, in turn, is $\eta/s=1/(4\pi)$ for a large class of (infinitely) strongly coupled theories, which can be shown with the help of holographic methods based on the gauge/gravity duality \cite{Kovtun:2003wp}.
Experimental data suggests that the shear viscosity in a quark-gluon plasma created 
in a heavy-ion collision is remarkably close to that value, which is difficult to explain by a naive extrapolation of the weak-coupling result to large values of $\alpha_s$. This has led to the conclusion that heavy-ion collisions produce a quark-gluon plasma which is strongly coupled. We may ask the same question in the context of 
dense QCD: to which extent are we allowed to extrapolate the weak-coupling result 
to more moderate densities present in neutron stars and should we rather be using 
non-perturbative methods? We know that Cooper pairing is one non-perturbative effect, 
which partially answers this question. If we ignore Cooper pairing for now, the entropy density of 
$N_f=N_c=3$ quark matter at low temperatures is $s\approx 2\mu^2 T$, 
and thus with Eq.\ (\ref{etaunpaired}) we find $\eta/s\approx 2.7\times 10^{-3}(\mu/T)^{8/3}\alpha_s^{-5/3}$. Interestingly, this result is qualitatively different 
from the $T\gg\mu$ result because of the appearance of the dimensionless ratio $\mu/T$. 
In particular, $\eta/s$ appears to become large for low temperatures and fixed $\alpha_s$, having no chance to 
approach $1/(4\pi)$, even when we boldly extrapolate to large values of $\alpha_s$. Holographic strong-coupling calculations  at large $N_c$ by \citet{Mateos:2006yd}
suggest that $\eta/s=1/(4\pi)$ does not receive corrections from a baryon chemical potential, although \citet{Myers:2009ij} found a $\mu/T$ dependence for more exotic theories, which 
were compared to and contrasted by Fermi-liquid theory by \citet{Davison:2013uha}. 
Putting these more exotic theories aside, one might be tempted to conclude that weak-coupling transport in dense QCD is even more different from strong-coupling transport
than it is in hot QCD. It is, however, conceivable that $\eta/s=1/(4\pi)$ is not a good benchmark for dense QCD, for instance because of the large-$N_c$ limit that underlies 
this holographic result. In any case, it would be very interesting to go beyond the  weak-coupling calculation of high-density transport properties of quark matter. Since the quasiparticle picture is no longer valid at strong coupling, the 
shear viscosity can then no longer been calculated from a collision integral, and the more general Kubo formalism should be employed, which allows for a general 
spectral density. First steps in this direction have been made by \citet{Iwasaki:2007iv,0954-3899-36-11-115012,Lang:2015nca,Harutyunyan:2017ttz},
who used this formalism to compute shear viscosity and thermal conductivity of quark matter at finite $T$ and $\mu$. These calculations where performed within the Nambu--Jona-Lasinio model and for temperatures larger 
than relevant for neutron stars (a calculation within the same model, but using the Boltzmann approach, 
was performed recently by \citet{Deb:2016myz}). 

\subsubsection{Color-superconducting quark matter}\label{S:quark_colsup}

The results (\ref{etakapsigQuarks}) were extended to the 2SC phase (without strange quarks) by \citet{Alford:2014doa}, also in the weak-coupling regime. In the 2SC phase, no global symmetry is broken and thus there are no Goldstone modes. Therefore, we expect the main contribution to 
come from ungapped fermionic modes. Besides the effect on the fermionic modes, we now have to 
take into account the effect of pairing on the gauge bosons, as discussed in general terms in Sec.\ \ref{S:cooper} and 
for lepton transport in nuclear matter in Sec.~\ref{S:core_nuc_pairing}. Remember that screening determines the range of the interaction, and weaker screening results in a more efficient relaxation mechanism. In the 2SC phase, different gluons are screened differently, depending on whether they couple to the unpaired blue quarks or to the paired red and green quarks. Let us first discuss the static screening of the gauge bosons. 
The three gluons which only see red and green quarks -- 
corresponding to the three generators $T_{1-3}$ of the $SU(3)$ gauge group -- are neither magnetically screened (as in unpaired quark matter) nor electrically screened (unlike in unpaired quark matter).  This is because the red and green charges are all confined in Cooper pairs and the Cooper pairs themselves carry color charge anti-blue. 
The gluons corresponding to $T_{4-7}$ acquire a Meissner mass and also are electrically screened. Since the relevant interactions 
with gluons $T_{1-7}$ involve at least one red or green quark, they do not matter for the shear viscosity at temperatures much smaller than the gap, $T\lesssim 10\, {\rm MeV}$. 
It remains the 8th gluon and the photon. Their behavior is complicated because they mix, and in the 2SC phase they do so differently in electric and magnetic sectors  \cite{Schmitt:2003aa}. In the electric sector, there is a screened gluon $T_8$, and a screened photon corresponding to the generator $Q$ of the electromagnetic gauge group $U(1)$. In the magnetic sector, there is a screened gluon $\tilde{T}_8$ (with a small admixture of the photon), and an {\it unscreened} photon $\tilde{Q}$ (with a small admixture of the 8th gluon). Because of this rotated photon with zero magnetic screening mass, the 2SC phase is not an electromagnetic superconductor, i.e., does not show an electromagnetic Meissner effect. It does show a {\it color} Meissner effect for 5 out of the 8 gluons. (The CFL phase also has such a magnetically unscreened
rotated photon and is thus no electromagnetic superconductor either; in CFL quark matter all 8 gluons -- one of them having a small admixture of the photon -- 
acquire a Meissner mass.) The rotated photon is screened dynamically in the form of Landau damping like the ordinary transverse photon in nuclear matter (Sec.~\ref{S:core_nuc_lepton}) or the transverse gluons in unpaired quark matter of the previous subsection.

\begin{table}[t]
\begin{center}
\begin{tabular}{|c||c|c|c|c|} 
\hline
\rule[-1.5ex]{0em}{4ex} 
  & $T_8$ & $Q$ & $\tilde{T}_8$ & $\tilde{Q}$  \\[1ex] \hline\hline
\rule[-1.5ex]{0em}{6ex} 
& \multicolumn{2}{c}{$\;\;$electric screening$\;\;$}\vline & \multicolumn{2}{c}{$\;\;$magnetic screening$\;\;$} \vline \\[2ex] \hline
\rule[-1.5ex]{0em}{6ex} 
$\;\;$screening mass $\;\;$& $3g^2$ & $2e^2$& $g^2/3$ &0 \\[2ex] \hline\hline
\rule[-1.5ex]{0em}{6ex} 
blue up & $\;\;-g/\sqrt{3}\;\;$ & $\;\;2e/3\;\;$ & $-g/\sqrt{3}$ & $e$ \\[2ex] \hline
\rule[-1.5ex]{0em}{6ex} 
blue down & $-g/\sqrt{3}$ & $-e/3$ & $-g/\sqrt{3}$ & 0 \\[2ex] \hline
\rule[-1.5ex]{0em}{6ex} 
electron  & 0 & $-e$ & $\;\;-e^2/(g\sqrt{3})\;\;$ &$\;\;-e\;\;$ \\[2ex] \hline
\end{tabular}
\caption{Static 2SC screening masses (squared) of the eighth gluon $T_8$ and the photon $Q$, in units of $\mu^2/(3\pi^2)$ (from Ref.~\cite{Schmitt:2003aa}, with $N_f=2$),
and charges of the unpaired fermions in the 2SC phase, 
 assuming the strong coupling constant to be much larger than the electromagnetic coupling $g\gg e$.  The unpaired fermions dominate the transport properties, and at sufficiently small temperatures shear viscosity, thermal and electrical conductivities are 
dominated by the 
blue down quark because it does not couple to the only unscreened gauge boson, the rotated photon $\tilde{Q}$. (Here the electric charge is given in Heaviside-Lorentz units, such that $e^2=4\pi\alpha_f$.)
 }
\label{table2SC}
\end{center}
\end{table}

The dominant contribution to the shear viscosity comes from unpaired fermions: blue quarks and electrons. We collect their charges with respect to the 8th gluon and the photon 
(and their rotated versions in the magnetic sector) in Table \ref{table2SC}. Since the rotated photon is weakly screened, it provides the dominant contribution to the collision frequencies, effectively suppressing the relaxation times and hence the transport coefficients for the species which interact via $\tilde{Q}$. The only unpaired fermion that 
does not couple to $\tilde{Q}$
is the blue down ($bd$) quark. 
Therefore -- and although it interacts via the strong interaction -- it has the longest relaxation time and gives the dominant contribution to the shear viscosity at sufficiently small temperatures \cite{Alford:2014doa},
\be   \label{eta2SC}
\eta_{bd}^{\rm 2SC} \approx 2.3 \times 10^{-3}\frac{\mu^4}{\alpha_s^{3/2}T(T/\mu)} \, .
\ee
This result is qualitatively different from the unpaired result (\ref{etaunpaired}) because in unpaired quark matter all quarks experience unscreened magnetic interactions. In that case, 
the electron contribution becomes important as well, for a short discussion see Ref.\ \cite{Alford:2014doa}. 
 
At larger temperatures, 
dynamical screening of $\tilde{Q}$ becomes stronger [recall Eq.~(\ref{eq:Pi_T_nonsf})] and the interaction via the rotated magnetic photon no longer dominates over the interaction via the screened gauge bosons. As a consequence, electrons become dominant and 
the result (\ref{eta2SC}) is no longer valid. In fact, it is only valid at very small temperatures, $T/\mu\sim 10^{-5}$. (The contribution of blue up quarks is never important since they have smaller Fermi momentum that blue down quarks.)
 For the numerical results for all temperatures, including
thermal and electrical conductivities, see Ref.\ \cite{Alford:2014doa}. These results show in particular that for the thermal conductivity the transition from quark-dominated to electron-dominated regime occurs at a much higher temperature than for shear viscosity, not unlike the competition of lepton and nucleon contributions to $\kappa$ and $\eta$ in nuclear matter.
The electrical conductivity has also been computed close to the 
critical temperature and taking into account an external magnetic field by \citet{Kerbikov:2014ofa}.

If the mismatch between the up and down quark Fermi momenta is large, isotropic pairing is no longer possible. The red and green quarks that participate in Cooper pairing
then develop ungapped quasiparticle excitations in certain regions in momentum space. Their contribution to the shear viscosity, which is dominated by transverse, Landau damped 
gluons $T_{1-3}$, has been computed for the (anisotropic, but not crystalline) Fulde-Ferrell phase by \citet{Sarkar:2016gib}. The result is small compared to the 
contribution of the completely unpaired blue quarks, but elements of the calculation may be transferred in future studies to LOFF phases in CFL, where there are no completely unpaired
quarks, only few electrons, and all non-abelian gauge bosons have nonzero 
electric and magnetic screening masses.
While this calculation has not been done yet, we briefly review the results from the Goldstone modes in the 
pure (isotropic) CFL phase. The calculation of the contribution from the superfluid mode is based on the effective Lagrangian 
\bea
{\cal L} &=&  \frac{3(D_\mu\psi D^\mu\psi)^2}{4\pi^2} \\[2ex]
&=& \frac{1}{2}[(\partial_0\phi)^2-v^2(\nabla\phi)^2]-\frac{\pi}{9\mu^2}\partial_0\phi(\partial_\mu\phi\partial^\mu\phi)
+\frac{\pi^2}{108\mu^4}(\partial_\mu\phi\partial^\mu\phi)^2 +\ldots \, ,
\eea
where $\psi$ is the phase of the condensate introduced below Eq.\ (\ref{dTmunu}), the covariant derivative acting on this phase is $D_\mu\psi =\partial_\mu\psi -A_\mu$ 
with $A_\mu=(\mu,0)$, the rescaled field of the superfluid mode is $\phi=3\mu\psi/\pi$, the velocity of the Goldstone mode is $v=1/\sqrt{3}$, and we have dropped the terms linear and constant in $\phi$ in the second line. The result from $\phi + \phi \leftrightarrow \phi + \phi $ scattering for the shear viscosity was computed by \citet{Manuel:2004iv},
\be \label{etaphiCFL}
\eta_{\phi}^{\rm CFL}\approx 1.3\times 10^{-4}\frac{\mu^4}{T(T/\mu)^4} \approx  6.96 \times 10^{22}
\left(\frac{\mu}{500\, {\rm MeV}}\right)^{8}\left(\frac{T}{1\, {\rm MeV}}\right)^{-5}\frac{\rm g}{{\rm cm}\,{\rm s}} \, .
\ee
\citet{Alford:2009jm} calculated the contribution of kaon scattering $K^0 + K^0 \leftrightarrow K^0 + K^0$ to shear viscosity in the CFL-$K^0$ phase, where the relevant excitation is the Goldstone mode $K^0$ from kaon condensation. It was found that this contribution 
is smaller than that of the superfluid mode $\phi$. However, the relevant mean free path (the `shear mean free path') of the phonons becomes of the order of or larger than the radius of the star 
at temperatures lower than about $1\,{\rm MeV}$. (In this ballistic regime, 
an `effective shear viscosity'  can be  induced from shear stresses at the boundary of the system, for instance in superfluid cold atoms in an optical trap \cite{Mannarelli:2012eg,Mannarelli:2013hm}.) 
This is not the case for the kaons, which therefore may provide the dominant contribution to shear viscosity in this regime. 

The thermal conductivity of CFL due to phonons was obtained from a simple mean free path estimate by \citet{Shovkovy:2002kv}. Later, \citet{Braby:2009dw} made this estimate  
more precise by a calculation within kinetic theory, and it was found  
\be\label{kappaCFL}
\kappa_{\phi}^{\rm CFL} \gtrsim  4.01 \times 10^{-2} \frac{\mu^8}{\Delta^6} \approx 1.04\times 10^{26}\left(\frac{\mu}{500\, {\rm MeV}}\right)^8\left(\frac{\Delta}{50\, {\rm MeV}}\right)^{-6}\frac{\rm erg}{{\rm cm}\,{\rm s}\, {\rm K}} \, .
\ee
This large thermal conductivity suggests that a CFL quark matter core of a neutron star becomes isothermal within a few seconds \cite{Braby:2009dw}.
In addition, \citet{Braby:2009dw} also computed the kaon contribution. This was done in the CFL, not the 
CFL-$K^0$, phase, i.e., from a massive kaon $m_{K^0}\sim 10 \, {\rm MeV}$ instead of the (approximately) massless Goldstone kaon in the CFL-$K^0$ phase, which was used in the 
calculation for the shear viscosity we just mentioned. It was found that the contribution from the kaons for typical parameter values is much smaller than the phonon contribution (\ref{kappaCFL}). Neither for the shear viscosity nor for the thermal conductivity, scattering processes due to interactions between the superfluid mode and the kaon have been 
taken into account so far.    

The shear viscosity of spin-1 color superconductors has not yet been computed. In most phases, the dominant contribution can be expected to come from unpaired quarks, and the
calculation would be similar as for instance in the 2SC phase, with possible complications from anisotropies and ungapped directions in momentum space, like in the 
case of the Fulde-Ferrell calculation mentioned above. Only in the fully gapped version of the 
CSL phase (which seems to be disfavored, at least at weak coupling, as mentioned above) Goldstone modes would become important. An effective theory for the massless modes has been worked out by \citet{Pang:2010wk}, which can be used to 
compute the shear viscosity in CSL quark matter.

\subsection{Axial anomaly in neutron stars}

\subsubsection{Anomaly-induced transport}

Transport in the presence of a chiral imbalance, i.e., in systems where there are more left-handed than right-handed fermions or 
vice versa, is qualitatively different from `usual' transport. The reason is the chiral anomaly, which leads to the non-conservation of 
the axial current due to quantum effects. Anomaly-induced transport (or short: anomalous 
transport\footnote{The term `anomalous transport' is used in various contexts with different meaning, for instance in plasma physics, where it refers to unusual diffusion behavior and has nothing to do with the quantum anomaly. Confusion can be avoided by using the more cumbersome, but less ambiguous, `anomaly-induced transport'. Also `chiral 
transport' or `anomalous chiral transport' is sometimes used.}) has been discussed extensively in the recent literature, with applications 
in a multitude of different systems, reviewed recently in a pedagogical article by \citet{Landsteiner:2016led}. One prominent manifestation 
of anomalous  transport is the `chiral magnetic effect', where a dissipationless electric current is induced in the direction of a background  
magnetic field. This effect has been predicted to occur in  non-central heavy-ion collisions \cite{Kharzeev:2007jp}, where large magnetic fields
are created and where a chiral imbalance can be generated by fluctuations of the gluon fields through the QCD anomaly (while
the anomaly of Quantum Electrodynamics (QED) then provides the mechanism for the creation of the electric current). Signatures of the chiral magnetic effect 
have been seen in the data, although the interpretation still leaves room for alternative explanations \cite{Kharzeev:2015znc}. 
An unambiguous manifestation of the chiral magnetic effect has been observed in so-called Weyl semi-metals, which exhibit  chiral quasiparticles \cite{2016NatPh..12..550L}. The chiral magnetic effect is one example among various anomaly-induced 
phenomena. Others are the `chiral vortical effect', where the role of the magnetic field is played by a nonzero vorticity, and the 
`chiral separation effect', where the role of the 
difference in left- and right-handed fermion densities is played by their sum and an axial current, not a vector current, is generated. 
In a hydrodynamic formulation, anomalous effects generate additional terms with new --  `anomalous' -- transport coefficients
\cite{Son:2009tf}. Also an anomalous version of kinetic theory has been formulated \cite{Stephanov:2012ki,Son:2012zy}.

It is natural to ask whether anomalous transport plays a role in dense matter and whether it has observable consequences 
for neutron stars. Sizable effects can only come from massless or very light particles because 
a mass breaks chiral symmetry explicitly and thus tends to suppress any effects from the chiral anomaly. It has been 
suggested by \citet{Ohnishi:2014uea} that a dynamical instability (`chiral plasma instability') due to the chiral magnetic effect for electrons occurs in 
core collapse supernovae, possibly producing the very strong magnetic fields in magnetars. 
However, the chiral imbalance for electrons created from the electron capture process is completely washed out by the nonzero electron 
mass \cite{Grabowska:2014efa,Kaplan:2016drz}, although one might naively think that this mass is negligible in the astrophysical context. 
The instability may nevertheless be realized if the electrons experience instead an effective chiral chemical potential from the fluid helicity generated in the neutrino gas through the chiral vortical effect  \cite{Yamamoto:2015gzz}. 
It has also been suggested that 
pulsar kicks originate from chiral imbalance in leptons, either from electrons, which however would require a very small crust
(possibly in quark stars) \cite{Charbonneau:2009ax,Charbonneau:2009hq}, or from neutrinos due to the chiral separation effect from the magnetic field, treating electrons and neutrinos as a single fluid \cite{Kaminski:2014jda}. One may also ask whether 
anomalous transport of neutrinos has an effect on the dynamics or even the very existence of core-collapse supernova explosions. This question is motivated by the different behavior of a chiral fluid with respect to magnetohydrodynamic turbulence, pointed out by  \citet{Yamamoto:2015gzz} and \citet{Pavlovic:2016gac}. These studies have only begun recently, and it remains to be seen whether (proto-)neutron stars or
supernova explosions, maybe also neutron star mergers and the hyper-massive neutron stars resulting from them, provide yet another   
system where effects of the quantum anomaly become manifest on macroscopic scales. 

\subsubsection{Axions}

Another anomaly-related effect with relevance to neutron stars, now specifically from the QCD anomaly, is the existence of axions. Axions, which are a promising hypothetical candidate for cold dark matter, arise from the most natural solution to the so-called strong CP problem: the axial anomaly effectively -- via axial rotations -- induces a CP-violating term proportional to $G_{\mu\nu}\tilde{G}^{\mu\nu}$ to the QCD Lagrangian, where $G_{\mu\nu}$ and $\tilde{G}^{\mu\nu}$ are the gluon field strength tensor and its dual. We know that the prefactor of this term, which is an angle $\theta\in[-\pi,\pi]$, must be extremely small because of very tight experimental constraints on the electric dipole moment of the neutron. Rather than viewing $\theta$ as a parameter, whose smallness then would be very 
difficult to understand, \citet{Peccei:1977hh} have suggested a dynamical mechanism that leads to extremely small values for $\theta$. This mechanism is based on the spontaneous breaking of a global anomalous $U(1)$ symmetry, and the axion is the corresponding (not exactly massless) Goldstone mode 
\cite{Weinberg:1977ma,Wilczek:1977pj}. The exact implementation of this mechanism in the Standard Model leaves room for different models, which essentially fall into two classes, 
introduced by \citet{kim1979weak,shifman1980can} on the one hand and \citet{dine1981simple,Zhitnitsky:1980tq} on the other hand.
Although axions are expected to couple to electrons, photons, and nucleons, so far a positive signal for the axion or 
any axion-like particle has remained elusive in experimental searches. Constraints on the coupling strengths (and thus on the axion mass) are obtained for instance from cooling of white dwarfs, from cosmology, and from solar physics \cite{Raffelt:2006cw}. In addition, supernova explosions \cite{Keil:1996ju} and the cooling of neutron stars \cite{Umeda:1997da} can potentially contribute to these constraints. 
To this end, the reaction rates for axions in a nuclear medium have to be
calculated. Many of these calculations are analogous to the calculations of the neutrino reaction rates reviewed 
in Sec.\ \ref{sec:urca} and share the same problems and uncertainties. Axions can be emitted from bremsstrahlung in electron scattering processes from ions in the crust \cite{Iwamoto:1984ir}, or 
from bremsstrahlung in nucleon-nucleon collisions $N+N\to N +N+a$ in the core, where $N$ can be a neutron or a proton and $a$ is the axion. The calculation of the latter process 
involves knowledge of the strong interaction between the nucleons, just like the analogous neutrino-emitting process $N+N\to N +N+\nu + \bar{\nu}$ and like the modified Urca process. Therefore, the rate 
contains significant uncertainties. It was first computed using the one-pion exchange interaction for neutrons by \citet{Iwamoto:1984ir} and later extended to the cases that involve
protons \cite{Iwamoto:1992jp,2009NuPhA.828..439S}. These results are expected to present upper limits since medium corrections to the interactions are likely to reduce the rates
\cite{Keil:1996ju,Hanhart2001PhLB,2016PhRvD..94h5012F}.
Recently, \citet{Keller:2012yr} computed the axion emissivity for superfluid nuclear matter from pair breaking and formation processes. In unpaired quark matter, 
the rate from the analogous process $q+q\to q+q+a$, where $q$ is an $u$, $d$, or $s$ quark has been computed by \citet{Anand:1990az}, using one-gluon exchange for the quark interaction.  
The axion emissivities can be used to study numerically the axion contribution to the cooling of neutron stars. Such a simulation is naturally prone to large uncertainties, 
but conservative estimates yield an upper bound for the axion mass of the order of  $0.1\,{\rm eV}$ \cite{Umeda:1997da,Sedrakian:2015krq}, consistent with limits set by the 
direct neutrino detection from supernova SN 1987A.

\section{Outlook}

We have seen that understanding transport in neutron star matter requires a variety of different techniques and theoretical results -- sometimes even if we ask for the explanation of a single, specific astrophysical observation. Current efforts combine nuclear and particle physics with elements of condensed matter and solid state physics, using and developing methods from 
hydrodynamics, kinetic theory, many-body physics, quantum field theory, and general relativity. In many ways, neutron stars are a unique laboratory, with matter under more extreme conditions than 
anywhere else. This laboratory is far away from us and we seem to have very limited access to 
the matter deep inside the star. It is thus easy to get discouraged regarding
precise tests of the transport properties that we predict theoretically. Nevertheless, as we have pointed out, transport properties do provide us with an important tool to interpret astrophysical data and eventually answer the question about what the interior of the neutron star is made of. 
And, most importantly for future studies, the current exciting results from gravitational-wave astronomy promise more, and more precise, data for the near future, especially if combined with 
electromagnetic signals as for the recently observed neutron star merger event \cite{PhysRevLett.119.161101,2041-8205-848-2-L12}. 
Neutron star mergers are sensitive to both the equation of state and transport of (relatively hot) 
dense matter. Moreover, a possible future detection of gravitational waves from 
isolated neutron stars \cite{Glampedakis2017}  would be another spectacular testing ground for transport in 
ultra-dense matter. The reason is that potential sources such as 
the $r$-mode instability and 
a sustained ellipticity of the star are intricately linked to transport properties, such 
as viscous effects and the formation and evolution of magnetic flux tube arrays. 

Throughout the review we have pointed out open questions and unsolved problems. Many of them are 
inevitably related to our limited quantitative grip on the strong interaction, i.e., on QCD at 
baryon densities significantly larger than nuclear saturation density. This concerns for example the modified Urca process or shear viscosity of ultra-dense matter, be it nuclear or quark matter. First-principle QCD calculations on the lattice exist for thermodynamic quantities at zero baryon density, and there are some promising attempts to extend these calculations, firstly, to finite baryon densities and, secondly, to transport properties. Nevertheless, both extensions are extremely difficult, let alone implementing  them simultaneously. Therefore, in the foreseeable future, the input from the strong interaction to transport properties of dense matter will most likely not go beyond the use of effective theories, phenomenological models, or extrapolations from perturbative calculations. 

Other open problems that we have mentioned are related to transport in a magnetic field and transport in the presence of Cooper pairing. This concerns for instance microscopic calculations of transport 
properties in the crust and the inhomogeneous nuclear pasta phases, which obviously become very cumbersome through the anisotropy induced by a magnetic field. It also concerns more macroscopic magnetohydrodynamic studies (which we did not discuss in detail), which currently do not yield a satisfactory picture of the magnetic field evolution if compared to observational data. Also in the case of Cooper pairing, microscopic calculations become much more complicated, and we have pointed out various approaches and approximations used for that case. Transport properties of many possible phases, in particular a 
large part of the multitude of possible color-superconducting phases, have already been discussed in the literature and significant progress has been made. Challenges for future studies 
are for instance the nature of interfaces between superfluid and superconducting phases, especially if there are rotational vortices and/or magnetic flux tubes, possibly even color-magnetic flux tubes in quark matter. Again, also more macroscopic studies are difficult, and 
many things remain to be understood, for instance multi-fluid effects due to nonzero temperatures, or the time evolution of rotating superfluids, say the neutron superfluid in the crust or the color-flavor locked phase in a possible quark matter core. 

Finally, let us emphasize the need of cross-disciplinary approaches for future efforts in the 
field of transport theory of dense matter. It is obvious that theoretical studies from nuclear and particle physics have to 
be combined with observational astrophysics. Maybe less obvious are parallels to other fields that 
deal with strongly coupled systems where transport properties can be measured. For instance, 
transport in unitary atomic Fermi gases has been studied in detail, including effects of superfluidity. One example is the study of critical velocities in two-component superfluids \cite{2015PhRvL.115z5303D}, which is of possible relevance to superfluid neutron star matter. It is even conceivable that future experiments with cold atoms can be `designed' to mimic, at least qualitatively, effects that we expect in neutron stars, such as unpinning of vortices from a lattice structure. Also experiments with more traditional superfluids such as liquid helium might 
shed some light on questions we encounter in neutron stars \cite{Graber:2016imq}. Transport also plays a prominent role in relativistic heavy-ion collisions, which provide a laboratory for strongly interacting matter
at larger temperatures and lower baryon densities. Future experiments aim, in fact, at increasing the densities in these collisions, possibly reaching beyond nuclear saturation density \cite{Friman:2011zz,refId0}. In any case, 
heavy-ion collisions raise various interesting fundamental questions about (relativistic)
hydrodynamics and its regime of applicability, and we can imagine that insights gained in these studies might, even if not being directly applicable, give interesting input and pose 
relevant questions also in the context of neutron stars.

\begin{acknowledgments}
We thank M.\ Alford, A.\ Chugunov, A.\ Kaminker, A.\ Rebhan, S.\ Reddy, A.\ Sedrakian, I.\ Shovkovy, and D.\ Yakovlev for useful comments and discussions and acknowledge support from the NewCompStar network, COST Action MP1304. A.S.\ is supported by the Science \& Technology Facilities Council (STFC) in the form of an Ernest Rutherford Fellowship. P.S.\ is supported by the Foundation for the Advancement of Theoretical Physics and Mathematics ``BASIS'' and the Russian Foundation for Basic Research grant \# 16-32-00507-mol-a.
\end{acknowledgments}

\bibliography{NStransp}

\end{document}